\documentclass[10pt]{scrartcl}

\usepackage[utf8]{inputenc}
\usepackage[margin=13mm]{geometry}
\usepackage[svgnames]{xcolor}

\usepackage{mathtools,tensor,amssymb,multicol}

\usepackage[colorlinks=true,allcolors=DarkBlue]{hyperref}
\usepackage{cleveref}

\usepackage[backend=biber,sorting=none,citestyle=numeric-comp,giveninits=true,doi=false]{biblatex}

\renewbibmacro{in:}{}
\AtEveryBibitem{\clearfield{number}}

\DeclareFieldFormat[article]{volume}{\mkbibbold{#1}}
\DeclareFieldFormat{pages}{#1}
\DeclareFieldFormat{eprint:arxiv}{\href{http://arxiv.org/abs/#1}{#1}}

\renewcommand{\vec}[1]{\boldsymbol{#1}}
\DeclareMathOperator{\sign}{sign}
\DeclareMathOperator{\Tr}{Tr}
\DeclareMathOperator{\diag}{diag}
\DeclareMathOperator{\transpose}{T}
\renewenvironment{eqnarray}{\begin{equation}\begin{aligned}}{\end{aligned}\end{equation}\ignorespacesafterend}

\setkomafont{section}{\large}
\setkomafont{subsection}{\normalsize}
\begin{document}

\title{Quantum correlations for the metric}
\author{C. Wetterich}
\date{\small\textit{Institut für Theoretische Physik, Universität Heidelberg, Philosophenweg 16, D-69120 Heidelberg\\[1ex]
\href{mailto:c.wetterich@thphys.uni-heidelberg.de}{c.wetterich@thphys.uni-heidelberg.de}}}
\maketitle

\begin{abstract}
    \small\noindent We discuss the correlation function for the metric for homogeneous and isotropic cosmologies. The exact propagator equation determines the correlation function as the inverse of the second functional derivative of the quantum effective action. This formulation relates the metric correlation function employed in quantum gravity computations to cosmological observables as the graviton power spectrum. In the Einstein-Hilbert approximation for the effective action the on-shell graviton correlation function can be obtained equivalently from a product of mode functions which solve the linearized Einstein equations. In contrast, the product of mode functions, often employed in the context of cosmology, does not yield the correlation function for the vector and scalar components of the metric fluctuations. We divide the metric fluctuations into ``physical fluctuations'', which couple to a conserved energy momentum tensor, and gauge fluctuations. On the subspace of physical metric fluctuations the relation to physical sources becomes invertible, such that the effective action and its relation to correlation functions no longer needs to involve a gauge fixing term. The physical metric fluctuations have a similar status as the Bardeen potentials, while being formulated in a covariant way. We compute the effective action for the physical metric fluctuations for geometries corresponding to realistic cosmologies.
\end{abstract}

\vfil

\begin{multicols}{2}

{\hypersetup{linkcolor=black}
\small\tableofcontents}
\end{multicols}

\newpage

\begin{multicols}{2}

\section{Introduction}

The correlation function for the metric is a central quantity in classical and quantum gravity. It permits to compute the (linear) response of the metric to a source, e.g. a moving body. The equal time correlation function contains the information on the power spectrum of fluctuations in the Gaussian approximation. The (two-point) correlation function or propagator for the metric plays also a central role for any computation in quantum gravity. A typical loop contribution from the metric fluctuations involves a trace over powers of the metric propagator, with appropriate vertices inserted. Within functional renormalization the exact flow equation for the effective average action $\Gamma_k$,
\begin{equation}\label{eqn:I1}
	\partial_k\Gamma_k
	= \frac{1}{2} \Tr\bigl\{(\partial_k R_k)G_k\bigr\},
\end{equation}
involves the propagator $G_k$ in presence of the infrared cutoff $R_k$ \cite{CWFE,CWMR,MR}.

While the metric correlation in flat space can be computed rather easily for a simple form of the effective action, much less is known for the metric correlation in curved space. For the linear response of the metric to some sources as galaxies, stars or other moving bodies one needs the metric propagator in some ``background cosmology'', i.e. for an appropriate homogeneous and isotropic solution of the gravitational field equations.The same holds for the determination of the fluctuation spectrum. The metric correlation in a background is needed if one wants to explore the dependence of the effective action on the metric, e.g. in \cref{eqn:I1}. Particularly interesting are backgrounds that obey the field equations.

Indeed, functional methods in quantum field theory work best if the ``background field'', for which expressions as \labelcref{eqn:I1} are evaluated, is close to an appropriate extremum of the action. For example, the functional renormalization group for scalar fields gives excellent results in simple truncations if one expands around the minimum of the effective potential \cite{CWFRG1,CWFRG2,CWFRG3,CWFRG4,AOK}. In contrast, expansions with few couplings around the origin in field space, $\varphi = 0$, fail to provide good results in case of spontaneous symmetry breaking. For quantum gravity computations often only a few couplings are kept. One therefore would like to evaluate the effective action in the vicinity of characteristic solutions of the cosmological field equations. Typically, these may be geometries close to de Sitter space as relevant for inflation. This is also the region in field space for which knowledge of the effective action $\Gamma$ is most useful. The exact field equations follow from the first functional variation of $\Gamma$ and employ therefore knowledge about its form in the vicinity of the relevant solution.

In gravity, the metric propagator evaluated for a background that solves the field equations (``on shell propagator'') shows particular properties that do not hold for general background geometries. One may define gauge invariant fluctuation quantities (Bardeen potentials \cite{Bar}). If the background obeys the field equations only the graviton mode corresponds to a propagating wave or particle, whereas the gauge invariant scalar and vector modes contained in the metric play the role of ``auxiliary fields'' that do not describe propagating waves or particles. Nevertheless the correlation function for the scalar and vector parts of the physical metric fluctuations does not vanish.

The consequences of the non-propagating character of scalar and vector modes for quantum gravity calculations are not much explored. The auxiliary field property does not hold away from solutions of the field equations. It is not easily visible on the level of the metric fluctuations in general covariant gauges, for which the difference between background geometries obeying the field equations or not is not very apparent at first sight. We clarify in this paper the relation between the non-propagating Bardeen potentials and the non-trivial metric correlation function in the vector and scalar sector.

More generally, it is the aim of the present paper to constitute a bridge between concepts typically used in cosmology, such as mode functions and particular assumptions about ``vacua'' on one side, and functional integral approaches for a quantum field theory of gravity on the other. For background geometries obeying the field equations the metric correlation contains directly the information about the power spectrum of cosmic fluctuations. For example, the amplitude and spectrum of the tensor fluctuations can be extracted directly from the equal time correlation function for the graviton component of the metric. The situation is similar for an additional scalar (inflaton) field for which amplitude and spectrum can be obtained from the (gauge invariant) scalar correlation. Since substantial work has been invested in the computation of the cosmic fluctuation spectrum for various interesting cosmological solutions \cite{MUK,STA,GP,BST,AW,STAT,RSV}, one may use this knowledge in order to gain information about the metric correlation for realistic cosmological solutions. In the other direction, a computation of the quantum correlation for the metric translates directly to important cosmological observables.

While the connection between existing computations of the cosmic fluctuation spectrum and the metric correlation is rather direct for the propagating graviton fluctuations (or an additional inflaton), this is no longer the case for the scalar and vector modes contained in the metric. First of all, the standard approach of using commutation relations for operators of free fields for the definition of a ``vacuum correlation'' is only meaningful for the fields describing propagating waves or particles. Second, the linearized field equations (mode equations) admit for the gauge invariant scalar fields or gravitational potentials $\Phi,\Psi$ (Bardeen potentials) only the solution $\Phi = \Psi = 0$ in the absence of additional matter fluctuations. The usual prescription for obtaining the correlation function as a product of mode functions (solutions of the mode equation) would then imply that the metric correlation in the scalar sector vanishes. This is, however, not the case.

For the gauge invariant vector fluctuation $\Omega_m$ the situation is similar. The only solution of the mode equation is $\Omega_m = 0$, while we find a rotation invariant correlation function in Fourier space
\begin{eqnarray}\label{eqn:I2}
	\langle\Omega^\ast_n(\vec{k},\eta)\Omega_m(\vec{k},\eta^\prime)\rangle_c
	= \frac{2i\delta(\eta-\eta^\prime)}{M^2 k^2 a^2(\eta)} \bigl(\delta_{mn}-\tfrac{k_m k_n}{k^2}\bigr)\\
	\times (2\pi)^3\delta^3(\vec{k}-\vec{k}^\prime).
\end{eqnarray}
Here $\eta$ is conformal time, $a(\eta)$ the scalar factor, $M$ the Planck mass and $\vec{k}$ the comoving wave number. This correlation function is ``instantaneous'', i.e. $\sim \delta(\eta-\eta^\prime)$, and reflects the role of $\Omega_m$ as an auxiliary field. It cannot be written as a product of mode functions.

The different properties of correlation functions in the graviton sector on one side, and the vector and scalar sector on the other side, seem related to the difference between ``propagating'' and ``auxiliary'' fields in the operator formalism. While the computation of the correlation functions for the propagating tensor mode is rather straightforward in the operator formalism, a computation for the scalar and vector correlations is presumably a rather involved exercise in this formalism. (See ref. \cite{AM} for structural aspects.)

For the computation of the metric correlation or Green's function we need a method that goes beyond mode functions for free quantum fields. We will directly employ the defining equation for the Green's function $G$
\begin{equation}\label{eqn:I3}
	\Gamma^{(2)} \, G
	= E,
\end{equation}
with $\Gamma^{(2)}$ involving a suitable differential operator and $E$ the unit matrix in the appropriate space of fields. Here $G$ is considered as a matrix with internal and space or momentum indices, and similar for $\Gamma^{(2)}$. For the quantum effective action $\Gamma$ the matrix $\Gamma^{(2)}$ is the second functional derivative, and \cref{eqn:I3} is an exact identity that follows from the basic definition of the effective action. The use of this identity for the computation of primordial cosmic fluctuations has been demonstrated in refs. \cite{CW2,CWneu,CWPFVG}. One recovers known results as special solutions, but also can discuss the most general solution as an initial value problem for a differential equation. Starting from the defining equation \labelcref{eqn:I3} we will discuss the conditions under which the correlation function can be represented as a product of mode functions.

The metric fluctuations around a given background metric can be divided into physical and gauge fluctuations. Only the propagator for the physical metric fluctuations matters for the response of the metric to a covariantly conserved energy-momentum tensor. Similarly, only the correlation function for the physical metric fluctuations leads to observable quantities such as the primordial fluctuation spectrum. An important aspect of the present paper is the clear separation between physical and gauge fluctuations of the metric. This can be achieved by imposing a constraint on the metric fluctuations which eliminates the gauge fluctuations. Alternatively, one can employ a particular ``physical'' gauge fixing.

This paper is organized as follows: we present the basic concepts in \cref{sec:basic concepts}. \Cref{sec:physical and gauge metric} deals with the distinction between physical metric fluctuations and gauge fluctuations. In \cref{sec:qea}, we introduce the quantum effective action for the physical metric fluctuations. It contains all the information needed for the computation of the quantum field equations and the correlations for physical metric fluctuations. \Cref{sec:correlation function} turns to the correlation function for the metric and the defining propagator equation. The on-shell metric correlation in flat space is addressed in \cref{sec:propagator in flat space}. This demonstrates several issues as projectors onto physical modes, irreducible representations of the rotation group, connection to Bardeen potentials, time dependence and gauge fixing in an explicit form, employing a language that can be directly used in the following sections. \Cref{sec:off-shell metric propagator} extends this discussion to the off-shell propagator for the metric fluctuations which is needed in quantum gravity computations.

In \cref{sec:homogeneous and isotropic cosmology} we turn to homogeneous and isotropic geometries and discuss, in particular, the role of mode functions, the linearized Einstein equations, projectors onto physical fluctuations, and the connection to gauge fixing. \Cref{sec:mode decomposition} proceeds to a decomposition of the physical metric fluctuations into representations of the rotation group. We obtain propagator equations for the individual modes which can be the basis for a future explicit computation of the correlation function for all components of the physical metric fluctuations. In \cref{sec:graviton correlation} we focus on the graviton correlation which is technically simplest. This makes a direct connection to the observable tensor modes in the primordial cosmic fluctuation spectrum. The results agree with the well known results obtained in the operator formalism \cite{STAT,RSV}. This section mainly serves the demonstration of equivalence of methods for the case of propagating fluctuations in a background solving the field equations, where the operator formalism is straightforward.

We specialize to de Sitter space in order to underline the equivalence by the explicit form of the graviton propagator. The full metric correlation has been discussed extensively for a de Sitter geometry \cite{AT,TW,HHT,HK,HG,MTW1,MTW2,XX}. The results of ref. \cite{MTW2} include the physical gauge fixing advocated here. Still, some work needs to be done to extract the explicit form of the propagator for physical scalar and vector fluctuations from the general structure described in ref. \cite{MTW2}. Geometries close to de Sitter space may avoid the singular behavior of propagators in de Sitter space, cf. \cite{XX} for a discussion. Only little is known \cite{YY} about the full metric propagator in general homogeneous and isotropic cosmologies.

Our conclusions are found in \cref{sec:conclusions}. Several more technical points, as the explicit connection to the Bardeen potentials or a more general mode decomposition can be found in the appendices.

\section{Basic concepts}
\label{sec:basic concepts}

For an effective action of gravity which is invariant under general coordinate transformations (diffeomorphisms) the second functional derivative is not invertible in the function space of arbitrary metric fluctuations. The local gauge symmetry implies that there are ``gauge modes'' for which $\Gamma^{(2)}$ vanishes. There are two possible ways to cope with this issue. The first reduces the field space for $G$ and $\Gamma^{(2)}$ to ``physical fluctuations'' by projecting out the ``gauge fluctuations''. In this case the inhomogeneous term $E$ on the r.h.s. of \cref{eqn:I3} is a projector onto the space of physical fluctuations. The second functional derivative becomes invertible on this restricted space if suitable boundary conditions are specified. (For massless fields the zero momentum mode may need a special regularization). The second alternative employs gauge fixing in a standard way. In the presence of gauge fixing $\Gamma$ is no longer gauge invariant. Thus $\Gamma^{(2)}$ becomes invertible on the full space of metric fluctuations and $E$ is the unit matrix in this space.

We will concentrate in this paper on the projection to physical metric fluctuations. We show that this is equivalent to a particular gauge fixing. For local gauge theories as gravity the source for the metric field is related to the energy momentum tensor $T^{\mu\nu}$. A central point of our formalism is the restriction to sources that reflect the most general covariantly conserved energy momentum tensors, $T^{\mu\nu}{_{;\nu}} = D_\nu T^{\mu\nu} = 0$, with $D_\nu$ a covariant derivative. Such sources couple only to covariantly conserved metric fluctuations, such that the quantum effective action will only involve these ``physical fluctuations'' of the metric.

A quantum field theory for gravity can be formulated as a functional integral over the ``fluctuating metric'' $g^\prime_{\mu\nu}$.
We decompose the metric $g^\prime_{\mu\nu}$ as
\begin{equation}\label{eqn:AA}
	g^\prime_{\mu\nu}
	= \bar{g}_{\mu\nu} + h^\prime_{\mu\nu}
	= \hat{g}^\prime_{\mu\nu}+a^\prime_{\mu;\nu}+a^\prime_{\nu;\mu},
\end{equation}
with ``background metric'' $\bar{g}_{\mu\nu}$ and
\begin{equation}\label{eqn:A1}
	\hat{g}^\prime_{\mu\nu}
	= \bar{g}_{\mu\nu}+f^\prime_{\mu\nu},
	\qquad
	f^\prime_{\mu\nu;}{^\nu}
	= 0.
\end{equation}
Here semicolons denote covariant derivatives that are formed with the background metric $\bar{g}_{\mu\nu}$ such that $\bar{g}_{\mu\nu;\rho} = 0$. Similarly, an arbitrary symmetric second rank contravariant tensor $B^{\mu\nu}$ is decomposed as
\begin{equation}\label{eqn:A2}
	B^{\mu\nu}
	= T^{\mu\nu}+T^\mu_{V;}{^\nu}+T^\nu_{V;}{^\mu},
	\qquad
	T^{\mu\nu}{_{;\nu}}
	= 0.
\end{equation}
For a source term (with $\bar{g} = \det \bar{g}_{\mu\nu}$)
\begin{eqnarray}\label{eqn:A3}
	S_B
	&= -\frac{1}{2} \int_x\sqrt{\bar{g}} g^\prime_{\mu\nu} B^{\mu\nu}\\
	&= -\frac{1}{2}\int \sqrt{\bar{g}}
	\left\{\hat{g}^\prime_{\mu\nu} T^{\mu\nu}+
	(a^\prime_{\mu;\nu}+a^\prime_{\nu;\mu})
	(T^\mu_{V;}{^\nu}+T^\nu_{V;}{^\mu})\right\}
\end{eqnarray}
one finds indeed that only the physical metric $\hat{g}^\prime_{\mu\nu}$ couples to $T^{\mu\nu}$. Restricting the source to $T_V = 0$, the argument of the effective action $\Gamma$ will be restricted to
\begin{equation}\label{eqn:A4}
	\hat{g}_{\mu\nu}
	= \bar{g}_{\mu\nu} + f_{\mu\nu},
	\qquad
	f_{\mu\nu}
	= \langle f^\prime_{\mu\nu}\rangle,
	\qquad
	f_{\mu\nu;}{^\nu}
	= 0.
\end{equation}

In order to avoid explicit constraints for the metric we may extend the argument of $\Gamma$ formally to arbitrary metrics $g_{\mu\nu}$, e.g. $\Gamma[\hat{g}_{\mu\nu}] \to \Gamma[g_{\mu\nu}]$. The fact that $\Gamma$ actually only depends on $\hat{g}_{\mu\nu}$ is then reflected by the local gauge symmetry of $\Gamma$. The local gauge symmetry corresponds to the statement that $\Gamma$ does not depend on the gauge fluctuations of the metric
\begin{equation}\label{eqn:8A2}
	a_{\mu\nu}
	= \left\langle a^\prime_{\mu;\nu} + a^\prime_{\nu;\mu} \right\rangle.
\end{equation}

In a gauge fixed version of gravity the metric correlation depends, in general, on the choice of the gauge. For a general gauge, this often obscures the relation between the effective action and the propagator for the physical metric fluctuations. Having identified the physical metric fluctuations it will be natural to choose for the fluctuating metric in the functional integral a gauge $a^\prime_\mu = 0$ or $h^\prime_{\mu\nu;}{^\nu} = 0$, corresponding to $g^\prime_{\mu\nu} = \hat{g}^\prime_{\mu\nu}$.
For the functional integral defining quantum gravity one may therefore employ a corresponding gauge fixing with the associated ghosts. For this type of gauge fixing the propagator equation \labelcref{eqn:I3} becomes block diagonal, decaying into separate sectors for the physical fluctuations and the gauge fluctuations. We can therefore compute the correlation function for the physical metric fluctuation on a restricted function space with appropriate projector $E$ in \cref{eqn:I3}. On the level of the relation between the effective action and the correlation function for the metric the gauge fixing and ghost terms are not needed if $G$ is restricted to the correlation function for physical metric fluctuations and $E$ is the appropriate projector. We can work directly with a diffeomorphism invariant quantum effective action $\Gamma$ and do not have to worry about gauge fixing and ghosts.

The physical metric fluctuations $f_{\mu\nu}$ are ``gauge invariant'' in the same sense as the Bardeen potentials. We explicitly construct the relation between the physical metric fluctuations and the Bardeen potentials, which turn out to be rather involved. In contrast to the Bardeen potentials the projection on $f_{\mu\nu}$ can be done in a manifestly covariant way. This is important for quantum gravity and flow equations where diffeomorphism invariance plays a crucial role in order to restrict the form of the effective action. There is, however, a price to pay for the covariant formulation. While the relation of the Bardeen potentials to metric fluctuations is simple in certain gauges as the Newtonian gauge, it gets more complex in a covariant setting.

The propagator equation \labelcref{eqn:I3} is a differential equation,
\begin{equation}\label{eqn:8A}
	D \, G
	= E,
\end{equation}
with differential operator $D = \Gamma^{(2)}$. This makes it manifest that $G$ is given by an initial value problem \cite{VF,AEH, BM1, PAD, BMR, PJS, BM2, JS1, AHR, DGPR}, and is not \textit{a priori} fixed for a given cosmological solution and a given time.
As a simple condition for a possible scaling solution \cite{CW2} we employ here the condition that the high momentum tail of the metric correlation is already at some early time given by the Lorentz invariant correlation function in flat space. This generalizes the Bunch-Davies initial condition \cite{BD} to interacting fields, arbitrary geometric backgrounds and non-propagating modes. It selects a particular correlation among several proposed ones \cite{EM,BA,AP,EA}. The complete discussion of the physical metric correlation function in flat space presented in this paper is therefore not only a very explicit example how the projection on physical fluctuations operates, but also sets the initial conditions for the solution of \cref{eqn:8A}.

The correlation function $G$ for the metric is an important quantity beyond its crucial role for quantum gravity computations on one side and the cosmic fluctuation spectrum on the other side. For example, it enters directly the computation of the bispectrum $B$ from the third functional derivative of the effective action $\Gamma^{(3)}$, that we may symbolically express as $B =\Gamma^{(3)} G^3$. In this paper we discuss $G$ in the Einstein frame. The determination of $G$ by the propagator equation \labelcref{eqn:I3} makes transformations between different frames straightforward \cite{CWPFVG}.

Besides the development of the formalism for computing $G$ from the propagator equation, and the direct relation between cosmological fluctuation observables and the covariant correlation for physical metric fluctuations that may be extracted from quantum gravity calculations, our paper also contains practical progress: we derive the explicit form of the propagator equation for the physical metric fluctuations for a homogeneous and isotropic cosmological background.

\section{Physical and gauge part of metric}
\label{sec:physical and gauge metric}

We formulate quantum gravity as a functional integral for the partition function
\begin{equation}\label{eqn:1}
	Z[K^{\mu\nu}]
	= \int\tilde{D}g^\prime_{\rho\sigma}\exp\Bigl\{-S[g^\prime_{\rho\sigma}] + \int_x g^\prime_{\mu\nu}(x)K^{\mu\nu}(x)\Bigr\}.
\end{equation}
The regularization of this functional integral as, for example, gauge fixing and ghost terms, are here formally included in the functional measure $\int\tilde{D}g^\prime_{\rho\sigma}$. The action $S$ is supposed to be invariant under general coordinate transformations or diffeomorphisms
\begin{equation}\label{eqn:2}
	\delta_{\xi}g^\prime_{\mu\nu}
	= -\partial_{\mu}\xi^{\rho}g^\prime_{\rho\nu}-\partial_{\nu}\xi^{\rho}g^\prime_{\mu\rho}-\xi^{\rho}\partial_{\rho}g^\prime_{\mu\nu}.
\end{equation}
The source $K^{\mu\nu} = K^{\nu\mu}$ transforms as a contravariant tensor density
\begin{equation}\label{eqn:3}
	\delta_{\xi}K^{\mu\nu}
	= \partial_{\rho}\xi^{\mu}K^{\rho\nu}+\partial_{\rho}\xi^{\nu}K^{\mu\rho}-\xi^{\rho}\partial_{\rho}K^{\mu\nu}-\partial_{\rho}\xi^{\rho}
	K^{\mu\nu},
\end{equation}
such that the source term is diffeomorphism invariant. For the functional measure we will employ a background field formalism such that the measure is invariant under a simultaneous diffeomorphism transformation of the background metric and the fluctuations, see below. Therefore $Z$ is invariant under this combined transformation.

We write the action in terms of a scalar function $L$,
\begin{equation}\label{eqn:4}
	S
	= \int_x\sqrt{g^\prime}L[g^\prime_{\rho\sigma}],
	\quad
	g^\prime
	= \det(g^\prime_{\mu\nu}),
	\quad
	\delta_\xi L
	= -\xi^{\rho}\partial_\rho L.
\end{equation}
For the example of Einstein gravity with reduced Planck mass $M$ and cosmological constant $V$ one has
\begin{equation}\label{eqn:5}
	L
	= -\frac{M^2}{2}R + V,
\end{equation}
with $R$ the curvature scalar of the metric $g^\prime_{\mu\nu}$. For purposes of analytic continuation we will admit complex values of $g^\prime_{\mu\nu}$, while coordinates remain fixed. For Minkowski signature $g^\prime_{\mu\nu}$ is real and one has $\sqrt{g^\prime} = i\sqrt{\smash[b]{-\det(g^\prime_{\mu\nu})}}$, accounting for the factor $i$ in the weight factor $e^{-S}$ of the functional integral.

As mentioned in \cref{sec:basic concepts}, we split the metric $g^\prime_{\mu\nu}$ into a ``physical metric'' $\hat{g}^\prime_{\mu\nu}$ and a ``gauge part'' $a^\prime_{\mu\nu}$ which can be obtained by covariant derivatives of a vector $a^\prime_\mu $,
\begin{equation}\label{eqn:6}
	g^\prime_{\mu\nu}
	= \hat{g}^\prime_{\mu\nu}+a^\prime_{\mu\nu},
	\qquad
	a^\prime_{\mu\nu}
	= a^\prime_{\mu;\nu}+a^\prime_{\nu;\mu}.
\end{equation}
Covariant derivatives, denoted by semicolons, are formed with the connection ${\bar{\Gamma}_{\mu\nu}}{^{\rho}}$ of a background metric $\bar{g}_{\mu\nu}$,
\begin{equation}\label{eqn:7}
	a^\prime_{\mu;\nu}= D_\nu  a^\prime_\mu = \partial_\nu  a^\prime_\mu -{\bar{\Gamma}_{\nu\mu}}{^{\rho}} a^\prime_\rho.
\end{equation}
In principle, the background metric is arbitrary. We will focus later on solutions of the field equations.

General sources $K^{\mu\nu}(x)$ are introduced in order to construct generating functionals as in \cref{eqn:I1}. This allows to probe the response of the metric expectation value to any given particular source, as the energy momentum tensor for radiation and matter in cosmology. We use the background metric $\bar{g}_{\mu\nu}$ to relate the source $K^{\mu\nu}$ to the energy momentum tensor $T^{\mu\nu}$,
\begin{equation}\label{eqn:8}
	K^{\mu\nu}
	= \frac{1}{2}{\bar{g}}^{\frac{1}{2}}T^{\mu\nu},
	\qquad
	\bar{g} = \det(\bar{g}_{\mu\nu}).
\end{equation}
Again, $T^{\mu\nu}$ is considered here as general source, with possible a posteriori specification of a ``physical source'' if appropriate.

For the effective action $\Gamma$ the source term in \cref{eqn:1} is reflected in the quantum field equation (for details see \cref{sec:qea})
\begin{equation}\label{eqn:18A}
	\frac{\partial\Gamma}{\partial g_{\mu\nu}}
	= K^{\mu\nu},
	\qquad
	\frac{2}{\sqrt{\bar{g}}}\frac{\partial\Gamma}{\partial g_\mu\nu}
	= T^\mu\nu.
\end{equation}
Identifying $g_\mu\nu$ and $\bar{g}_\mu\nu$, such that $\Gamma$ depends only on $g_\mu\nu$, the second equation \labelcref{eqn:18A} is the usual defining equation for the energy momentum tensor. (See ref. \cite{CWGFE} for a discussion and modifications of the identification $g_\mu\nu = \bar{g}_\mu\nu$.) If one considers extended field theories, for example with an additional scalar inflaton field, the metric variation of the effective matter action would contribute (with negative sign) to $T^\mu\nu$. The precise nature of $K^\mu\nu$ and $T^\mu\nu$ will not be important for our discussion. We only will employ the structural aspect of a conserved energy momentum tensor.

We will focus on sources $K^\mu\nu$ corresponding to a conserved energy momentum tensor, $T^{\mu\nu};{_\nu}$. They obey
\begin{equation}\label{eqn:9}
	\partial_\nu K^{\mu\nu}+\bar{\Gamma}_{\nu\rho}{^{\mu}}K^{\rho\nu}
	= 0.
\end{equation}
These sources couple only to the physical metric, motivating the naming,
\begin{equation}\label{eqn:10}
	\int_x g^\prime_{\mu\nu}K^{\mu\nu}
	= \int_x\hat{g}^\prime_{\mu\nu}K^{\mu\nu}.
\end{equation}
Indeed, partial integration and the relation \labelcref{eqn:9} imply
\begin{eqnarray}\label{eqn:11}
	\int_x a^\prime_{\mu\nu}K^{\mu\nu}
	&= \int_x(a^\prime_{\mu;\nu}+a^\prime_{\nu;\mu})K^{\mu\nu} \\
	&= -2\int_x a^\prime_\mu (\partial_\nu K^{\mu\nu}+{\bar{\Gamma}_{\nu\rho}}{^{\mu}}K^{\rho\nu}) = 0.
\end{eqnarray}
The constraint \labelcref{eqn:9} is invariant under simultaneous gauge transformations of $K^{\mu\nu}$ and $\bar{g}_{\mu\nu}$.

With respect to diffeomorphisms all three objects $\hat{g}^\prime_{\mu\nu}$, $\bar{g}_{\mu\nu}$ and $a^\prime_{\mu\nu}$ transform as tensors according to \cref{eqn:2}. Writing
\begin{equation}\label{eqn:11A}
	\hat{g}^\prime_{\mu\nu}
	= \bar{g}_{\mu\nu}+f^\prime_{\mu\nu},
\end{equation}
and observing
\begin{equation}\label{eqn:12}
	\delta_\xi \bar{g}_{\mu\nu}
	= -(\xi_{\mu;\nu}+\xi_{\nu;\mu}),
\end{equation}
one sees that the transformation of $g^\prime_{\mu\nu}$ can also be realized for a fixed background metric $\bar{g}_{\mu\nu}$ if the transformation of $a^\prime_{\mu\nu}$ obtains an additional inhomogeneous part,
\begin{eqnarray}\label{eqn:13}
	\hat{\delta}_\xi a^\prime_{\mu\nu}
	&= \delta_\text{inh}a^\prime_{\mu\nu}+\delta_\xi a^\prime_{\mu\nu}, \\
	\delta_\text{inh}a^\prime_{\mu\nu} &= -(\xi_{\mu;\nu}+\xi_{\nu;\mu}).
\end{eqnarray}
For $a^\prime_\mu \rightarrow0$ the inhomogeneous part dominates and becomes
\begin{equation}\label{eqn:13A}
	\delta_\text{inh}a^\prime_\mu =-\xi_\mu.
\end{equation}
This identifies infinitesimal $a^\prime_{\mu\nu}$ with the infinitesimal change of the background metric $\bar{g}_{\mu\nu}$ under a diffeomorphism transformation. By a suitable gauge transformation one can always achieve $a^\prime_\mu =0$. This justifies the naming of $a^\prime_{\mu\nu}$ as the gauge part of the metric $g^\prime_{\mu\nu}$.

Strictly speaking, the classification of physical and gauge fluctuations is exact only on the linear level, e.g. for infinitesimal $f^\prime_{\mu\nu}$ and $a^\prime_{\mu\nu}$. Beyond, the non-linear construction of the notion of a ``physical metric'' $\hat{g}_{\mu\nu}$ is more involved \cite{CWNN}. Beyond the linear level one would also like to replace $\bar\Gamma_{\nu\rho}{^\mu}$ in \cref{eqn:9} by the connection formed with the macroscopic metric $g_\mu\nu$. Then this equation, together with the first equation \labelcref{eqn:18A}, guarantees invariance of the effective action with respect to gauge transformations acting only on $g_\mu\nu$ \cite{CWNN}. Linear fluctuations are sufficient for the computation of propagators and field equations. We therefore stick to the linear definition \labelcref{eqn:6,eqn:7}, leaving non-linear extensions aside.

Formally, we can obtain the physical metric fluctuations by applying a suitable projector $P^{(f)}$,
\begin{equation}\label{eqn:20a}
	h^\prime_{\mu\nu}
	= g^\prime_{\mu\nu}-\bar{g}_{\mu\nu},
	\qquad
	f^\prime_{\mu\nu}
	= P^{(f)\rho\tau}_{\mu\nu}h^\prime_{\rho\tau},
\end{equation}
where the product includes a product in position space
\begin{equation}\label{eqn:21A}
	f^\prime_{\mu\nu}(x) = \int_y P^{(f)}_{\mu\nu}{^{\rho\tau}}(x,y)h^\prime_{\rho\tau}(y).
\end{equation}
The projector $P^{(f)\rho\tau}_{\mu\nu}$ is symmetric in $\mu\to\nu$ and $\rho\leftrightarrow \tau$ and obeys
\begin{eqnarray}\label{eqn:20B}
	&P^{(f)}_{\mu\nu}{^{\alpha\beta}}P^{(f)}_{\alpha\beta}{^{\rho\tau}}
	= P^{(f)}_{\mu\nu}{^{\rho\tau}},\\
	&D^\mu P^{(f)}_{\mu\nu}{^{\rho\tau}}
	= 0,\\
	&P^{(f)}_{\mu\nu}{^{\rho\tau}}D_\tau
	= 0.
\end{eqnarray}
We discuss this projector in more detail in \cref{sec:projectors and gauge fixing} as well as in \cref{sec:propagator in flat space,sec:homogeneous and isotropic cosmology}. The properties \labelcref{eqn:20B} guarantee that $f^\prime_{\mu\nu}$ is divergence free
\begin{equation}\label{eqn:20C}
	D^\mu f^\prime_{\mu\nu}
	= 0,
\end{equation}
and invariant under the inhomogeneous gauge transformation $\delta_\textbf{inh}$. Indeed, applying $P^{(f)}$ on the transformed fluctuations yields again $f^\prime_{\mu\nu}$
\begin{equation}\label{eqn:20D}
	P^{(f)}_{\mu\nu}{^{\rho\tau}}(f^\prime_{\rho\tau}+a^\prime_{\rho\tau}-D_\rho\xi_\tau -D_\tau\xi_\rho) = f^\prime_{\mu\nu}.
\end{equation}

The regularization of the functional integral is done by using only the objects $\hat{g}^\prime_{\mu\nu}$ and $a^\prime_{\mu\nu}$, preserving the gauge transformation $\delta_\xi $ which acts on both objects. It will not be invariant under the inhomogeneous transformation $\hat{\delta}_\xi $. We note that $\delta_\xi $ and $\hat{\delta}_\xi $ can be related by a ``split transformation'' $\delta_s\hat{g}^\prime_{\mu\nu} = s_{\mu\nu}$, $\delta_s a^\prime_{\mu\nu} = -s_{\mu\nu}$, for the particular case $s_{\mu\nu} = \xi_{\mu;\nu}+\xi_{\nu;\mu}$. The split symmetry of objects formed only with $g^\prime_{\mu\nu}$ is broken by the regularization which involves $\hat{g}^\prime_{\mu\nu}$ and $a^\prime_{\mu\nu}$ separately. The regularized functional integral employs gauge fixing
\begin{eqnarray}\label{eqn:14}
	Z[K^{\mu\nu};\bar{g}_{\mu\nu}]
	&= \int D g^\prime_{\mu\nu} J[a^\prime_{\mu\nu},\hat{g}^\prime_{\mu\nu}]\exp \{-S_\text{gf}[a^\prime_{\mu\nu},\hat{g}^\prime_{\mu\nu}]\}\\
	&\hphantom{={}}\times\exp \{-S[g^\prime_{\mu\nu}]\}\exp \{\int_x\hat{g}^\prime_{\mu\nu} K^{\mu\nu}\},
\end{eqnarray}
with $S_\text{gf}$ a gauge fixing term in the action and $J[a^\prime_{\mu\nu},\hat{g}^\prime_{\mu\nu}]$ the associated Faddeev-Popov determinant. As usual, $J$ can be represented by a functional integral over ghost degrees of freedom.

In the setting of the present paper we form covariant derivatives and the source constraint \labelcref{eqn:9} with the background metric $\bar{g}_{\mu\nu}$. For a fixed $\bar{g}_{\mu\nu}$ this maintains the discussion of metric fluctuations within the standard approach. In particular, the source term remains linear in the fluctuating metric $g^\prime_{\mu\nu}$. As a shortcoming of this formalism $T^{\mu\nu}$ is covariantly conserved only with respect to the background of the metric $\bar{g}_{\mu\nu}$, and not with respect to the macroscopic metric $g_{\mu\nu}$ as one would like it to be.

One may wish to find a formulation where \cref{eqn:8,eqn:9} employ the macroscopic metric $g_{\mu\nu}$ which is the argument of the effective action, such that for all $g_{\mu\nu}$ the energy momentum tensor is covariantly conserved. This possibility is described elsewhere \cite{CWGFE}. In this case $\bar{g}_{\mu\nu}$ may be replaced by a dynamical macroscopic field $g_{\mu\nu}$, e.g.
\begin{equation}\label{eqn:30A}
	g_{\mu\nu} = \frac{\partial \text{ln} Z}{\partial K^{\mu\nu}}.
\end{equation}
The source term is then no longer linear since $g_\mu\nu$ depends implicitly on $Z$. As a consequence, \cref{eqn:A4} holds only for infinitesimal $f_{\mu\nu}$, while the general form of the physical metric $\hat{g}_{\mu\nu}$ receives corrections. In this paper we keep a fixed $\bar{g}_{\mu\nu}$ different from $g_{\mu\nu}$, and the present setting can be viewed as an approximation to the formulation which uses the macroscopic metric.

\section{Quantum effective action}\label{sec:qea}

For the construction of the effective action we have two options. The first one restricts the sources to those obeying the constraint \labelcref{eqn:9}. In consequence, the effective action will only depend on fields that couple to the constrained source, i.e.
\begin{equation}
	\hat{g}_{\mu\nu}
	= \langle\hat{g}^\prime_{\mu\nu}\rangle.
\end{equation}
These fields will be constrained according to
\begin{align}\label{eqn:constraint32}
	&\hat{g}_{\mu\nu}
	= \bar{g}_{\mu\nu} + f_{\mu\nu},
	&f_{\mu\nu}
	= \langle f^\prime_{\mu\nu} \rangle,
	&&{f_{\mu\nu;}}^\nu
	= 0.
\end{align}
In this formulation the effective action contains no gauge modes such that the second functional derivative $\Gamma^{(2)}$ is typically invertible once projected on the appropriate space of physical fluctuations. If a possible gauge fixing term vanishes for $g_{\mu\nu} = \hat{g}_{\mu\nu}$, it needs not to be included on the level of the effective action. This is the option we will mainly pursue in this paper. The second option considers instead of the constrained sources $K^{\mu\nu}$ arbitrary sources $L^{\mu\nu}$, and therefore arbitrary $g_{\mu\nu}$. Then typically a gauge fixing term is present in $\Gamma$. One can subsequently project onto the space of physical metric fluctuations. If the gauge fixing term is projected out by this procedure, it no longer appears in the projected quantities. In our case we will see that the two options are equivalent.

\subsection{Effective action for constrained fields}

Let us now formulate the effective action in the presence of constraints on sources and fields. Our starting point is the partition function \labelcref{eqn:14} where we have indicated explicitly that $Z$ depends on the background metric $\bar{g}_{\mu\nu}$. This dependence arises from the constraint \labelcref{eqn:9} for $K^{\mu\nu}$ which involves the connection formed with $\bar{g}_{\mu\nu}$. Also the definition of the split of $g^\prime_{\mu\nu}$ into $\hat{g}^\prime_{\mu\nu}$ and $a^\prime_{\mu\nu}$ involves $\bar{g}_{\mu\nu}$. It is worthwhile to note, however, that for our construction the background metric only enters indirectly through the projections on physical sources and fields, i.e. $K^{\mu\nu}$ and $\hat{g}^\prime_{\mu\nu} = P^{(f)\rho\tau}_{\mu\nu} g^\prime_{\rho\tau}$. By construction, $Z[K^{\mu\nu};\bar{g}_{\mu\nu}]$ is diffeomorphism invariant if both $K^{\mu\nu}$ and $\bar{g}_{\mu\nu}$ are transformed simultaneously. The invariant partition function is the basis for the construction of the quantum effective action.

We first define the generating functional for the connected correlation functions
\begin{equation}\label{eqn:35A}
	W[K^{\mu\nu};\bar{g}_{\mu\nu}] = \ln Z [K^{\mu\nu};\bar{g}_{\mu\nu}],
\end{equation}
with
\begin{equation}\label{eqn:15}
	\frac{\delta W}{\delta K^{\mu\nu}}
	= \langle\hat{g}^\prime_{\mu\nu}\rangle
	= \hat{g}_{\mu\nu}.
\end{equation}
The second functional derivative $W^{(2)}$ defines the connected two-point correlation function (Green's function, propagator)
\begin{eqnarray}\label{eqn:30}
	W^{(2)}_{\rho\tau\sigma\lambda}(x,y)
	&= \langle f^\prime_{\rho\tau}(x)f^\prime_{\sigma\lambda}(y)\rangle_c\\
	&= \langle f^\prime_{\rho\tau}(x)f^\prime_{\sigma\lambda}(y)\rangle - \langle f^\prime_{\rho\tau}(x)\rangle \langle f^\prime_{\sigma\lambda}(y)\rangle.
\end{eqnarray}
(Note that the background metric $\bar{g}_\mu\nu$ in \cref{eqn:11A} drops out in the connected correlation function.) Below we will identify the correlation function $W^{(2)}$ with the propagator for the physical metric fluctuations.

In \cref{eqn:15} the expectation value $\hat{g}_{\mu\nu}$ obeys the same constraint as $\hat{g}^\prime_{\mu\nu}$, namely
\begin{eqnarray}\label{eqn:16}
	\hat{g}_{\mu\nu;}{^\nu}
	= 0.
\end{eqnarray}
(Recall that this constraint is not trivial since covariant derivatives are formed with $\bar{g}_{\mu\nu}$). Due to the presence of the constraint \labelcref{eqn:16} we can invert \cref{eqn:15} and obtain the constrained source $K^{\mu\nu}$ as a functional of $\hat{g}_{\mu\nu}$.

We can make the constraint \labelcref{eqn:30} more explicit by employing the general decomposition
\begin{equation}\label{eqn:18}
	g_{\mu\nu}
	= \bar{g}_{\mu\nu}+b_{\mu\nu}+\frac{1}{4}\sigma \bar{g}_{\mu\nu}+v_{\mu;\nu}+v_{\nu;\mu}+2\tau_{;\mu\nu}.
\end{equation}
With
\begin{equation}\label{eqn:20A}
	b_\mu{^\nu}{_{;\nu}}
	= -\frac{1}{4}\partial_\mu\sigma
	,
	\qquad
	b_\mu{^\mu}
	= 0,
	\qquad
	v^\mu{_{;\mu}}
	= 0,
\end{equation}
the constraint \labelcref{eqn:16} is realized for
\begin{equation}\label{eqn:20}
	v_\mu
	= 0,
	\qquad
	\tau
	= 0.
\end{equation}
Indeed, we have chosen the basis \labelcref{eqn:18} such that for $v_\mu = 0,\tau = 0$ one has $g_{\mu\nu;}{^\nu} = 0$, according to \cref{eqn:16}. Due to the restriction \labelcref{eqn:9} for the sources, which corresponds to a conserved energy momentum tensor, no ``gauge part'' of $g_{\mu\nu}$ is present. The metric $\hat{g}_{\mu\nu}$ contains therefore only the ``physical excitations'' around the background, namely $b_{\mu\nu}$ and $\sigma$, while $v_\mu$ and $\tau$ are set to zero,
\begin{equation}\label{eqn:37A}
	f_{\mu\nu} = b_{\mu\nu} + \frac{1}{4} \sigma \bar{g}_{\mu\nu}.
\end{equation}

The effective action obtains by a Legendre transform
\begin{equation}\label{eqn:21}
	\Gamma[\hat{g}_{\mu\nu};\bar{g}_{\mu\nu}]
	= -W[K^{\mu\nu};\bar{g}_{\mu\nu}] + \int_x\hat{g}_{\mu\nu}K^{\mu\nu},
\end{equation}
with $K^{\mu\nu}[\hat{g}_{\mu\nu};\bar{g}_{\mu\nu}]$ obtained by solving \cref{eqn:15}. As usual, one has the exact quantum field equation
\begin{equation}\label{eqn:22}
	\frac{\delta\Gamma}{\delta\hat{g}_{\mu\nu}}
	= K^{\mu\nu}
	= \frac{1}{2}\sqrt{\bar{g}}T^{\mu\nu}.
\end{equation}

Our setting is realized by considering $\Gamma$ as a functional of $f_{\mu\nu}$ (as well as $\bar{g}_{\mu\nu}$), while $W$ depends on sources $K^\mu\nu$ that correspond to $T^{\mu\nu}$ in the general decomposition \labelcref{eqn:A2}. We will work within an approximation where $\Gamma$ is a gauge invariant functional only of the metric $\hat{g}_\mu\nu$. This can formally be achieved by setting $\bar{g}_\mu\nu = \hat{g}_\mu\nu$ in \cref{eqn:21}. Gauge invariance permits us to drop the explicit constraint on $\hat{g}_\mu\nu$ since the $\Gamma$ is independent of the gauge fluctuations. We can therefore consider well known approximations to the effective action as the Einstein-Hilbert action. A justification of our approximation and a detailed discussion of the issue of diffeomorphism invariance can be found in \cref{sec:local gauge symmetries}.

\subsection{Expansion around a cosmological background}

Let us consider some particular ``physical source'' $K^\mu\nu_0$ that corresponds to a homogeneous and isotropic energy momentum tensor $T^\mu\nu_0$. Examples are radiation or dust in cosmology. We choose the background metric such that for the physical source $K_0^{\mu\nu}$ the field equation is obeyed if $\hat{g}_{\mu\nu} = \bar{g}_{\mu\nu}$,
\begin{equation}\label{eqn:23}
	\frac{\delta\Gamma}{\delta\hat{g}_{\mu\nu}}(\hat{g}_{\mu\nu}
	= \bar{g}_{\mu\nu}) = K_0^{\mu\nu}.
\end{equation}

General (inhomogeneous) sources can be written as an expansion around $K^\mu\nu$,
\begin{equation}\label{eqn:60A}
	K^\mu\nu = K^\mu\nu_0 + \Delta K^\mu\nu.
\end{equation}
We consider small $\Delta K^\mu\nu$ such that linearization is valid. We may expand $W$ around $K_0^{\mu\nu}$,
\begin{eqnarray}\label{eqn:27}
	W
	&= W^{(0)}+\int_x\bar{g}_{\mu\nu}(x)\Delta K^{\mu\nu}(x)\\
	&\hphantom{={}}+\frac{1}{2} \int_x\int_y \Delta K^{\mu\nu}(x)W^{(2)}_{\mu\nu\rho\sigma}(x,y)\Delta K^{\rho\sigma}{(y)}+\dots,
\end{eqnarray}
with $W^{(0)}$ and $W^{(2)}$ depending on $K^{\mu\nu}_0$. Eq. \labelcref{eqn:15} reads
\begin{equation}\label{eqn:60B}
	\frac{\delta W}{\delta \Delta K^\mu\nu}
	= \bar{g}_\mu\nu + f_\mu\nu
\end{equation}
and comparison with \cref{eqn:27} yields
\begin{equation}\label{eqn:28}
	f_{\mu\nu}(x)
	= \int_y W^{(2)}_{\mu\nu\rho\sigma}(x,y) \, \Delta K^{\rho\sigma}(y) + \dots
\end{equation}
This equation expresses the response of the metric to sources in the linear approximation. It involves the metric correlation function $W^{(2)}$.

As an example we may consider Einstein gravity in flat space with $K^\mu\nu_0 = 0$ and $\bar{g}_\mu\nu = \eta_\mu\nu$. A small
\begin{equation}\label{eqn:NL1}
	\Delta K^{00}(y) = \frac{i}{2} m\delta^3(\vec{y})
\end{equation}
may represent a static test mass $m$ or a star at position $\vec{y} = 0$, with $T^{00} = m\delta^3(\vec{y})$. For the component $f_{00}$ \cref{eqn:28} reads
\begin{equation}\label{eqn:NL2}
	f_{00}(t,\vec{x}) = \frac{i}{2} m\int_{t^\prime} W^{(2)}_{0000}(t,\vec{x};t^\prime,0).
\end{equation}
Thus the correlation function $W^{(2)}_{0000}$ is related to the Newtonian potential
\begin{equation}\label{eqn:NL3}
	\Phi_N
	= -\frac{1}{2} f_{00}
	= -\frac{im}{4}\int_{t^\prime} W^{(2)}_{0000}
	(t,\vec{x};t^\prime0)
	= -\frac{m}{8\pi M^2 |\vec{x}|}.
\end{equation}

The linear relation \labelcref{eqn:28} accounts for the response of the metric to arbitrary small ``perturbations'' or inhomogeneous sources $\Delta K^\mu\nu$. For this purpose the cosmological background $K^\mu\nu_0$ and $\bar{g}_\mu\nu$ is arbitrary. The relation \labelcref{eqn:28} encodes one of the central properties of the metric propagator.

\subsection{Expansion in physical metric fluctuations}

For a given background metric $\bar{g}_{\mu\nu}$ we can expand the effective action in terms of the physical metric fluctuations $f_{\mu\nu}$. Expanding in second order in
\begin{equation}\label{eqn:24}
	f_{\mu\nu}
	= \hat{g}_{\mu\nu}-\bar{g}_{\mu\nu}
	= b_{\mu\nu} + \frac{1}{4} \sigma \bar{g}_{\mu\nu}
\end{equation}
yields
\begin{eqnarray}\label{eqn:25}
	\Gamma
	&= \Gamma^{(0)}+\int_x\Gamma^{(1)\mu\nu}(x)f_{\mu\nu}(x)\\
	&\hphantom{={}}+\frac{1}{2}\int_x\int_y f_{\mu\nu}(x)\Gamma^{(2)\mu\nu\rho\sigma}(x,y)f_{\rho\sigma}(y) + \dots
\end{eqnarray}
with $\Gamma^{(0)}$, $\Gamma^{(1)}$ and $\Gamma^{(2)}$ depending on $\bar{g}_{\mu\nu}$ and obeying
\begin{equation}\label{eqn:62A}
	\Gamma^{(1)\mu\nu}(x) = K_0^{\mu\nu}(x).
\end{equation}

An expansion of the quantum field equation \labelcref{eqn:22}, combined with \cref{eqn:23}, yields in linear order
\begin{eqnarray}\label{eqn:26}
	\int_y &\Gamma^{(2)\mu\nu\rho\sigma}(x,y)f_{\rho\sigma}(y) + \dots\\
	&= K^{\mu\nu}(x)-
	K^{\mu\nu}_0(x) = \Delta K^{\mu\nu}(x).
\end{eqnarray}
Comparison of \cref{eqn:28} with \cref{eqn:26} shows already that the metric correlation function is related to an appropriately projected inverse of $\Gamma^{(2)}$. This relation will be discussed in the next section.

Both $K^{\mu\nu}_0$ and $\Delta K^{\mu\nu}$ obey separately \cref{eqn:9}. Thus \cref{eqn:26} yields a constraint on $\Gamma^{(2)}$,
\begin{eqnarray}\label{eqn:27A}
	\int_y
	\biggl(&\frac{\partial}{\partial x^\nu}\Gamma^{(2)\mu\nu\rho\sigma}(x,y)\\
	&+\bar{\Gamma}_{\nu\tau}{^\mu}(x)\Gamma^{(2)\tau\nu\rho\sigma}(x,y)\biggr)
	f_{\rho\sigma}(y) = 0,
\end{eqnarray}
where higher orders in $f_{\mu\nu}$ are omitted.

The formulation of the effective action in terms of ``physical metrics'' obeying a constraint, due to the use of constrained physical sources, may seem somewhat unfamiliar. In \cref{sec:projectors and gauge fixing} we relate this formulation to the more common approach with gauge fixing. It corresponds to the limit of an infinite gauge parameter $\beta$ that enforces the constraint $f_{\mu\nu;}{^\nu} = 0$.

\section{Correlation function}
\label{sec:correlation function}

In this section we discuss the defining equation for the correlation function, namely the exact propagator equation \labelcref{eqn:I3} based on the second functional derivative of the effective action. For the Einstein-Hilbert action with a cosmological constant we display the inverse propagator both for unconstrained metric fluctuations $h_{\mu\nu}$ and for physical metric fluctuations $f_{\mu\nu}$.

\subsection{Propagator equation}

We may interpret the second functional derivatives $\Gamma^{(2)}$ and $W^{(2)}$ as matrices. They obey the usual matrix identity
\begin{equation}\label{eqn:29}
	\Gamma^{(2)} \, W^{(2)}
	= 1,
\end{equation}
that follows directly from the defining relations for $\Gamma$. In position space this reads
\begin{equation}\label{eqn:40A}
	\int_y\Gamma^{(2)\mu\nu\rho\tau}(x,y)W^{(2)}_{\rho\tau\sigma\lambda}(y,z)
	= 
	E^{\mu\nu}{_{\sigma\lambda}}(x,z),
\end{equation}
where $E^{\mu\nu}{_{\sigma\lambda}}$ is the unit matrix in the space of appropriate functions. For unconstrained $h_{\sigma\lambda}$ the unit matrix reads $E^{^\prime\mu\nu}{_{\sigma\lambda}} = \frac{1}{2}(\delta^\mu_\sigma\delta^\nu_\lambda + \delta^\mu_\lambda\delta^\nu_\sigma)\delta(x-z)$, while in the presence of a constraint for $f_{\mu\nu}$ it becomes the projector $P^{(f)}{^{\mu\nu}}{_{\sigma\lambda}}$, which obeys the defining relations \labelcref{eqn:20B}. \Cref{eqn:29} is the exact ``propagator equation'' for the Green's function
\begin{equation}\label{eqn:37AA}
	G_{\rho\tau\sigma\lambda}(x,y) = W^{(2)}_{\rho\tau\sigma\lambda}(x,y).
\end{equation}
In the presence of a constraint on physical sources and fluctuations we recall the connection \labelcref{eqn:30} to the two-point correlation function.

If $\Gamma^{(2)}$ contains time-derivatives \cref{eqn:40A} is an evolution equation which describes the time dependence of the Green's function. Typically, $\Gamma^{(2)}$ is of the form
\begin{equation}\label{eqn:37AB}
	\Gamma^{(2)\mu\nu\rho\tau}(x,y) = \delta(x-y) \, \Gamma^{(2)\mu\nu\rho\tau}(y),
\end{equation}
where $\Gamma^{(2)\mu\nu\rho\tau}(y)$ contains derivatives with respect to $y$. The resulting propagator equation reads
\begin{equation}\label{eqn:37AC}
	\Gamma^{(2)\mu\nu\rho\tau}(x) \, G_{\rho\tau\sigma\lambda}(x,y)
	= 
	E^{\mu\nu}{_{\sigma\lambda}}(x,y).
\end{equation}

\subsection{Inverse propagator for unconstrained metric fluctuations}

We will next assume a simple form of $\Gamma$ based on an expansion in the number of derivatives. The first two lowest invariants are given by
\begin{equation}\label{eqn:36}
	\Gamma
	= \int_x\sqrt{g}\left(V-\frac{M^2}{2}R[g_{\mu\nu}]\right).
\end{equation}
Expanding an unconstrained metric $g_{\mu\nu} = \bar{g}_{\mu\nu}+h_{\mu\nu}$ in second order in $h_{\mu\nu}$ one has
\begin{equation}\label{eqn:A14}
	(g^{\frac{1}{2}})_{(2)}
	= \bar{g}^{\frac{1}{2}}
	\left(\frac{1}{8} h^2 -\frac{1}{4} h^\rho_\mu h^\mu_\rho\right),
\end{equation}
and
\begin{eqnarray}\label{eqn:B14}
	(g^{\frac{1}{2}}R)_{(2)}
	&= \frac{1}{2}\bar{g}^{\frac{1}{2}}
	\Bigl\{\bar{R}\left(\frac{1}{4} h^2 -\frac{1}{2} h^\rho_\mu h^\mu_\rho\right)-\bar{R}^{\mu\nu}h h_{\mu\nu}\\
	&+h h^{\mu\nu}{_{;\mu\nu}}-hh;^\mu{_\mu}+2R_{(2)}\Bigr\},
\end{eqnarray}
with
\begin{eqnarray}\label{eqn:B14A}
	R_{(2)}
	&= \bar{R}^{\mu\nu} h_{\nu\rho}h^\rho_\mu
	 + h^{\mu\nu}h_{;\mu\nu}+h^\mu_\nu h^\nu_{\mu;}{^\rho}{_\rho}\\
	&\hphantom{={}}-h^{\mu\nu}(h^\rho_{\nu;\rho\mu}+h^\rho_{\nu;\mu\rho})-\frac{1}{2} h^{\mu\nu}{_{;\rho}} h^\rho_{\nu;\mu}+\frac{3}{4} h^\mu_{\nu;\rho}h^\nu_{\mu;}{^\rho}\\
	&\hphantom{={}}-h^{\mu\nu}{_{;\nu}}h^\rho_{\mu;\rho}+h^{\mu\nu}{_;{_\nu}}h_{;\mu}-\frac{1}{4} h_;{^\mu}h_{;\mu}\bigr\}.
\end{eqnarray}

For unconstrained $h_{\mu\nu}$ the second functional derivative of the effective action \labelcref{eqn:36} is given by
\begin{eqnarray}\label{eqn:47A}
	\Gamma^{(2)\mu\nu\rho\tau}
	&= -\frac{M^2}{8}\sqrt{\bar{g}}
	\Bigl\{\bigl(\bar{g}^{\mu\rho}\bar{g}^{\nu\tau}+\bar{g}^{\nu\rho}\bar{g}^{\mu\tau}-2\bar{g}^{\mu\nu}\bar{g}^{\rho\tau}\bigr)D^2\\
	&\hphantom{={}}+\bar{g}^{\mu\nu}(D^\tau D^\rho + D^\rho D^\tau) + \bar{g}^{\rho\tau}(D^\mu D^\nu + D^\nu D^\mu)\\
	&\hphantom{={}}-(\bar{g}^{\mu\rho}D^\tau D^\nu + \bar{g}^{\nu\rho}D^\tau D^\mu\\
	&\hphantom{
	= (+}+\bar{g}^{\mu\tau}D^\rho D^\nu + \bar{g}^{\nu\tau}D^\rho D^\mu)\\
	&\hphantom{={}}+\bar{R}(\bar{g}^{\mu\nu}\bar{g}^{\rho\tau}-\bar{g}^{\mu\rho}\bar{g}^{\nu\tau}-\bar{g}^{\mu\tau}\bar{g}^{\nu\rho})\\
	&\hphantom{={}}+2(\bar{R}^{\mu\rho}\bar{g}^{\nu\tau}+\bar{R}^{\nu\rho}\bar{g}^{\mu\tau}
	 + \bar{R}^{\mu\tau}\bar{g}^{\nu\rho}+\bar{R}^{\nu\tau}\bar{g}^{\mu\rho})\\
	&\hphantom{={}}-2(\bar{R}^{\mu\nu}\bar{g}^{\rho\tau}
	+ \bar{R}^{\rho\tau}\bar{g}^{\mu\nu})\Bigr\}\\
	&\hphantom{={}}+\tfrac{V}{4}\sqrt{\bar{g}}(\bar{g}^{\mu\nu}\bar{g}^{\rho\tau}-\bar{g}^{\mu\rho}\bar{g}^{\nu\tau}-\bar{g}^{\mu\tau}\bar{g}^{\nu\rho}).
\end{eqnarray}
Applying a suitable projection of this operator in \cref{eqn:37AC} constitutes the basic equation of this paper. Correlation functions are obtained as solutions to this differential equation with appropriate initial values.

We can take account of the constraint to physical metrics in different equivalent ways. One method projects the second functional derivative \labelcref{eqn:47A} onto the space of physical metrics. A second one inserts the constraint $f^\nu_{\mu;\nu} = 0$ already into the expansion of $\Gamma$. If the physical metric fluctuations $f_{\mu\nu}$ are expressed in terms of independent fields one can directly obtain $\Gamma^{(2)}$ in the space of these fields by functional variation. While the second method is often technically simpler, we will also use occasionally the first method in order to make the role of projections apparent.

\subsection{Physical metric fluctuations}

According to the second method we directly investigate the effective action \labelcref{eqn:36} in quadratic order in the physical metric fluctuations $f_{\mu\nu},f^\nu_{\mu;\nu} = 0$, i.e.
\begin{eqnarray}\label{eqn:46A}
	\Gamma_2
	&= \int_x\left\{V\left(g^{\frac{1}{2}}\right)_{(2)}-\frac{M^2}{2}(g^{\frac{1}{2}}R)_{(2)}\right\}\\
	&= \Gamma^{(V)}_2 + \Gamma^{(R)}_2.
\end{eqnarray}
In \cref{sec:metric fluctuations decomposition} we decompose $\Gamma_2$ into parts from the trace and traceless metric fluctuations. This decomposition simplifies considerably if we restrict the background geometries to the ones with a vanishing Weyl tensor,
\begin{eqnarray}\label{eqn:46B}
	\Gamma^{(V)}_2
	&= \frac{V}{16}\int_x\bar{g}^{1/2}(\sigma^2 -4b^{\mu\nu}b_{\mu\nu}),\\
	\Gamma^{(R)}_2
	&= \frac{M^2}{8}\int_x\bar{g}^{\frac{1}{2}}
	\left\{b^{\mu\nu}\left(-D^2 + \frac{2}{3}\bar{R}\right)b_{\mu\nu}+\frac{3}{4}\sigma D^2\sigma\right\},
\end{eqnarray}
where $D^2 = D_\mu D^\mu$ and
\begin{eqnarray}\label{eqn:46C}
	&\sigma
	= f_{\mu\nu}\bar{g}^{\mu\nu},
	\qquad
	&&b_{\mu\nu}
	= f_{\mu\nu}-\frac{1}{4}\sigma\bar{g}_{\mu\nu},\\
	&\bar{g}^{\mu\nu}b_{\mu\nu}
	= 0,
	\qquad
	&&b^\nu_{\mu;\nu}
	= -\frac{1}{4}\partial_\mu\sigma.
\end{eqnarray}
Due to the last relation the trace $\sigma$ and the traceless part $b_{\mu\nu}$ are not independent.

A decomposition of $f_{\mu\nu}$ into independent fields can be done as
\begin{equation}\label{eqn:46D}
	f_{\mu\nu}
	= t_{\mu\nu} + s_{\mu\nu}
\end{equation}
where $t_{\mu\nu}$ is traceless and divergence free
\begin{equation}\label{eqn:46E}
	t_{\mu\nu}\bar{g}^{\mu\nu}
	= 0,
	\qquad
	t^\nu_{\mu;\nu}
	= 0,
\end{equation}
while $s_{\mu\nu}$ is a linear function of $\sigma$
\begin{equation}\label{eqn:46EA}
	s_{\mu\nu}
	= \hat{S}_{\mu\nu}\sigma,
\end{equation}
with
\begin{equation}\label{eqn:46F}
	\hat{S}_{\mu\nu}\bar{g}^{\mu\nu}
	= 1,
	\qquad
	D^\mu\hat{S}_{\mu\nu}
	= 0,
	\qquad
	\hat{S}_{\mu\nu}
	= \hat{S}_{\nu\mu}.
\end{equation}
This entails the relation
\begin{equation}\label{eqn:46G}
	b_{\mu\nu}
	= t_{\mu\nu}+\tilde{s}_{\mu\nu},
\end{equation}
with
\begin{equation}\label{eqn:46H}
	\tilde{s}_{\mu\nu}
	= s_{\mu\nu}-\frac{1}{4}\sigma\bar{g}_{\mu\nu},
	\qquad
	\tilde{s}^\nu_{\mu;\nu}
	= -\frac{1}{4}\partial_\mu\sigma.
\end{equation}
The construction of the operator $\hat{S}_{\mu\nu}$ needs, however, some care due to the non-commuting properties of the covariant derivatives. It is not unique, since we can always add a divergence free and traceless tensor to $s_{\mu\nu}$.

For important simple cases we can easily find $\hat{S}_{\mu\nu}$. Consider geometries with a constant curvature scalar, $\partial_\mu\bar{R} = 0$. In this case we can choose
\begin{equation}\label{eqn:46I}
	\hat{S}_{\mu\nu}
	= (\bar{g}_{\mu\nu}D^2 -D_\mu D_\nu + \bar{R}_{\mu\nu})
	(3D^2 + \bar{R})^{-1}.
\end{equation}
Indeed, one has
\begin{eqnarray}\label{eqn:46J}
	D^\mu\hat{S}_{\mu\nu}
	&= \bigl([D_\nu, D^2] + D_\mu\bar{R}^\mu_\nu \bigr)
	(3D^2 + \bar{R})^{-1}\\
	&= (D_\mu\bar{R}^\mu_\nu-\bar{R}^\mu_\nu D_\mu)(3D^2 + R)^{-1}\\
	&= R^\mu_{\nu;\mu}(3D^2 + \bar{R})^{-1}
	= 0,
\end{eqnarray}
where we use the commutator relation (acting on a scalar)
\begin{equation}\label{eqn:46K}
	[D_\nu,D^2]
	= -\bar{R}^\mu_\nu D_\mu.
\end{equation}
The Bianchi identity $\bar{R}^\mu_{\nu;\mu}-\partial_\nu \bar{R}/2 = 0$ implies $\bar{R}^\mu_{\nu;\mu}$ for geometries with constant $\bar{R}$. The other two relations \labelcref{eqn:46F} are easily verified. One infers
\begin{equation}\label{eqn:46L}
	\tilde{s}_{\mu\nu}
	= -\bigl[(D_\mu D_\nu-\frac{1}{4} D^2\bar{g}_{\mu\nu})-
	(\bar{R}_{\mu\nu}-\frac{1}{4}\bar{R}\bar{g}_{\mu\nu})\bigr](3D^2 + \bar{R})^{-1}\sigma.
\end{equation}

\subsection{Inverse propagator for physical metric fluctuations}

The inverse propagator for the physical metric fluctuations can be extracted directly from the expansion of the effective action in second order in $f_{\mu\nu}$. We therefore compute $\Gamma_2$ in terms of the independent fields $t_{\mu\nu}$ and $\sigma$,
\begin{equation}\label{eqn:IA}
	\Gamma_2 = \Gamma^{(t)}_2 + \Gamma^{(\sigma)}_2 + \Gamma^{(t\sigma)}_2.
\end{equation}
For the transversal traceless tensor $t_{\mu\nu}$ one finds
\begin{equation}\label{eqn:46JA}
	\Gamma_2^{(t)}
	= -\frac{M^2}{8}\int_x\bar{g}^{1/2}t^{\mu\nu}
	\left(D^2 -\frac{2}{3}\bar{R} + \frac{2V}{M^2}\right)t_{\mu\nu},
\end{equation}
while $\Gamma_2^{(\sigma)}$ and $\Gamma_2^{(t\sigma)}$ are computed in \cref{sec:metric fluctuations decomposition}, given explicitly by \cref{eqn:A.15G,eqn:A.15H}.
The mixed term $\Gamma^{(t\sigma)}_2$ is proportional to
\begin{equation}\label{eqn:46KA}
	\int_x\bar{g}^{1/2}t^{\mu\nu}\tilde{s}_{\mu\nu}
	= \int_x\bar{g}^{1/2}t^{\mu\nu}\bar{R}_{\mu\nu}
	(3D^2 + \bar{R})^{-1}\sigma.
\end{equation}
It vanishes for $\bar{R}_{\mu\nu} = c\bar{g}_{\mu\nu}$.

For the computation of $\Gamma^{(\sigma)}_2$ we need
\begin{eqnarray}\label{eqn:LA}
	D^2\tilde{s}_{\mu\nu}
	= -\Bigl\{&D_\mu D^2 D_\nu-\frac{1}{4} D^4\bar{g}_{\mu\nu}+\frac{7}{12}\bar{R} D^2\bar{g}_{\mu\nu}\\
	&-\frac{1}{3}\bar{R} D_\mu D_\nu-2\bar{R}_{\mu\nu}D^2 + 2\bar{R}^\rho_\mu D_\rho D_\nu\\
	&+\bar{R}^\rho_\nu D_\rho D_\mu-\bar{R}^{\rho\lambda}D_\rho D_\lambda\bar{g}_{\mu\nu}\\
	&-\frac{5}{2}\bar{R}_{\mu\nu;}{^\rho}D_\rho + \frac{1}{2}\bar{R}^\rho_{\mu;\nu}D_\rho\Bigr\}(3D^2 + \bar{R})^{-1}\sigma.
\end{eqnarray}
Here we use the commutator relation
\begin{equation}\label{eqn:46M}
	[D^2,D_\mu]D_\nu = R^\rho_\mu D_\rho D_\nu-2 R_\mu{^\rho}{_\nu}{^\lambda}
	D_\rho D_\lambda-R_\mu{^\rho}{_\nu}{^\lambda}{_{;\rho}}D_\lambda
\end{equation}
and assume a vanishing Weyl tensor.

\subsection{Propagator equation for maximally symmetric spaces}

We next specialize to geometries with
\begin{equation}\label{eqn:46N}
	\bar{R}_{\mu\nu}
	= \frac{1}{4}\bar{R}\bar{g}_{\mu\nu},
	\qquad
	\partial_\mu\bar{R}
	= 0,
\end{equation}
where the r.h.s of \cref{eqn:46KA} vanishes and
\begin{eqnarray}\label{eqn:46O}
	D^2\tilde{s}_{\mu\nu}
	= \biggl\{&\frac{1}{4} D^4 g_{\mu\nu}-D_\mu D_\nu D^2 + \frac{1}{6}\bar{R} D^2\bar{g}_{\mu\nu}\\
	&\-\frac{2}{3}\bar{R} D_\mu D_\nu\biggr\}(3D^2 + \bar{R})^{-1}\sigma.
\end{eqnarray}
We conclude that for such spaces the fluctuations $t_{\mu\nu}$ and $\sigma$ decouple
\begin{equation}\label{eqn:46P}
	\Gamma^{(t\sigma)}_2
	= 0.
\end{equation}
Geometries with vanishing Weyl tensor and obeying \cref{eqn:46N} are maximally symmetric,
\begin{equation}\label{eqn:46Q}
	\bar{R}_{\mu\rho\nu\lambda}
	= \frac{\bar{R}}{12}(\bar{g}_{\mu\nu}\bar{g}_{\rho\lambda}-\bar{g}_{\mu\lambda}\bar{g}_{\nu\rho}).
\end{equation}
They describe de Sitter, anti-de Sitter or flat space.

For maximally symmetric spaces $\tilde{s}_{\mu\nu}$ simplifies
\begin{equation}\label{eqn:46R}
	\tilde{s}_{\mu\nu}
	= -(D_\mu D_\nu-\frac{1}{4} D^2\bar{g}_{\mu\nu})
	(3D^2 + \bar{R})^{-1}\sigma.
\end{equation}
One finds
\begin{equation}\label{eqn:46S}
	\int_x\bar{g}^{\frac{1}{2}}\tilde{s}^{\mu\nu}\tilde{s}_{\mu\nu}
	= \frac{1}{4}\int_x\bar{g}^{1/2}\sigma D^2(3D^2 + \bar{R})^{-1}\sigma
\end{equation}
and
\begin{equation}\label{eqn:46T}
	\int_x\bar{g}^{1/2}\tilde{s}^{\mu\nu}D^2\tilde{s}_{\mu\nu}
	= \frac{1}{4}\int \bar{g}^{1/2}\sigma D^2(D^2 + \frac{2}{3}\bar{R})(3D^2 + \bar{R})^{-1}\sigma.
\end{equation}
This yields the scalar part of the quadratic part of the expansion of the effective action,
\begin{eqnarray}\label{eqn:46U}
	\Gamma^{(\sigma)}_2
	= \frac{M^2}{4}\int_x &\bar{g}^{1/2}\sigma
	\left\{(D^2 + \frac{3}{8}\bar{R})D^2 + \frac{V}{2M^2}(D^2 + \frac{1}{2}\bar{R})\right\}\\
	&\times(3D^2 + \bar{R})^{-1}\sigma.
\end{eqnarray}

For background geometries that solve the field equations, e.g. $\bar{R} = 4V/M^2$, this simplifies further
\begin{equation}\label{eqn:46V}
	\Gamma^{(\sigma)}_2 = \frac{M^2}{12}\int_x\bar{g}^{1/2}\sigma
	\frac{(D^2 + \frac{1}{4}\bar{R})^2}{D^2 + \frac{1}{3} \bar{R}}\sigma.
\end{equation}
Correspondingly, $\Gamma^{(t)}_2$ takes the form
\begin{equation}\label{eqn:46W}
	\Gamma^{(t)}_2
	= -\frac{M^2}{8}\int_x\bar{g}^{1/2}t^{\mu\nu}(D^2 -\frac{1}{6}\bar{R})t_{\mu\nu}.
\end{equation}
The corresponding second functional derivative $\Gamma^{(2)}$ is block diagonal in the fields $t_{\mu\nu}$ and $\sigma$.

For maximally symmetric geometries obeying the field equations the correlation functions for $t_{\mu\nu}$ and $\sigma$,
\begin{eqnarray}\label{eqn:46X1}
	&G^{tt}_{\mu\nu\rho\tau}(x,y)
	= \langle t_{\mu\nu}(x)t_{\rho\tau}(y)\rangle_c,\\
	&G^{\sigma\sigma}(x,y)
	= \langle\sigma(x)\sigma(y)\rangle_c,
\end{eqnarray}
obey the propagator equations
\begin{eqnarray}\label{eqn:46X2}
	&-\frac{M^2}{4}\sqrt{\bar{g}}\left(D^2 -\frac{\bar{R}}{6}\right)
	G^{tt}_{\mu\nu\rho\tau}(x,y) = P^{(t)}_{\mu\nu\rho\tau}(x,y),\\
	&\frac{M^2}{6}\sqrt{\bar{g}}\left(D^2 + \frac{\bar{R}}{4} \right)^2
	\left(D^2 + \frac{\bar{R}}{3}\right)^{-1}G^{\sigma\sigma}(x,y) = \delta(x-y),\\
\end{eqnarray}
where $D^2$ acts on the variable $x$ and the first two indices $(\mu\nu)$ of $G^{tt}_{\mu\nu\rho\tau}$. Here $P^{(t)}$ projects onto $t_{\mu\nu}$
\begin{eqnarray}\label{eqn:46X3}
	&t_{\mu\nu}(x) = P^{(t)}_{\mu\nu}{^{\alpha\beta}}(x)h_{\alpha\beta}(x),\\
	&P^{(t)}_{\mu\nu}{^{\alpha\beta}}P^{(t)}_{\alpha\beta}{^{\rho\tau}}
	= P^{(t)}_{\mu\nu}{^{\rho\tau}}.
\end{eqnarray}
The full Green's function obtains as
\begin{equation}\label{eqn:46X4}
	G_{\mu\nu\rho\tau}(x,y) = G^{tt}_{\mu\nu\rho\tau}(x,y) + \hat{S}_{\mu\nu}(x)\hat{S}_{\rho\tau}(y)G^{\sigma\sigma}(x,y).
\end{equation}

The projector $P^{(t)}_{\mu\nu}{^{\rho\tau}}$ obeys the defining relations
\begin{equation}\label{eqn:46X5}
	D^\mu P^{(t)}_{\mu\nu}{^{\rho\tau}}
	= 0,
	\qquad
	\bar{g}^{\mu\nu}P^{(t)}_{\mu\nu}{^{\rho\tau}}
	= 0.
\end{equation}
In terms of the projector $P^{(f)}$ given by \cref{eqn:20a,eqn:20B} it obtains as
\begin{equation}\label{eqn:46X6}
	P^{(t)}_{\mu\nu}{^{\rho\tau}}
	= P^{(f)}_{\mu\nu}{^{\rho\tau}}- P^{(\sigma)}_{\mu\nu}{^{\rho\tau}},
\end{equation}
with
\begin{eqnarray}\label{eqn:46X7}
	P^{(\sigma)}_{\mu\nu}{^{\rho\tau}}
	&= \hat{S}_{\mu\nu}\bar{g}^{\alpha\beta}P^{(f)}_{\alpha\beta}{^{\rho\tau}}\\
	P^{(\sigma)}_{\mu\nu}{^{\rho\tau}}h_{\rho\tau}
	&= \hat{S}_{\mu\nu}\sigma.
\end{eqnarray}
One verifies the conditions \labelcref{eqn:46X5}, while the projector property of $P^{(t)}$ follows from the projector properties of $P^{(f)}$ and $P^{(\sigma)}$. (For the latter we use $P^{(f)}_{\mu\nu}{^{\rho\tau}}\hat{S}_{\rho\tau} = \hat{S}_{\mu\nu}$.)

\section{Metric correlation in flat space}
\label{sec:propagator in flat space}

In this section we discuss the metric correlation function in a flat space background. This has the advantage that all projections can be made in a simple explicit form, and $D^2$ becomes a block-diagonal differential operator. The flat space correlation function describes the limiting high momentum or short distance behavior of the metric correlation in an arbitrary background geometry. This holds in the range where terms involving the curvature tensor can be neglected as compared to the squared momentum.

For a non-compact space as Minkowski space a unique specification of the Green's function involves boundary conditions. They are typically set at some initial time, that may go to minus infinity. We do not discuss here the general solution of the evolution equation for the propagator \cite{CW2}. We rather impose Lorentz symmetry on the correlation function which fixes it uniquely, if the effective action is taken to be the Einstein-Hilbert action, cf.ref. \cite{CW2}. We concentrate in this section on the ``on-shell propagator'' for which the background metric obeys the field equations. For a flat background the cosmological constant $V$ in \cref{eqn:36} is therefore set to zero. The off-shell propagator in flat space, with $V\neq 0$, will be discussed in the next section.

In cosmology one is often interested in the time evolution of the propagator which we display explicitly. For more general geometries the correlation functions are best formulated in dependence on conformal time $\eta$, i.e. $G = G(\eta, \eta^\prime)$. The flat space conformal time $\eta$ coincides with Minkowski time $t$. For easy comparison with the following sections we use $\eta$ as time argument. The propagator equation amounts to a differential equation for the time evolution. The explicit form of the time-dependent metric correlation function \labelcref{eqn:86Q} in three-dimensional Fourier space constitutes for arbitrary geometries the ``initial value'' for $\eta \to -\infty$, $\eta^\prime \to -\infty$ and small $|\eta - \eta^\prime|$. Indeed, for geometries close to de Sitter space, as appropriate for inflationary cosmology, the flat space correlations describe the limiting behavior of the metric correlations for $|k \eta| \gg 1$, $|k \eta^\prime| \gg 1$, $|k (\eta - \eta^\prime)| \ll 1$, with $k$ the comoving wave number. For fixed $k$ and initial values set at minus infinity, the flat space correlations are therefore well suited as initial conditions for the evolution towards larger $\eta$ or $\eta^\prime$. For the propagating modes these initial conditions are equivalent to the Bunch Davies vacuum in the operator formalism for free quantum fields.

The Lorentz invariant flat space metric propagator is known since a long time, see for example ref. \cite{CAP}. We describe it nevertheless in some detail in this section. The reason is that several important characteristic features of our approach to compute the correlation function of the physical metric fluctuations from the inversion of the second functional derivative of the effective action can be seen very explicitly in flat space. A first point concerns the use of ``mode functions'' as familiar in cosmology. Mode functions are solutions of the linearized field equations for fluctuations. In cosmology, the propagator is often constructed as the ``square of mode functions''. We discuss the applicability of this procedure in \cref{sec:homogeneous and isotropic cosmology} and find that the on-shell correlation for the propagating modes can indeed be expressed in terms of mode functions. For the metric, this applies to the graviton, i.e. the traceless divergence free tensor mode. For the scalar and vector parts of the physical metric fluctuations the mode functions vanish while the propagator differs from zero. It is therefore important to understand the origin of the scalar and vector parts of the metric correlation. In flat space this can be done in a straightforward way. Related to this issue is the role of the gauge invariant Bardeen potentials. For this purpose we display explicitly the propagator for the different representations of the rotation group.

A second issue concerns the relation between the use of constrained physical metric fluctuations and a procedure with unconstrained metric fluctuations and explicit gauge fixing. For a particular ``physical gauge fixing'' the two approaches are equivalent. This can be seen rather explicitly in flat space. Functional variation for unconstrained fields and the corresponding evolution equation for the propagator have to take the constraints properly into account. This can be facilitated by the use of representations of the rotation group for which the physical scalar fluctuations are unconstrained, while the constraints for the graviton and vectors take a time independent form.

\subsection{Inverse propagator}

For Einstein gravity $(V = 0$ in \cref{eqn:36}) the expansion of the effective action in second order in the metric fluctuations is given in flat space, $\bar{g}_{\mu\nu} = \eta_{\mu\nu} = \diag(-1,1,1,1)$, by
\begin{eqnarray}\label{eqn:F1a}
	\Gamma_{2}
	&= -\frac{iM^2}{8}\int_x(h^{\mu\nu}D^2 h_{\mu\nu}-hD^2 h\\
	&-2h^{\mu\nu}h^\rho_{\nu;\rho\mu}+2h h^{\mu\nu}{_{;\nu\mu}}).
\end{eqnarray}
In four-dimensional Fourier space we can replace $\partial_\mu \to i q_\mu$, $q_\mu = (-\omega,\vec{k})$, $q^\mu = (\omega,\vec{k})$, $D^2 = -q^2 = -q^\mu q_\mu = \omega^2 -k^2$, such that
\begin{equation}\label{eqn:F2a}
	\Gamma_{2}
	= \frac{1}{2} \int_q h_{\mu\nu}(-q)\Gamma^{(2)\mu\nu\rho\tau}(q)h_{\rho\tau}(q),
\end{equation}
where $\int_q = \int d^4 q/(2\pi)^4$. The second functional derivative reads
\begin{equation}\label{eqn:86Aa}
	\Gamma^{(2)\mu\nu\rho\tau}(q,q^\prime) = \Gamma^{(2)\mu\nu\rho\tau}(q)\delta(q-q^\prime),
\end{equation}
with
\begin{eqnarray}\label{eqn:F3a}
	\Gamma^{(2)\mu\nu\rho\tau}(q) = \frac{iM^2}{8}\Bigl\{&(\eta^{\mu\rho}\eta^{\nu\tau}+\eta^{\nu\rho}\eta^{\mu\tau}-2\eta^{\mu\nu}\eta^{\rho\tau})q^2\\
	&+ 2\eta^{\rho\tau}q^\mu q^\nu + 2\eta^{\mu\nu}q^\rho q^\tau\\[1ex]
	&- \eta^{\mu\rho}q^\nu q^\tau - \eta^{\nu\rho}q^\mu q^\tau\\
	&- \eta^{\mu\tau} q^\nu q^\rho - \eta^{\nu\tau}q^\mu q^\rho)\Bigr\},
\end{eqnarray}
and $\delta(q-q^\prime) = (2\pi)^4\delta^4(q_\mu-q^\prime_\mu)$.
Acting on the gauge part of the metric fluctuation, $a_{\rho\tau} = i(q_\rho a_\tau + q_\tau a_\rho)$, one has
\begin{equation}\label{eqn:F4a}
	\Gamma^{(2)\mu\nu\rho\tau}a_{\rho\tau}
	= 0.
\end{equation}

We introduce the projector onto ``transversal components'',
\begin{equation}\label{eqn:F11a}
	\tilde{P}_\mu{^\rho}
	= \delta^\rho_\mu-\frac{q_\mu q^\rho}{q^2},
\end{equation}
obeying
\begin{eqnarray}\label{eqn:F11A}
    &\tilde{P}_\mu{^\rho}q_\rho
    = 0,
    \qquad
    &&q^\mu\tilde{P}_\mu{^\rho}
    = 0,\\
	&\tilde{P}_\mu{^\rho}\tilde{P}_\rho{^\nu}
	= \tilde{P}_\mu{^\nu},
	\qquad
	&&\tilde{P}_{\mu\nu}
	= \tilde{P}_{\nu\mu},
	\qquad
	\tilde{P}^\mu_\mu
	= 3.
\end{eqnarray}
We can write $\Gamma^{(2)}$ in \cref{eqn:F3a} in terms of this transversal projector as
\begin{equation}\label{eqn:F11B}
	\Gamma^{(2)\mu\nu\rho\tau}
	= \frac{iM^2 q^2}{8}
	\{\tilde{P}^{\mu\rho}\tilde{P}^{\nu\tau}+\tilde{P}^{\mu\tau}\tilde{P}^{\nu\rho}-2\tilde{P}^{\mu\nu}\tilde{P}^{\rho\tau}\}.
\end{equation}

\subsection{projector onto physical metric fluctuations}

In flat space we may define the projector onto the physical part of the metric
\begin{eqnarray}\label{eqn:F8a}
	&f_{\mu\nu}
	= P^{(f)}_{\mu\nu}{^{\rho\tau}}h_{\rho\tau},
	\qquad
	q^\mu P^{(f)}_{\mu\nu}{^{\rho\tau}}
	= 0,
	\qquad
	P^{(f)}_{\mu\nu}{^{\rho\tau}}q_\tau
	= 0,\\
	&P^{(f)}_{\mu\nu}{^{\rho\tau}}P^{(f)}_{\rho\tau}{^{\lambda\sigma}}
	= P^{(f)}_{\mu\nu}{^{\lambda\sigma}},
	\qquad
	P^{(f)}_{\mu\nu}{^{\rho\tau}}
	= P^{(f)}_{\nu\mu}{^{\rho\tau}}
	= P^{(f)}_{\mu\nu}{^{\tau\rho}}.
\end{eqnarray}
The projector is diagonal in momentum space,
\begin{equation}\label{eqn:92A}
	P^{(f)}_{\mu\nu}{^{\rho\tau}}(q,q^\prime) = P^{(f)}_{\mu\nu}{^{\rho\tau}}(q)\delta(q-q^\prime),
\end{equation}
and we will often omit the $\delta$-function for simplicity of notation. The projector onto physical metric fluctuations has a simple expression in terms of the transversal projector,
\begin{equation}\label{eqn:F10aa}
	P^{(f)}_{\mu\nu}{^{\rho\tau}}
	= \frac{1}{2}(\tilde{P}_\mu{^\rho}\tilde{P}_\nu{^\tau}+\tilde{P}_\mu{^\tau}\tilde{P}_\nu{^{\rho}}).
\end{equation}

The orthogonal projector $P^{(a)}_{\mu\nu}{^{\rho\tau}}$ projects onto the gauge fluctuations and obeys
\begin{equation}\label{eqn:103AA}
	P^{(a)\rho\tau}_{\mu\nu}h_{\rho\tau}
	= a_{\mu\nu}.
\end{equation}
It is given by
\begin{eqnarray}\label{eqn:54A}
	P^{(a)}_{\mu\nu}{^{\rho\tau}}
	&= \frac{1}{2}(\delta_\mu^{\rho}\delta_\nu^{\tau}+\delta_\mu^{\tau}\delta_\nu^{\rho})-P^{(f)}_{\mu\nu}{^{\rho\tau}} \\
	&= -q_\mu N_\nu{^\rho}q^\tau + (\mu\leftrightarrow\nu) + (\rho\leftrightarrow\tau),
\end{eqnarray}
where
\begin{equation}\label{eqn:104AA}
	N_\nu{^\rho}
	= -\frac{1}{q^2}\delta^\rho_\nu + \frac{1}{2q^4}q_\nu q^\rho.
\end{equation}
With
\begin{equation}\label{eqn:F9a}
	a_\nu
	= -\frac{i}{q^2}
	\left(\delta^\tau_\nu-\frac{q_\nu q^\tau}{2q^2}\right)q^\rho h_{\rho\tau}
\end{equation}
and
\begin{eqnarray}\label{eqn:F10a}
	a_{\mu\nu}
	&= i(q_\mu a_\nu + q_\nu a_\mu)\\
	&= \frac{1}{q^2}
	\left(q_\mu\delta^\tau_\nu + q_\nu\delta^\tau_\mu-\frac{q_\mu q_\nu q^\tau}{q^2}
	\right)q^\rho h_{\rho\tau}
\end{eqnarray}
one easily verifies \cref{eqn:103AA}.

The explicit form of $P^{(a)}$ reads
\begin{eqnarray}\label{eqn:111A}
	P^{(a) \rho \tau}_{\mu\nu}
	&= \frac{1}{2 q^2} \left( q_{\mu} q^{\rho} \delta_{\nu}^{\tau} + q_{\nu} q^{\rho} \delta_{\mu}^{\tau} +q_{\mu} q^{\tau} \delta_{\nu}^{\rho} +q_{\nu} q^{\tau} \delta_{\mu}^{\rho} \right)\\
    &\hphantom{={}} - \frac{1}{q^4}q_{\mu} q_{\nu} q^{\rho} q^{\tau}.
\end{eqnarray}
From there the explicit form of $P^{(f)}$ obtains easily as
\begin{equation}\label{eqn:111B}
	P^{(f) \rho \tau}_{\mu\nu} = \frac{1}{2} \left( \delta_{\mu}^{\rho} \delta_{\nu}^{\tau} + \delta_{\mu}^{\tau} \delta_{\nu}^{\rho} \right) - P^{(a) \rho \tau}_{\mu\nu}.
\end{equation}

\subsection{Correlation function}

The Green's function for the physical metric fluctuations takes the form
\begin{eqnarray}\label{eqn:F12a}
	G_{\rho\tau\sigma\lambda}(q,q^\prime) &=\langle f_{\rho \tau}(q)f_{\sigma\lambda}(-q^\prime)\rangle_c\\
	&=P^{(f)}_{\rho\tau}{^{\rho^\prime\tau^\prime}}(q)\langle h_{\rho^\prime\tau^\prime}(q) h_{\sigma^\prime\lambda^\prime}(-q^\prime) \rangle_c P^{(f)}_{\sigma\lambda}{^{\sigma^\prime\lambda^\prime}}(q^\prime).
\end{eqnarray}
It vanishes when contracted with $q^\rho,q^\tau,q^{\prime\sigma}$ or $q^{\prime\lambda}$. The defining propagator equation \labelcref{eqn:37AC} reads
\begin{equation}\label{eqn:100A}
	\Gamma^{(2)\mu\nu\rho\tau}G_{\rho\tau\sigma\lambda}
	= P^{(f)\mu\nu}{_{\sigma\lambda}}.
\end{equation}
The propagator on the r.h.s of \cref{eqn:100A} corresponds to the unit matrix in the space of the physical metric fluctuations $f_{\mu\nu}(q)$. From \cref{eqn:100A,eqn:F11B} we infer the propagator equation
\begin{equation}\label{eqn:F13a}
	\frac{iM^2}{4}A^{\mu\nu\rho\tau}(q)G_{\rho\tau\sigma\lambda}(q,q^\prime) = P^{(f)\mu\nu}{_{\sigma\lambda}}(q)\delta(q-q^\prime),
\end{equation}
where
\begin{equation}\label{eqn:F14a}
	A^{\mu\nu\rho\tau}(q) = \frac{q^2}{2}(\tilde{P}^{\mu\rho}\tilde{P}^{\nu\tau}+\tilde{P}^{\mu\tau}\tilde{P}^{\nu\rho}-2\tilde{P}^{\mu\nu}\tilde{P}^{\rho\tau})
\end{equation}
obeys
\begin{equation}\label{eqn:F14Aa}
	P^{(f)\alpha\beta}{_{\mu\nu}}(q)A^{\mu\nu\rho\tau}(q) = A^{\alpha\beta\rho\tau}(q).
\end{equation}

We impose translation symmetry in space and time
\begin{equation}\label{eqn:F15a}
	G_{\rho\tau\sigma\lambda}(q,q^\prime) = G_{\rho\tau\sigma\lambda}(q)\delta(q-q^\prime),
\end{equation}
such that for every $q$ we have the matrix-type equation
\begin{equation}\label{eqn:F16a}
	A^{\mu\nu\rho\tau}(q)G_{\rho\tau\sigma\lambda}(q)
	= -\frac{4i}{M^2}P^{(f)\mu\nu}{_{\sigma\lambda}}(q).
\end{equation}
The solution reads
\begin{eqnarray}\label{eqn:F17a}
	G_{\rho\tau\sigma\lambda}
	&= -\frac{2i}{M^2 q^2}
	\left( \tilde{P}_{\rho\sigma}\tilde{P}_{\tau\lambda}+\tilde{P}_{\rho\lambda}\tilde{P}_{\tau\sigma}-\tilde{P}_{\rho\tau}\tilde{P}_{\sigma\lambda}\right)\\
	&= -\frac{2i}{M^2 q^6}
	\Bigl\{ q^4(\eta_{\rho\sigma}\eta_{\tau\lambda}+\eta_{\rho\lambda}\eta_{\tau\sigma}-\eta_{\rho\tau}\eta_{\sigma\lambda})\\
	&-q^2(\eta_{\rho\sigma}q_\tau q_\lambda + \eta_{\tau\lambda}q_\rho q_\sigma + \eta_{\rho\lambda}q_\tau q_\sigma + \eta_{\tau\sigma}q_\rho q_\lambda\\
	&-\eta_{\rho\tau}q_\sigma q_\lambda-\eta_{\sigma\lambda}q_\rho q_\tau)
	 + q_\rho q_\tau q_\sigma q_\lambda\Bigr\}.
\end{eqnarray}
It is manifestly Lorentz covariant, with both sides of \cref{eqn:F17a} transforming as appropriate tensors. The ingredient that makes the inversion of $\Gamma^{(2)}$ unique is four-dimensional translation invariance.

The components of $G$ are
\begin{eqnarray}\label{eqn:67A}
    & G_{0000}
    =-\frac{2ik^4}{M^2 q^6},\\
    & G_{m000}
	= \frac{2i\omega k^2 k_m}{M^2 q^6},\\
    & G_{m0n0}
	= \frac{2i}{M^2}\left\{\frac{k^2}{q^4}\delta_{mn}-\frac{k^2 k_m k_n}{q^6}\right\},\\
    & G_{mn00}
	= -\frac{2i}{M^2}
    \left\{\frac{k^2\delta_{mn}-2k_m k_n}{q^4}+\frac{k^2 k_m k_n}{q^6}\right\}, \\
    &\begin{aligned}
        G_{mnp0}
        = - \frac{2 i \omega}{M^2} \Bigl\{& \frac{1}{q^4} \left( \delta_{m p} k_n + \delta_{n p} k_m - \delta_{m n} k_p \right)\\
        &- \frac{k_m k_n k_p}{q^6} \Bigr\},
    \end{aligned}\\
    & G_{mnpq} = - \frac{2 i}{M^2} \Bigl\{ \frac{1}{q^2} \left( \delta_{m p} \delta_{n q} + \delta_{m q} \delta_{n p} - \delta_{m n} \delta_{p q} \right)\\
    & - \frac{1}{q^4} ( \delta_{m p} k_n k_q + \delta_{n q} k_m k_p + \delta_{m q} k_n k_p + \delta_{n p} k_m k_q\\
    & - \delta_{m n} k_p k_q - \delta_{p q} k_m k_n ) + \frac{k_m k_n k_p k_q}{q^6} \Bigr\},
\end{eqnarray}
with all other components obtained by appropriate index permutations using the symmetries of $G$. For fixed $k_m$ we observe that the divergence for $\omega^2\to k^2$ can be up to $q^{-6}$.

\subsection{Irreducible representations of Lorentz symmetry}

In flat space the irreducible representations of the Lorentz group are given for $f_{\mu\nu}$ by a scalar $\sigma$ and a traceless divergence free tensor $t_{\mu\nu}$,
\begin{eqnarray}\label{eqn:67B}
	&\sigma
	= f^\mu_\mu,
	\qquad
	f_{\mu\nu}
	= t_{\mu\nu}+\frac{1}{4}\sigma \eta_{\mu\nu}+\tilde{s}_{\mu\nu},\\
	&\tilde{s}_{\mu\nu}
	= \left(\frac{1}{12}\eta_{\mu\nu}-\frac{q_\mu q_\nu}{3q^2}\right)\sigma,
\end{eqnarray}
with $t_{\mu\nu}$ traceless and divergence free,
\begin{equation}\label{eqn:67C}
	t^\mu_\mu = 0,
	\qquad
	q^\nu t_{\mu\nu}
	= 0.
\end{equation}
This is a special case of \cref{eqn:46R}. Using
\begin{equation}\label{eqn:67D}
	f_{\mu\nu}
	= t_{\mu\nu} + s_{\mu\nu} = t_{\mu\nu}+\frac{1}{3}\tilde{P}_{\mu\nu}\sigma
\end{equation}
one has $\hat{S}_{\mu\nu} = \tilde{P}_{\mu\nu}/3$. Contractions with the transversal projector obey the simple properties
\begin{equation}\label{eqn:67E}
	\tilde{P}^{\mu\nu}t_{\mu\nu}
	= 0,
	\qquad
	\tilde{P}^{\mu\nu}f_{\mu\nu}
	= \sigma.
\end{equation}

The metric correlation \labelcref{eqn:F17a}, \labelcref{eqn:67A} can be decomposed into contributions from the different irreducible Lorentz representations. For the scalar part we infer
\begin{eqnarray}\label{eqn:67F}
	G^{\sigma\sigma}_{\mu\nu\rho\tau}
	&= \frac{1}{9}\tilde{P}_{\mu\nu}\tilde{P}_{\rho\tau}G^{\sigma\sigma},\\
	G^{\sigma\sigma}
	&= \tilde{P}^{\rho\tau}\tilde{P}^{\sigma\lambda}G_{\rho\tau\sigma\lambda}
	= \frac{6i}{M^2 q^2}\sim \langle\sigma\sigma\rangle_c.
\end{eqnarray}
Here $\langle\sigma\sigma\rangle_c$ symbolizes the relation
\begin{equation}\label{eqn:74A}
	\langle\sigma(q)\sigma(-q^\prime)\rangle_c = G^{\sigma\sigma}\delta(q-q^\prime).
\end{equation}
The mixed term vanishes
\begin{equation}\label{eqn:67G}
	G^{t\sigma}_{\rho\tau}
	= G_{\rho\tau\sigma\lambda}\tilde{P}^{\sigma\lambda}-\frac{1}{3}\tilde{P}_{\rho\tau}G^{\sigma\sigma}
	= 0,
\end{equation}
and similar for $G^{\sigma t}_{\rho\tau}$. The transversal traceless correlation function therefore reads
\begin{eqnarray}\label{eqn:67H}
	G^{tt}_{\mu\nu\rho\tau}
	&= G_{\mu\nu \rho\tau}-\frac{1}{9}\tilde{P}_{\mu\nu}\tilde{P}_{\rho\tau}G^{\sigma\sigma}\\
	&= -\frac{2i}{M^2 q^2}
	(\tilde{P}_{\mu\rho}\tilde{P}_{\nu\tau}+\tilde{P}_{\mu\tau}\tilde{P}_{\nu\rho}-\frac{2}{3}\tilde{P}_{\mu\nu}\tilde{P}_{\rho\tau}),\\
	&\sim\langle t_{\mu\nu} t_{\rho\tau}\rangle_c.
\end{eqnarray}
These Lorentz invariant Green's functions correspond to particular solutions of the propagator equation \labelcref{eqn:46X2}.

We may employ the projector onto the traceless and divergence free part of the metric fluctuations,
\begin{equation}\label{eqn:86B4}
	P^{(t)}_{\mu\nu\rho\tau}
	= \frac{1}{2}(\tilde{P}_{\mu\rho}\tilde{P}_{\nu\tau}+\tilde{P}_{\mu\tau}\tilde{P}_{\nu\rho})-\frac{1}{3}\tilde{P}_{\mu\nu}\tilde{P}_{\rho\tau}.
\end{equation}
It obeys
\begin{eqnarray}\label{eqn:128A}
	&P^{(t) \rho \tau}_{\mu\nu} h_{\rho \tau}
	= t_{\mu\nu},\\
	&P^{(t) \rho \tau}_{\mu\nu} P^{(t) \sigma \lambda}_{\rho \tau}
	= P^{(t) \sigma \lambda}_{\mu\nu}.
\end{eqnarray}
In terms of this projector the Green's function reads
\begin{equation}\label{eqn:128B}
	G^{tt}_{\mu\nu\rho\tau} = - \frac{4 i}{M^2 q^2} P^{(t)}_{\mu\nu\rho\tau}.
\end{equation}
We also use the projector onto the $\sigma$-mode
\begin{equation}\label{eqn:128C}
	P^{(\sigma)}_{\mu\nu\rho\tau}= \frac{1}{3} \tilde{P}_{\mu\nu} \tilde{P}_{\rho \tau}
\end{equation}
with
\begin{equation}\label{eqn:128D}
	P^{(f)}_{\mu\nu\rho\tau} = P^{(t)}_{\mu\nu\rho\tau} + P^{(\sigma)}_{\mu\nu\rho\tau}
\end{equation}
and
\begin{equation}\label{eqn:128E}
	P^{(\sigma) \rho \tau}_{\mu\nu} h_{\rho \tau} = s_{\mu\nu}.
\end{equation}
In terms of this projector one has
\begin{equation}\label{eqn:128F}
	G_{\mu\nu\rho\tau}^{\sigma\sigma} = \frac{2 i}{M^2 q^2} P^{(\sigma)}_{\mu\nu\rho\tau}.
\end{equation}

We observe the relations
\begin{eqnarray}\label{eqn:128H}
	&P^{(t) \rho \tau}_{\mu\nu} P^{(f) \lambda \sigma}_{\rho \tau}
	= P^{(t) \lambda \sigma}_{\mu\nu},      \\
	&P^{(\sigma) \rho \tau}_{\mu\nu} P^{(f) \lambda \sigma}_{\rho \tau}
	= P^{(\sigma) \lambda \sigma}_{\mu\nu}, \\
	&P^{(t) \rho \tau}_{\mu\nu} P^{(\sigma) \lambda \sigma}_{\rho \tau}
	= 0.
\end{eqnarray}
The metric correlation function \labelcref{eqn:F17a} can be written in terms of the projectors $P^{(t)}$ and $P^{(\sigma)}$ as
\begin{equation}\label{eqn:128G}
	G_{\mu\nu\rho\tau} = G_{\mu\nu\rho\tau}^{t t} + G_{\mu\nu\rho\tau}^{\sigma \sigma} = - \frac{4 i}{M^2 q^2} \left( P_{\mu\nu\rho\tau}^{(t)} - \frac{1}{2} P_{\mu\nu\rho\tau}^{(\sigma)} \right).
\end{equation}
Similarly, the second functional derivative \labelcref{eqn:F3a} obeys
\begin{equation}\label{eqn:128I}
	\Gamma^{(2) \mu\nu\rho\tau} = \frac{i M^2 q^2}{4} \left( P^{(t) \mu\nu\rho\tau} -  2 P^{(\sigma) \mu\nu\rho\tau} \right).
\end{equation}
Using the projector properties \labelcref{eqn:128H} one verifies easily that \cref{eqn:128I,eqn:128G} obey the propagator equation.

\subsection{Irreducible representations of rotation symmetry}

With respect to the subgroup of space-rotations the trace $\sigma$ transforms as a scalar. The part $t_{\mu\nu}$ can be reduced to tensor, vector and scalar components, $\gamma_{mn}, W_m$ and $\kappa$,
\begin{eqnarray}\label{eqn:67I}
	t_{00}
	&= \kappa,
	\qquad
	t_{m0}
	= W_m -\frac{\omega k_m}{k^2}\kappa,\\
	t_{mn}
	&= \gamma_{mn}-\frac{\omega}{k^2}(k_m W_n + k_n W_m)\\
	&+\left(\frac{q^2\delta_{mn}}{2k^2}+\frac{2k^2 -3q^2}{2k^4}k_m k_n\right)\kappa,
\end{eqnarray}
with
\begin{equation}\label{eqn:67J}
	\delta^{mn}\gamma_{mn}
	= 0,
	\qquad
	k^m\gamma_{mn}
	= 0,
	\qquad
	k^m W_m = 0.
\end{equation}
We further decompose the metric correlation function into contributions from different representations of the rotation symmetry. This will be useful for the matching with more general geometries that represent homogeneous and isotropic cosmologies, but no longer exhibit Lorentz symmetry.

For the scalar part of the transversal correlation function we define
\begin{equation}\label{eqn:67K}
	G^{\kappa\kappa}
	= G^{\kappa\kappa}_{0000}
	= G^{tt}_{0000}
	= -\frac{8ik^4}{3M^2 q^6}\sim \langle \kappa\kappa\rangle_c.
\end{equation}
The scalar contribution to other index combinations of the transversal traceless correlation function \labelcref{eqn:67H} can be obtained by employing relations of the type
\begin{eqnarray}\label{eqn:151A}
	G^{\kappa\kappa}_{m000}
	&= \langle t^{(\kappa)}_{m0} t^{(\kappa)}_{00}\rangle_c = \langle t^{(\kappa)}_{m0}\kappa\rangle_c\\
	&= -\frac{\omega k_m}{k^2}\langle \kappa\kappa\rangle_c
	= -\frac{\omega k_m}{k^2}G^{\kappa\kappa},
\end{eqnarray}
with $t^{(\kappa)}_{m0}$ the part in $t_{m0}$ proportional to $\kappa$. This yields
\begin{eqnarray}\label{eqn:67L}
	G^{\kappa\kappa}_{m000}
	&= -\frac{\omega k_m}{k^2}G^{\kappa\kappa}
	= \frac{8i\omega k^2 k_m}{3M^2 q^6},\\
	G^{\kappa\kappa}_{m0n0}
	&= \frac{\omega^2 k_m k_n}{k^4}G^{\kappa\kappa}
	= -\frac{8i\omega^2 k_m k_n}{3M^2 q^6},\\
	G^{\kappa\kappa}_{mn00}
	&= \left(\frac{q^2\delta_{mn}}{2k^2}+\frac{8k^2 -3q^2}{3k^4}k_m k_n\right)
	G^{\kappa\kappa},
\end{eqnarray}
and similarly for the other components.

The transverse vector component obtains as
\begin{eqnarray}\label{eqn:67M}
	G^{WW}_{m0n0}
	&= G^{tt}_{m0n0}-G^{\kappa\kappa}_{m0n0}
	= G^{WW}_{mn}\\
	&= \frac{2i}{M^2 q^4}(k^2\delta_{mn}-k_m k_n)\sim \langle W_m W_n\rangle_c,
\end{eqnarray}
while
\begin{eqnarray}\label{eqn:67N}
	G^{WW}_{mnpq}
	&= \frac{\omega^2}{k^4}
	(k_m k_p G^{WW}_{nq}+k_m k_q G^{WW}_{np}\\
	&+k_n k_p G^{WW}_{mq}+k_n k_q G^{WW}_{mp})\\
	&= \frac{2ik^2}{M^2 q^4}
	(B_m B_p Q_{nq}+B_m B_q Q_{np}\\
	&+B_n B_p Q_{mq}+B_n B_q Q_{mp}),
\end{eqnarray}
and
\begin{eqnarray}\label{eqn:154A}
	G^{WW}_{mnp0}
	&= B_m G^{WW}_{np}+B_n G^{WW}_{mp}\\
	&= \frac{2ik^2}{M^2 q^4}(B_m Q_{np}+B_n Q_{mp}).
\end{eqnarray}
Here we employ the three dimensional projector
\begin{eqnarray}\label{eqn:670}
	Q_{mn}
	&= \delta_{mn}-\frac{k_m k_n}{k^2},\\
	\sum_n Q_{mn}Q_{np}
	&= Q_{mp},\\
	\sum_m k_m Q_{mn}
	&= 0,
	\qquad
	\sum_m Q_{mm}
	= 2.
\end{eqnarray}
We also introduce the shorthand
\begin{equation}\label{eqn:67P}
	B_m
	= -\frac{\omega k_m}{k^2}.
\end{equation}
For the objects carrying only space indices as $Q_{mn}$ or $B_m$ we will raise indices with $\delta^{mn}$, such that $Q^m_m = 2$, $Q^n_m B_n = 0$, etc.

We can write the decomposition of $t_{\mu\nu}$ in terms of $Q_{mn}$ and $B_m$
\begin{eqnarray}\label{eqn:67Q}
	t_{m0}
	&= W_m + B_m\kappa,\\
	t_{mn}
	&= \gamma_{mn}+B_m W_n + B_n W_m\\
	&+\left(\frac{q^2}{2k^2}Q_{mn}+B_m B_n\right)\kappa.
\end{eqnarray}
This simplifies the explicit representation of particular components, as
\begin{eqnarray}\label{eqn:67R}
	G^{\kappa\kappa}_{mnpq}
	&= \Bigl\{\frac{q^4}{4k^4}Q_{mn}Q_{pq}+\frac{q^2}{2k^2}
	(Q_{mn}B_p B_q + B_m B_n Q_{pq})\\
	&+B_m B_n B_p B_q\Bigr\}G^{\kappa\kappa}.
\end{eqnarray}

The transversal tensor part can now be extracted as
\begin{eqnarray}\label{eqn:67S}
	G^{\gamma\gamma}_{mnpq}
	&= G^{tt}_{mnpq}-G^{WW}_{mnpq}-G^{\kappa\kappa}_{mnpq}\\
	&= -\frac{2i}{M^2 q^2}
	(Q_{mq}Q_{np}+Q_{mq}Q_{np}-Q_{mn}Q_{pq})\\
	&\sim \langle\gamma_{mn}\gamma_{pq}\rangle_c.
\end{eqnarray}
We observe that $G^{\gamma\gamma}_{\mu\nu\rho\tau}$ vanishes if at least one of the indices equals zero, and $G^{WW}_{\mu\nu\rho\tau}$ is nonzero only for the index combinations $\mu\nu\rho\tau \in \{m0n0,\, m0np,\, mpn0,\, mpnq\}$, with permutations according to the symmetries $\mu\to\nu,\rho\leftrightarrow\tau$ and $(\mu\nu)\leftrightarrow(\rho\tau)$.
All mixed terms vanish, such that
\begin{equation}\label{eqn:86A}
	G_{\mu\nu\rho\tau}
	= G^{\sigma\sigma}_{\mu\nu\rho\tau}+G^{\kappa\kappa}_{\mu\nu\rho\tau}+G^{WW}_{\mu\nu\rho\tau}+G^{\gamma\gamma}_{\mu\nu\rho\tau}.
\end{equation}
It is straightforward to verify this by a direct computation.

The different pieces of the Green's function can also be obtained from \cref{eqn:128B} by the use of suitable projectors,
\begin{eqnarray}\label{eqn:PPA}
	P^{(t)}_{\mu\nu\rho\tau}
	= P^{(\gamma)}_{\mu\nu\rho\tau}
	 + P^{(W)}_{\mu\nu\rho\tau}+P^{(\kappa)}_{\mu\nu\rho\tau},
\end{eqnarray}
with
\begin{eqnarray}\label{eqn:PPB}
	G^{\gamma\gamma}_{\mu\nu\rho\tau}
	&= -\frac{4i}{\mu^2 q^2}P^{(\gamma)}_{\mu\nu\rho\tau},\\
	G^{WW}_{\mu\nu\rho\tau}
	&= -\frac{4i}{M^2 q^2}P^{(W)}_{\mu\nu\rho\tau},\\
	G^{\kappa\kappa}_{\mu\nu\rho\tau}
	&= -\frac{4i}{M^2 q^2}P^{(\kappa)}_{\mu\nu\rho\tau}.
\end{eqnarray}
The projector onto the graviton mode
\begin{equation}\label{eqn:86B9}
	P^{(\gamma) pq}_{mn}
	= \frac{1}{2}(Q^p_m Q^q_n +Q^q_m Q^p_n -Q_{mn}Q^{pq})
\end{equation}
obeys
\begin{equation}\label{eqn:86B9A}
	P^{(\gamma)pq}_{mn}h_{pq}
	= \gamma_{mn},
	\qquad
	P^{(\gamma)}_{mn}{^{pq}}P^{(\gamma)}_{pq}{^{rs}}
	= P^{(\gamma)}_{mn}{^{rs}},
\end{equation}
and
\begin{equation}\label{eqn:173A}
	P^{(\gamma)pq}_{mn}Q_{pq}
	= 0,
	\qquad
	P^{(\gamma)pq}_{mn}k_q = 0.
\end{equation}
As for $Q_{mn}$ the indices of $P^{\left( \gamma \right)}_{mnpq}$ are raised and lowered with $\delta^{mn}$, $\delta_{mn}$.
All components pf $P^{(\gamma)}$ with at least one index equal to zero vanish.

The projector onto the vector part obeys
\begin{eqnarray}\label{eqn:PPC}
	P^{(W)}_{mnpq}
	= -\frac{k^2}{2q^2}
	(B_m B_p Q_{nq}+B_m B_q Q_{np}\\
	 + B_n B_p Q_{mq}+B_n B_q Q_{mp}),
\end{eqnarray}
and
\begin{equation}\label{eqn:PPD}
	P^{(W)}_{mnp0}
	= -\frac{k^2}{2q^2}(B_m Q_{np}+B_n Q_{mp}),
\end{equation}
with similar components obtained by index-symmetries. One also has
\begin{equation}\label{eqn:PPE}
	P^{(W)}_{m0n0}
	= -\frac{k^2}{2q^2}Q_{mn},
	\qquad
	P^{(W)}_{mn00}
	= 0,
\end{equation}
while the components of $P^{(W)}$ with three of four indices zero vanish. One verifies the relations
\begin{eqnarray}\label{eqn:PPF}
	P^{(W)\rho\tau}_{m0}t_{\rho\tau}
	&= W_m,\\
	P^{(W)\rho\tau}_{mn}t_{\rho\tau}
	&= B_m W_n + B_n W_m,
\end{eqnarray}
as well as the projector property
\begin{equation}\label{eqn:PPG}
	P^{(W)\alpha\beta}_{\mu\nu}P^{(W)\rho\tau}_{\alpha\beta}+P^{(W)\rho\tau}_{\mu\nu},
\end{equation}
and the orthogonality
\begin{equation}\label{eqn:PPH}
	P^{(\gamma)\alpha\beta}_{\mu\nu}P^{(W)\rho\tau}_{\alpha\beta}
	= 0,
	\qquad
	P^{(W)\alpha\beta}_{\mu\nu}P^{(\gamma)\rho\tau}_{\alpha\beta}
	= 0.
\end{equation}
Finally, the projector $P^{(\kappa)}$ can be extracted from \cref{eqn:PPA}, employing \cref{eqn:PPC,eqn:PPD,eqn:PPE,eqn:86B9A,eqn:128A}. It is orthogonal to $P^{(\gamma)}$ and $P^{(W)}$ and obeys $P^{(\kappa)})^2 = P^{(\kappa)}$.

\subsection{Effective action for physical metric fluctuations}

So far we have computed the correlation function by first deriving the form \labelcref{eqn:F3a} of $\Gamma^{(2)}$ for arbitrary metric fluctuations $h_{\mu\nu}$, and subsequently inverting it on the space of functions $f_{\mu\nu}$. The resulting Green's function was then decomposed into irreducible representations of the symmetry groups. One may also proceed more directly by inserting $h_{\mu\nu} = t_{\mu\nu} + s_{\mu\nu}$ directly in $\Gamma_2$, decomposing into irreducible representations, taking functional derivatives with respect to these independent representations and performing the inversion at the end. We briefly show here that the two procedures are equivalent.

We first employ that the pieces for $t_{\mu\nu}$ and $\sigma$ decouple in $\Gamma_2$,
\begin{equation}\label{eqn:86B1}
	\Gamma_2 = \Gamma^{(t)}_2 + \Gamma^{(\sigma)}_2,
\end{equation}
with ($t_{\mu\nu}(-q) = t^\ast_{\mu\nu}(q)$)
\begin{eqnarray}\label{eqn:86B2}
	\Gamma^{(t)}_2
	&= \frac{iM^2}{8}\int_q q^2 t_{\mu\nu}(-q)t^{\mu\nu}(q)\\
	&= \frac{iM^2}{8}\int_q q^2 t^{\mu\nu}(-q)P^{(t)}_{\mu\nu\rho\tau}t^{\rho\tau}(q)
\end{eqnarray}
and
\begin{eqnarray}\label{eqn:86B3}
	\Gamma^{(\sigma)}_2
	&= \frac{iM^2}{8}\int_q q^2
	\left[ s_{\mu\nu}(-q)s^{\mu\nu}(q)-\frac{3}{4}\sigma(-q)\sigma(q)\right]\\
	&= -\frac{iM^2}{12}\int_q q^2\sigma(-q)\sigma(q).
\end{eqnarray}
Variation with respect to the independent fluctuations $t_{\mu\nu}$ and $\sigma$ yields
\begin{equation}\label{eqn:86B5}
	\Gamma^{(2)}_{\sigma\sigma}
	= -\frac{iM^2 q^2}{6}
\end{equation}
and
\begin{equation}\label{eqn:86B6}
	\Gamma^{(2)\mu\nu\rho\tau}_{tt}
	= \frac{iM^2 q^2}{4}P^{(t)\mu\nu\rho\tau}.
\end{equation}
These expressions are easily inverted. The corresponding correlation functions $G^{\sigma\sigma}$ and $G^{tt}_{\mu\nu\rho\tau}$ coincide with \cref{eqn:67F,eqn:67H}.

For comparison it is also instructive to decompose $f_{\mu\nu}$ into a trace and traceless part

\begin{eqnarray}\label{eqn:F5a}
	h_{\mu\nu}
	&= b_{\mu\nu}+\frac{1}{4}\sigma\eta_{\mu\nu}+a_{\mu\nu}
	= f_{\mu\nu}+a_{\mu\nu},\\
	\eta^{\mu\nu}b_{\mu\nu}
	&= 0,
	\qquad
	b_{\mu\nu}q^\nu
	= -\frac{1}{4} \sigma q_\mu.
\end{eqnarray}
One finds
\begin{eqnarray}\label{eqn:F6a}
	\Gamma^{(2)\mu\nu\rho\tau}h_{\rho\tau}
	&= \Gamma^{(2)\mu\nu\rho\tau}(b_{\rho\tau}+\frac{1}{4}\sigma\eta_{\rho\tau})\\
	&= \frac{iM^2}{4}\Bigl\{q^2 b^{\mu\nu}+(q^\mu q^\nu-\frac{3}{4}\eta^{\mu\nu}q^2)\sigma\Bigr\}
\end{eqnarray}
and therefore
\begin{equation}\label{eqn:F7a}
	\Gamma_{2}
	= \frac{iM^2}{8}\int_q q^2
	\bigl[b_{\mu\nu}(-q)b^{\mu\nu}(q)-\frac{3}{4}\sigma(-q)\sigma(q)\bigr].
\end{equation}
One should recall, however, that $b_{\mu\nu}$ is not unconstrained, cf. \cref{eqn:46C}, such that the $\sigma$-propagator cannot be obtained by variation of \cref{eqn:F7a} at fixed $b_{\mu\nu}$.

\subsection{Effective action for scalar, vector and graviton modes}

We can further decompose $\Gamma^{(t)}_2$ into pieces corresponding to the irreducible representations of the rotation group,
\begin{equation}\label{eqn:86B7}
	\Gamma^{(t)}_2 = \Gamma^{(\gamma)}_2 + \Gamma^{(W)}_2 + \Gamma^{(\kappa)}_2.
\end{equation}
With
\begin{equation}\label{eqn:86C}
	t_{\mu\nu}t^{\mu\nu}
	= \gamma^{mn}\gamma_{mn}-\frac{2q^2}{k^2}W^m W_m + \frac{3q^4}{2k^2}\kappa^2
\end{equation}
one obtains
\begin{eqnarray}\label{eqn:86B8}
	\Gamma^{(\gamma)}_2
	&= \frac{iM^2}{8}\int_q q^2\gamma^{mn}(-q)P^{(\gamma)}_{mnpq}\gamma^{pq}(q),\\
	\Gamma^{(W)}_2
	&= -\frac{iM^2}{4k^2}\int_q q^4 W^m(-q) Q_{mn}W^n(q),\\
	\Gamma^{(\kappa)}_2
	&= \frac{3iM^2}{16k^4}\int_q {q^6} \kappa(-q)\kappa(q).
\end{eqnarray}
The corresponding pieces of the second functional derivative are
\begin{eqnarray}\label{eqn:86B10}
	\Gamma^{(2)mnpq}_{\gamma\gamma}
	&= \frac{iM^2 q^2}{4}P^{(\gamma)mnpq},\\
	\Gamma^{(2)mn}_{WW}
	&= -\frac{iM^2 q^4}{2k^2}Q^{mn}\\
	\Gamma^{(2)}_{\kappa \kappa}
	&= \frac{3iM^2 q^6}{8k^4}.
\end{eqnarray}
The Green's functions $G^{\kappa\kappa},G^{WW}$ and $G^{\gamma\gamma}$ follow by simple inversion and coincide with \cref{eqn:67K,eqn:67M,eqn:67S}.

\subsection{Gauge invariant Bardeen potentials}

The physical metric fluctuations $f_{\mu\nu}$ or $ \gamma_{m n}, W_m, \kappa$ and $\sigma$ are ``gauge invariant'' in the same sense as the well known Bardeen potentials, i.e. that they are not affected by an infinitesimal diffeomorphism transformation of $\bar{g}_{\mu\nu}$. It is instructive to express the scalars $\sigma$ and $\kappa$ in terms of the ``gauge invariant'' Bardeen potentials \cite{Bar} $\Phi$ and $\Psi$, and to employ the gauge invariant vector fluctuation
\begin{equation}\label{eqn:86D}
	\Omega_m = \frac{q^2}{k^2}W_m.
\end{equation}
The correlation function for $\Omega_m$,
\begin{equation}\label{eqn:86E}
	G^{\Omega\Omega}_{mn}
	= \frac{q^4}{k^4}G^{WW}_{mn}
	= \frac{2i}{M^2 k^2}Q_{mn}\sim\langle \Omega_m\Omega_n\rangle_c
\end{equation}
is independent of $\omega$. This shows that $\Omega_m$ is not a propagating degree of freedom, but rather a constrained field.

The Bardeen potentials are given by (cf. \cref{sec:mode equation})
\begin{eqnarray}\label{eqn:86F}
	\Phi
	&= \frac{q^2}{4k^2}\kappa + \frac{1}{6}\sigma,\\
	\Psi
	&= \frac{q^2(k^2 -3q^2)}{4k^4}\kappa + \frac{1}{6}\sigma,
\end{eqnarray}
such that
\begin{eqnarray}\label{eqn:86G}
	\kappa
	&= \frac{4k^4}{3q^4}(\Phi-\Psi),\\
	\sigma
	&= \frac{2}{q^2}\bigl[k^2\Psi + (3q^2 -k^2)\Phi\bigr].
\end{eqnarray}
The correlation functions for the Bardeen potentials read
\begin{eqnarray}\label{eqn:86H}
	G^{\phi\phi}
	&= \frac{q^4}{16k^4}G^{\kappa\kappa}+\frac{1}{36}G^{\sigma\sigma}
	= 0\\
	G^{\Psi\Psi}
	&= \frac{q^4(k^2 -3q^2)^2}{16k^8}G^{\kappa\kappa}+\frac{1}{36}G^{\sigma\sigma} \\
	&= \frac{i}{M^2 k^2}\left(1-\frac{3q^2}{2k^2}\right)
	= -\frac{i}{2M^2 k^2}\left(1-\frac{3\omega^2}{k^2}\right),\\
	G^{\Phi\Psi}
	&= G^{\Psi\Phi}
	= \frac{q^4(k^2 -3q^2)}{16k^6}
	G^{\kappa\kappa}+\frac{1}{36}G^{\sigma\sigma}
	= \frac{i}{2M^2 k^2}.
\end{eqnarray}
Again, the propagator matrix in the $(\Phi,\Psi)$-space has no pole for $k\neq 0$ such that $\Phi$ and $\Psi$ are not propagating.

The fact that the Bardeen-potentials are not propagating implies that their mode functions vanish in the absence of sources. This does not imply that the correlation function of the metric in the scalar and vector channel vanishes, as obvious from our explicit computations. This simple finding tells us that correlations functions cannot always be constructed as products of mode functions.

\subsection{Time-dependent correlation functions}

For cosmology one needs the metric correlation as a function of time. More precisely, the correlation function is bilinear in the fields and therefore involves two time arguments. The power spectrum of primordial fluctuations is given by the equal-time correlation function where the two time arguments coincide. In general, geometries relevant for cosmology do not show time translation invariance. Correlation functions are specified by initial conditions. Under many circumstances these initial values can be given by the Lorentz-invariant correlation function in flat space. This holds if the relevant momentum of the mode is much larger than all geometric scales given by curvature etc.

For the time translation invariant correlation in flat space $G$ only depends on the difference of the two time arguments. Starting from the Green's function in four-dimensional Fourier space derived previously, the dependence of the correlation function on time obtains by a Fourier transform
\begin{equation}\label{eqn:86I}
	G(\eta-\eta^\prime,\vec{k}) = \int\limits^\infty_{-\infty}\frac{d\omega}{2\pi}e^{-i\omega(\eta-\eta^\prime)}G(\omega,\vec{k}).
\end{equation}
Here we use the symbol $\eta$ for conformal time in view of later comparison with a homogeneous and isotropic background. (For flat space with $a = 1$ one has $t = \eta$.)
The time translation symmetry is reflected by a time dependence only involving the difference $\eta-\eta^\prime$. Analytic continuation replaces $\omega\to \omega(1 + i\epsilon)$, $q^2\to k^2 -\omega^2 -2i\epsilon\omega^2$, and the determinant $\sqrt{g} = i(1 + i\epsilon)$. We define the $\omega$-integration as the limit $\epsilon\to 0$ of the analytically continued integration. This fixes the integration contour around the poles of the propagator. (See ref. \cite{CW2} for a motivation of this procedure for the context of cosmology.) For example, one has
\begin{eqnarray}\label{eqn:86J}
	-\int_\omega\frac{i}{q^2}e^{-i\omega(\eta-\eta^\prime)}
	&= -\lim_{\epsilon\to 0}\int\limits^\infty_{-\infty}\frac{d\omega}{2\pi}
	\frac{i(1 + i\epsilon)e^{-i\omega(\eta-\eta^\prime)}}{k^2 -\omega^2 -2i\epsilon\omega^2}\\
	&= \frac{1}{2k} e^{-ik|\eta-\eta^\prime|},
\end{eqnarray}
where $\int_\omega = \int d\omega/2\pi$ and $k>0$. We infer from \cref{eqn:67S} the flat space graviton propagator
\begin{eqnarray}\label{eqn:86K}
	G^{\gamma\gamma}_{mnpq}
	&= G_\text{grav}
	P^{(\gamma)}_{mnpq},\\
	G_\text{grav}(k,\eta,\eta^\prime)
	&= \frac{2}{M^2 k}e^{-ik|\eta-\eta^\prime|}.
\end{eqnarray}

The Fourier transforms of $q^{-4}$ and $q^{-6}$ follow from
\begin{eqnarray}\label{eqn:86L}
	-\int_\omega\frac{i}{q^4}e^{-i\omega(\eta-\eta^\prime)}
	&= 
	\frac{1}{2k}\frac{\partial}{\partial k}\int_\omega\frac{i}{q^2}e^{-i\omega(\eta-\eta^\prime)}\\
	&= \frac{1}{4k^3}\bigl[1 + ik|\eta-\eta^\prime|\bigr]
	e^{-ik|\eta-\eta^\prime|},
\end{eqnarray}
and
\begin{eqnarray}\label{eqn:86M}
	&-\int_\omega\frac{i}{q^6}e^{-i\omega(\eta-\eta^\prime)}
	= \frac{1}{4k}\frac{\partial}{\partial k}\int_\omega
	\frac{i}{q^4}e^{-i\omega(\eta-\eta^\prime)}\\
	&= \frac{3}{16k^5}\bigl[1 + ik|\eta-\eta^\prime|-\frac{k^2}{3}(\eta-\eta^\prime)^2\bigr]
	e^{-ik|\eta-\eta^\prime|}.
\end{eqnarray}
(We omit in these results a factor $\exp (-\epsilon|\eta-\eta^\prime|)$ that will be needed for a well defined transformation from three-dimensional Fourier space to position space.) We infer the scalar correlation functions
\begin{eqnarray}\label{eqn:86N}
	G^{\sigma\sigma}
	&= -\frac{3}{M^2 k}e^{-ik|\eta-\eta^\prime|},\\
	G^{\kappa\kappa}
	&= \frac{1}{2M^2 k}
	\left[1 + ik|\eta-\eta^\prime|-\frac{k^2}{3}(\eta-\eta^\prime)^2\right]
	e^{-ik|\eta-\eta^\prime|},
\end{eqnarray}
and the vector correlation
\begin{equation}\label{eqn:86O}
	G^{WW}_{mn}
	= -\frac{Q_{mn}}{2M^2 k}[1 + ik|\eta-\eta^\prime|\bigr]
	e^{-ik|\eta-\eta^\prime|}.
\end{equation}
We observe the negative value of the equal time correlation functions $G^{\sigma\sigma}$ and $G^{WW}$.

We also may employ the relation
\begin{equation}\label{eqn:86P}
	\int_\omega\omega f(\omega,k)e^{-i\omega(\eta-\eta^\prime)}
	= i\partial_\eta\int_\omega f(\omega,k)
	e^{-i\omega(\eta-\eta^\prime)}.
\end{equation}
One infers the Fourier transforms
\begin{eqnarray}\label{eqn:162A}
	\int_{\omega} \frac{\omega}{q^4} e^{- i \omega (\eta - \eta^\prime)} &= - \frac{(\eta - \eta^\prime)}{4 k} e^{-i k |\eta -\eta^\prime|},                                  \\
	\int_{\omega} \frac{\omega}{q^6} e^{- i \omega (\eta - \eta^\prime)} &= - \frac{(\eta - \eta^\prime)}{16 k^3} [1 + i k |\eta -\eta^\prime|] e^{-i k |\eta -\eta^\prime|}.
\end{eqnarray}
This allows us to compute all components of the full metric correlation \labelcref{eqn:F17a,eqn:67A}, e.g.
\begin{eqnarray}\label{eqn:86Q}
	G_{0000}
	&= \frac{3}{8M^2 k}
	\bigl[1 + ik|\eta-\eta^\prime|-\frac{k^2}{3}(\eta-\eta^\prime)^2\bigr]
	e^{-ik|\eta-\eta^\prime|}, \\
	G_{m000}
	&= -\frac{i k_m (\eta - \eta^\prime)}{8M^2 k} \left[ 1 + i k |\eta -\eta^\prime| \right] e^{-ik|\eta-\eta^\prime|},\\
	G_{m0n0}
	&= -\frac{1}{2M^2 k}\Bigl\{Q_{mn}\bigl[1 + ik|\eta-\eta^\prime|\bigr] \\
	&+\frac{1}{4}\frac{k_m k_n}{k^2}
	\bigl[1 + ik|\eta-\eta^\prime| + k^2(\eta-\eta^\prime)^2\bigr]\Bigr\}
	e^{-ik|\eta-\eta^\prime|}, \\
	G_{m n 00}
	&= \frac{1}{2 M^2 k} \Bigl\{ \left( 1 + i k |\eta - \eta^\prime| \right) \delta_{m n}\\
	&- \frac{k_m k_n}{4 k^2} \left[ 5 \left( 1 + i k |\eta - \eta^\prime| \right) + (\eta - \eta^\prime)^2 \right] \Bigr\} e^{- i k |\eta -\eta^\prime|}, \\
	G_{m n p 0}
	&= \frac{i (\eta - \eta^\prime)}{2 M^2 k} \Bigl\{ \delta_{m p} k_n + \delta_{n p} k_m\\
	&- \delta_{m n} k_p - \frac{k_m k_n k_p}{4 k^2} \left( 1 + i k |\eta - \eta^\prime| \right) \Bigr\} e^{-i k |\eta -\eta^\prime|}, \\
	G_{m n p q} &= \frac{1}{M^2 k} \Bigl\{ \delta_{m p} \delta_{n q} + \delta_{m q} \delta_{n p} - \delta_{m n} \delta_{p q}  \\
	&- \frac{1}{2 k^2} ( \delta_{m p} k_n k_q + \delta_{n q} k_m k_p + \delta_{m q} k_n k_p + \delta_{n p} k_m k_q \\
	&- \delta_{m n} k_p k_q - \delta_{p q} k_m k_n ) \left( 1 + i k |\eta - \eta^\prime|\right) \\
	&+ \frac{3 k_m k_n k_p k_q}{8 k^4} \Big ( 1 + i k |\eta -\eta^\prime| \\
	&- \frac{k^2}{3}(\eta -\eta^\prime)^2 \Big ) \Bigr\} e^{-i k |\eta -\eta^\prime|}.
\end{eqnarray}

As a check, we may compute the Newton potential from $G_{0000} = W^{(2)}_{0000}$ according to \cref{eqn:NL3}. From \cref{eqn:86Q} one obtains
\begin{equation}\label{eqn:NLA1}
	\int_{\eta^\prime}G_{0000}(\eta-\eta^\prime,\vec{k})
	= -\frac{2i}{M^2 k^2}.
\end{equation}
The three dimensional Fourier transform yields indeed the familiar form \labelcref{eqn:NL3}. We can also relate the Newton potential to the correlation functions for the Bardeen potentials using
\begin{equation}\label{eqn:NLA2}
	f_{00}
	= \frac{2k^2}{q^4}(\omega^2\Phi-k^2\Psi)
\end{equation}
and \cref{eqn:86H},
\begin{equation}\label{eqn:NLA3}
	G_{0000}
	= \frac{4k^6}{q^8}(k^2 G^{\Psi\Psi}-2\omega^2 G^{\Phi\Psi}).
\end{equation}
This demonstrates in a simple way that the correlation function for the Bardeen potentials cannot vanish.

The equal time correlation $G_{0000}$ is positive,
\begin{equation}\label{eqn:NLAB}
	G_{0000}(k,\eta,\eta) = \frac{3}{8M^2 k},
\end{equation}
while we observe negative values of the equal time correlation in the vector channel, e.g.
\begin{equation}\label{eqn:86R}
	\delta^{mn}G_{m0n0}(k,\eta,\eta)
	= -\frac{9}{8M^2 k}.
\end{equation}

The correlation functions \labelcref{eqn:86Q} show discontinuities at $\eta = \eta^\prime$. Applying $\eta$-derivatives may produce singular terms $\sim \delta (\eta -\eta^\prime)$. For first and second derivatives one finds that the only such term arises from
\begin{equation}\label{eqn:164A}
	\partial^2_{\eta} G_{m n p q}^{\left( sing \right)} = - \frac{2 i}{M^2} \left( \delta_{m p} \delta_{n q} + \delta_{m q} \delta_{n p} - \delta_{m n} \delta_{p q} \right) \delta (\eta - \eta^\prime).
\end{equation}
We also note the ``secular'' increase of $G$ for increasing $|\eta - \eta^\prime|$. This is essentially due to the presence of projectors. \Cref{eqn:100A} is an inhomogeneous second order differential equation. The projector on the r.h.s. shows itself secular behavior.

Indeed, the projector onto the physical metric fluctuations does not vanish for $\eta \neq \eta^\prime$,
\begin{equation}\label{eqn:155A}
	P^{(f)\mu\nu}{_{\rho\tau}}(\vec{k},\eta,\eta^\prime) = \int _\omega e^{-i\omega(\eta-\eta^\prime)}
	P^{(f)\mu\nu}{_{\rho\tau}}(\omega,\vec{k}).
\end{equation}
For example, the component $P^{(f)00}{_{00}}$ is the Fourier transform of $k^4/q^4$, e.g.
\begin{equation}\label{eqn:155B}
	P^{(f)00}{_{00}}
	= \frac{ik}{4}
	\left[1 + ik|\eta-\eta^\prime|\right]e^{-ik|\eta-\eta^\prime|}.
\end{equation}
It shows a secular increase for large $|\eta - \eta^\prime|$. For $\eta>\eta^\prime$ one has
\begin{eqnarray}\label{eqn:155C}
	(\partial^2_\eta + k^2)P^{(f)00}{_{00}}
	&= \frac{ik^3}{2}e^{-ik(\eta-\eta^\prime)},\\
	(\partial^2_\eta + k^2)^2 P^{(f)00}{_{00}}
	&= 0.
\end{eqnarray}
Only expressions as $(\partial^2_\eta + k^2)(\partial_{\eta^\prime}^2 + k^2) P^{(f)00}{_{00}}(\eta,\eta^\prime)$ or $(\partial_\eta^2 + k^2)^2 P^{(f)00}{_{00}}(\eta,\eta^\prime)$ are proportional to $\delta(\eta-\eta^\prime)$.

\subsection{Propagator with gauge fixing}

So far we have discussed the correlation function for the metric by restricting to the physical metric fluctuations and employing the corresponding projector $P^{(f)}$ in the propagator equation. An equivalent description of this correlation function can be found in a gauge fixed version if the gauge fixing enforces vanishing gauge fluctuations, $a_{\mu\nu} = 0$. (This does not hold for arbitrary gauge fixing.) For our purpose we may employ the gauge fixing
\begin{eqnarray}\label{eqn:GFA}
	\Gamma_\text{gf}
	&= \frac{1}{2\beta} \int \bar{g}^{1/2}h^\nu_{\mu;\nu}h^{\mu\rho}{_{;\rho}}\\
	&= \frac{i}{2\beta}\int_q q^\nu q^\rho \eta^{\mu\tau} h^\ast_{\mu\nu}(q)h_{\rho\tau}(q),
\end{eqnarray}
and take the limit $\beta\to 0$ at the end.

In the presence of gauge fixing we can consider unconstrained metric fluctuations $h_{\mu\nu}$. The metric correlation is defined for arbitrary $h_{\mu\nu}$ and depends, in general, on the gauge fixing. The gauge fixing \labelcref{eqn:GFA} adds to $\Gamma^{(2)}$ in \cref{eqn:F3a} a term
\begin{equation}\label{eqn:GFB}
	\Gamma^{(2)\mu\nu\rho\tau}_\text{gf}
	= \frac{i}{4\beta}
	(q^\mu q^\rho \eta^{\nu\tau}+q^\nu q^\rho \eta^{\mu\tau}+q^\mu q^\tau \eta^{\nu\rho}+q^\nu q^\tau \eta^{\mu\rho}).
\end{equation}
The propagator equation has now the unit matrix on the r.h.s. and no longer a projector
\begin{equation}\label{eqn:GFC}
	(\Gamma^{(2)}_{ph}+\Gamma^{(2)}_\text{gf})^{\mu\nu\rho\tau}
	G_{\rho\tau\sigma\lambda}
	= \frac{1}{2}
	(\delta^\rho_\mu\delta^\tau_\nu + \delta^\tau_\mu\delta^\rho_\nu).
\end{equation}
Here $\Gamma^{(2)}_{ph}$ is given by \cref{eqn:F3a} or \labelcref{eqn:F11B}. The operator $\Gamma^{(2)} = \Gamma^{(2)}_{ph} + \Gamma^{(2)}_\text{gf}$ can be inverted on the full space of arbitrary metric fluctuations.

For the solution of \cref{eqn:GFC} we make the ansatz
\begin{equation}\label{eqn:GFD}
	G_{\rho\tau\sigma\lambda}
	= G^{ph}_{\rho\tau\sigma\lambda}+\beta G^\text{gf}_{\rho\tau\sigma\lambda},
\end{equation}
with $G^{ph}$ given by \cref{eqn:F17a}. With
\begin{equation}\label{eqn:GFE}
	\Gamma^{(2)}_\text{gf}G^{ph}
	= 0
\end{equation}
\cref{eqn:GFC} becomes
\begin{eqnarray}\label{eqn:GFF}
	\biggl[\frac{i}{2}& (q^\mu q^\rho \eta^{\nu\tau}+q^\nu q^\rho\eta^{\mu\tau}) + \beta\Gamma^{(2)\mu\nu\rho\tau}_{ph}\biggr] G^\text{gf}_{\rho\tau\sigma\lambda}\\
	&= \frac{1}{2}(\delta^\mu_\sigma\delta^\nu_\lambda + \delta^\mu_\lambda\delta^\nu_\sigma)-P^{(f)\mu\nu}{_{\sigma\lambda}}
	= P^{(a)\mu\nu}{_{\sigma\lambda}}\\
	&= \frac{1}{2q^2}
	(q^\mu q_\sigma \delta^\nu_\lambda + q^\nu q_\sigma\delta^\mu_\lambda + q^\mu q_\lambda\delta^\nu_\sigma + q^\nu q_\lambda\delta^\mu_\sigma)\\
	&\hphantom{={}}- \frac{q^\mu q^\nu q_\sigma q_\lambda}{q^4}.
\end{eqnarray}
The solution reads
\begin{eqnarray}\label{eqn:GFG}
	G^\text{gf}_{\rho\tau\sigma\lambda}
	&= -\frac{i}{q^4} \Bigl( q_\rho q_\sigma \eta_{\tau\lambda}+q_\tau q_\sigma \eta_{\rho\lambda}+q_\rho q_\lambda \eta_{\tau\sigma}\\
	& +q_\tau q_\lambda \eta_{\rho\sigma}-\frac{3}{q^2} q_\rho q_\tau q_\sigma q_\lambda \Bigr),
\end{eqnarray}
and obeys
\begin{eqnarray}\label{eqn:GFH}
	\Gamma^{(2)}_{ph}G^\text{gf}
	= 0.
\end{eqnarray}
In the limit $\beta\to0$ the contribution from $G^\text{gf}$ to the metric correlation \labelcref{eqn:GFD} vanishes. We recover the result based on a formulation in terms of constrained physical metric fluctuations.

We can decompose $a_{\mu\nu}$ into two representations of the Lorentz group, a divergence free vector $c_{\mu}$ and a scalar $d$,
\begin{eqnarray}\label{eqn:178A}
	&a_{\mu}
	= c_{\mu} + \partial_{\mu} d,
	\qquad
	\partial^{\mu} c_{\mu}
	= 0,\\
	&a_{\mu\nu}
	= i \left(q_{\mu} c_{\nu} + q_{\nu} c_{\mu}\right) - q_{\mu} q_{\nu} d.
\end{eqnarray}
For this purpose we may employ projectors
\begin{equation}\label{eqn:178B}
	P_{\mu\nu}^{(d) \rho \tau}
	= \frac{1}{q^4} q_{\mu} q_{\nu} q^{\rho} q^{\tau},
	\qquad
	P_{\mu\nu}^{(d)\rho\tau} h_{\rho \tau}
	= - q_{\mu} q_{\nu} d,
\end{equation}
and
\begin{eqnarray}\label{eqn:178C}
	P_{\mu\nu}^{(c) \rho \tau}
	&= \frac{1}{2 q^2} \left( q_{\mu} q^{\rho} \delta_{\nu}^{\tau} + q_{\nu} q^{\rho} \delta_{\mu}^{\tau} + q_{\mu} q^{\tau} \delta_{\nu}^{\rho} + q_{\nu} q^{\tau} \delta_{\mu}^{\rho} \right)\\
	&- \frac{2}{q^4} q_{\mu} q_{\nu} q^{\rho} q^{\tau}, \\
	P_{\mu\nu}^{(c) \rho \tau} h_{\rho \tau} &= i \left( q_{\mu} c_{\nu} + q_{\nu} c_{\mu} \right).
\end{eqnarray}
They obey
\begin{equation}\label{eqn:178C2}
	P_{\mu\nu}^{(c)\rho\tau} P_{\rho \tau}^{(d) \lambda \sigma}
	= 0,
	\qquad
	P_{\mu\nu}^{(a) \rho \tau}
	= P_{\mu\nu}^{(c) \rho \tau} + P_{\mu\nu}^{(d) \rho \tau}.
\end{equation}

In terms of these projectors we may write
\begin{equation}\label{eqn:178E}
	\Gamma_{g f}^{(2) \mu\nu\rho\tau} = \frac{i q^2}{2 \beta} \left( P^{(c) \mu\nu\rho\tau} + 2 P^{(d) \mu\nu\rho\tau} \right),
\end{equation}
and
\begin{equation}\label{eqn:178F}
	G_{\mu\nu\rho\tau}^{g f} = - \frac{2 i}{q^2} \left( P^{ (c)}_{\mu\nu\rho\tau} + \frac{1}{2}P^{ (d)}_{\mu\nu\rho\tau} \right).
\end{equation}
Using the orthogonality of the projectors the propagator equation \labelcref{eqn:GFC} decays into four separate equations for the Lorentz representations $t_{\mu\nu}, \sigma, c_{\mu\nu}$ and $d$. For arbitrary $\beta$ the propagator equation for $G_{ph}$ is given by \cref{eqn:100A}.

\section{Off-shell metric propagator}
\label{sec:off-shell metric propagator}

In quantum gravity one needs the effective action for arbitrary values of the metric, at least in the vicinity of the final cosmological solution. This permits to get the response to arbitrary conserved sources by \cref{eqn:22}. Functional derivatives of the effective action yield field equations, inverse propagator and interactions. It is not sufficient to evaluate $\Gamma$ only for a given cosmological solution. A computation of the effective action for quantum gravity is an off-shell problem, and one therefore needs the off-shell propagator for the metric fluctuations.

If one employs the exact flow equation \labelcref{eqn:I1} one may use a given cosmological solution for the background metric that is used in the definition of the constraint on physical metric fluctuations or the physical gauge fixing, as well as in the definition of the IR-cutoff $R_k$. This cosmological solution refers to $k = 0$ or some fixed value $k_0$. Evaluating the flow of $\Gamma_k$ for metrics equal to the background metric will involve the on-shell propagator only if $k = k_0$. During the flow with $k\neq k_0$ the propagator $G_k$ will be off-shell. For $k\neq k_0$ the on-shell propagator would correspond to solutions of the field equations derived from $\Gamma_k + \Delta \Gamma_k$, where $\Delta\Gamma_k$ contains the cutoff term \cite{CWFE}. These solutions differ from the solutions of the field equations derived from $\Gamma_{k_0}$. The flow equation therefore involves the off-shell propagator even if we evaluate it for a solution of the field equations derived from $\Gamma_{k_0}$.

At the end one is interested in on-shell quantities as the power spectrum of fluctuations which can be extracted from the on-shell propagator. It is therefore interesting to understand how the particular properties of on-shell propagators arise within the extended space of off-shell propagators. Our approach treats off-shell propagators and on-shell propagators on equal footing. In this section we explicitly discuss the off-shell metric propagator in flat space. It obtains by admitting $V \neq 0$ in the action \labelcref{eqn:36}.

Expanding around a flat background in the presence of a nonvanishing cosmological constant $V$ adds to $\Gamma_2$ in \cref{eqn:F1a} a term
\begin{equation}\label{eqn:MP1}
	\Gamma^{(V)}_2 = iV\int_x\left(\frac{1}{8} h^2 -\frac{1}{4} h^\rho_\mu h^\mu_\rho\right).
\end{equation}
In turn, this supplements a constant piece to the second functional derivative
\begin{equation}\label{eqn:MP2}
	\Gamma^{(2)\mu\nu\rho\tau}_{(V)}
	= \frac{iV}{4}
	(\eta^{\mu\nu}\eta^{\rho\tau}-\eta^{\mu\rho}\eta^{\nu\tau}-\eta^{\mu\tau}\eta^{\nu\rho}).
\end{equation}
In momentum space this piece adds to \cref{eqn:F3a}.

Applying the inverse propagator on the gauge fluctuations yields
\begin{equation}\label{eqn:MP3}
	\Gamma^{(2)\mu\nu\rho\tau}_{(V)}a_{\rho\tau}
	= \frac{V}{2}
	(q^\mu a^\nu + q^\nu a^\mu -q^\rho a_\rho \eta^{\mu\nu}).
\end{equation}
In contrast to \cref{eqn:F4a} the off-shell inverse propagator is therefore no longer acting only in the space of physical fluctuations. For constraint physical metric fluctuations, or for a ``physical gauge fixing'' with $\alpha\to 0$, we therefore have to project onto the physical fluctuations. This is most easily done by insertion of the decomposition
\begin{eqnarray}\label{eqn:MP4}
	h_{\mu\nu}
	&= f_{\mu\nu}+a_{\mu\nu}\\
	&= t_{\mu\nu}+\frac{1}{3}\sigma\tilde{P}_{\mu\nu}+i(q_\mu a_\nu + q_\nu a_\mu)
\end{eqnarray}
into \cref{eqn:MP1},
\begin{equation}\label{eqn:MP5}
	\Gamma^{(V)}_2 = iV
	\left(\frac{\sigma^2}{24}-\frac{1}{4} t^\nu_\mu t^\mu_\nu + \Delta_V\right),
\end{equation}
with
\begin{equation}\label{eqn:MP6}
	\Delta_V = \frac{1}{4}(q_\mu a^\mu)^2 + \frac{i\sigma}{2}q^\mu a_\mu + \frac{1}{2} q^2 a_\mu a^\mu.
\end{equation}
The projection on the physical metric fluctuations eliminates $\Delta_V$. In the gauge fixed version this is achieved by enforcing $q_\mu a_\nu + q_\nu a_\mu = 0$ through the gauge condition. The projection on physical fluctuations replaces $h_{\mu\nu}$ by $f_{\mu\nu}$ in \cref{eqn:MP1}. This results in multiplication of $\Gamma^{(2)}_V$ in \cref{eqn:MP2} by $P^{(f)}$ from left and right.

Combining the part for $t_{\mu\nu}$ in \cref{eqn:MP5} with \cref{eqn:86B2} one recovers \cref{eqn:46JA}. Correspondingly, the second functional derivative \labelcref{eqn:86B6} is extended to
\begin{equation}\label{eqn:MP7}
	\Gamma^{(2)\mu\nu\rho\tau}_{tt}
	= \frac{iM^2}{4}
	\left(q^2 -\frac{2V}{M^2}\right)P^{(t)\mu\nu\rho\tau}.
\end{equation}
Similarly, the combination of the $\sigma$-dependent part in \cref{eqn:MP5} with \cref{eqn:86B3} is consistent with \cref{eqn:46U}. This modifies the second functional derivative \labelcref{eqn:86B5},
\begin{equation}\label{eqn:MP8}
	\Gamma^{(2)}_{\sigma\sigma}
	= -\frac{iM^2}{6}
	\left(q^2 -\frac{V}{2M^2}\right).
\end{equation}

For the representations of the rotation group \cref{eqn:MP7} can be taken over by using the decomposition \labelcref{eqn:PPA} of the projector $P^{(t)}$. Correspondingly, the off-shell correlation functions $G^{\gamma\gamma},G^{WW}$ and $G^{\kappa\kappa}$ obtain from their on-shell counterparts \labelcref{eqn:67K,eqn:67M,eqn:67S} by multiplication with $q^2/(q^2 -2V/M^2)$. For $G^{\sigma\sigma}$ the multiplicative factor is $q^2/(q^2 -V/2M^2)$. The index structure and $k$-dependence arising from the projectors remains unchanged.

These simple observations have important consequences for quantum gravity. For $V < 0$ the negative cosmological constant acts like a mass term for the graviton, with mass given by $\sqrt{-2V/M^2}$. It provides for an infrared cutoff for the graviton fluctuations. In contrast, a positive cosmological constant $V>0$ makes the graviton fluctuations tachyonic. The negative mass term $-2V/M^2$ leads to a strong infrared instability. This concerns the momentum modes with $q^2\approx 2V/M^2$.

One infers that the on-shell metric propagator has a very special place in the space of metric propagators. It is located at the boundary between stable and unstable behavior. The particular property that the graviton is massless is realized only on-shell.

The case of a positive cosmological constant is of particular interest. We concentrate on the graviton propagator, where for $G^{\gamma\gamma} = G_\text{grav} P^{(\gamma)}$ one has now in Fourier space
\begin{equation}\label{eqn:GAa}
	G_\text{grav}
	= -\frac{4i}{M^2}
	\left(q^2 -\frac{2V}{M^2}\right)^{-1}.
\end{equation}
The Fourier transform \labelcref{eqn:86I} to position space in time,
\begin{equation}\label{eqn:GB}
	G_\text{grav}(\eta-\eta^\prime,\vec{k})
	= -
	\frac{4i}{M^2}\int_\omega
	\frac{e^{-i\omega(\eta-\eta^\prime)}}{k^2 -2V/M^2 -\omega^2 -2i\epsilon\omega^2},
\end{equation}
obeys
\begin{equation}\label{eqn:GC}
	G_\text{grav}
	= \begin{cases}
		\frac{2}{M^2\bar{k}(V)} e^{-i\bar{k}(V)|\eta-\eta^\prime|} & \text{for } k^2 >\frac{2V}{M^2}\\
		\frac{2i}{M^2\bar{k}(V)}e^{-\bar{k}(V)|\eta-\eta^\prime|} & \text{for } k^2 <\frac{2V}{M^2}
	\end{cases}.
\end{equation}
where
\begin{equation}\label{eqn:GD}
	\bar{k}(V) = \sqrt{|k^2 -2V/M^2|}.
\end{equation}
One may check that $G_\text{grav}$ obeys the propagator equation, which reads for $k^2 <2V/M^2$
\begin{equation}\label{eqn:NA}
	-\frac{iM^2}{4}
	\bigl(\partial^2_\eta + \bar{k}^2(V)\bigr)
	G_\text{grav}(\eta,\eta^\prime) = \delta(\eta-\eta^\prime).
\end{equation}

The solution \labelcref{eqn:GC} is not the only solution of the propagator equation \labelcref{eqn:NA}. For $k^2 <2V/M^2$ another solution is
\begin{equation}\label{eqn:NB}
	G_\text{grav}
	= -\frac{2i}{M^2\bar{k}(V)}e^{\bar{k}(V)|\eta-\eta^\prime|}.
\end{equation}
This solution grows exponentially with $|\eta-\eta^\prime|$, instead of the exponential decay in \cref{eqn:GC}. The general solution can be constructed from mode functions $w^\pm$ that obey for $k^2 <2V/M^2$
\begin{equation}\label{eqn:NC}
	\bigl(\partial^2_\eta + \bar{k}^2(V)\bigr)w_\pm(\eta) = 0.
\end{equation}
We normalize the mode functions according to
\begin{equation}\label{eqn:ND}
	w^\pm_k(\eta)
	= [2 \bar{k}(V)]^{-1/2}	e^{\mp i\pi/4} 
	e^{\pm \bar{k}(V)\eta}.
\end{equation}
Taking into account the symmetry of the propagator one finds for the general solution to the propagator equation \labelcref{eqn:NA} (for $k^2 <2V/M^2)$,
\begin{eqnarray}\label{eqn:NE}
	G&_\text{grav}(k,\eta,\eta^\prime)
	= G_\text{grav}(k,\eta^\prime,\eta)
	= \frac{4}{M^2}\Bigl\{\\
	&c^-_k \bigl[w^-_k(\eta)w^{+\ast}_k(\eta^\prime)\theta(\eta-\eta^\prime)
	 + w^{+\ast}_k(\eta) w^-_k(\eta^\prime)\theta(\eta^\prime-\eta)\bigr]\\
	&+c^+_k\bigl[w^+_k(\eta) w^{-\ast}_k(\eta^\prime)\theta(\eta-\eta^\prime)
	 + w^{-\ast}_k(\eta) w^+_k(\eta^\prime)\theta(\eta^\prime-\eta)\bigr]\\
	&+d^-_k\bigl[w^-_k(\eta) w^{+\ast}_k(\eta^\prime) + w^{+\ast}_k(\eta) w^-_k(\eta^\prime)\bigr]\\
	&+d^+_k\bigl[w^+_k(\eta) w^{-\ast}_k(\eta^\prime) + w^{-\ast}_k(\eta) w^+_k(\eta^\prime)\bigr]\\
	&+e^+ w^+_k(\eta) w^{+\ast}_k(\eta^\prime) + e^- w^-_k(\eta) w^{-\ast}_k(\eta^\prime)\Bigr\}.
\end{eqnarray}
The inhomogeneous term on the r.h.s of \cref{eqn:NA} yields a constraint for the coefficients,
\begin{equation}\label{eqn:NF}
	c^-_k + c^+_k = 1.
\end{equation}
The other coefficients are free. Time translation symmetry is obeyed for $e^+ = e^- = 0$. The solution \labelcref{eqn:GC} corresponds to $c^-_k = 1$, with all other coefficients vanishing, while \cref{eqn:NB} is realized for $c^+_k = 1$ as the only non-vanishing coefficient.

Finding the ``correct propagator'' amounts to an initial value problem \cite{CW2}, with $G_\text{grav}^{in}(k)$ an ``initial correlation'', typically given for $\eta^\prime\to -\infty, \eta\to-\infty$, with $\eta$ close to $\eta^\prime$. In the approximation \labelcref{eqn:36} for the effective action the coefficients $c^\pm_k$, $d^\pm_k$, $e^\pm_k$ do not depend on $\eta$ or $\eta^\prime$. Realizing the correlation function \labelcref{eqn:GC} requires a particular initial value for which only $c^-_k = 1$ differs from zero. In a more complete treatment one expects that the propagator equation no longer remains linear \cite{CW2}. For $V\leq 0$ this may induce an approach to the correlation function \labelcref{eqn:GC} for rather general initial conditions. For $V>0$, $k^2 <2V/M^2$, however, the nonlinearities typically induce non-zero $d^\pm_k$. Even if $c^-_k$ remains one (and $c^+_k$ therefore zero), the parts $\sim d^\pm_k$ are exponentially growing and will overwhelm the exponentially decaying part $\sim c^-_k$. At the end the exponentially growing parts will win. For non-zero exponentially growing parts no Fourier transform to momentum space exists - this is the reason why the growing parts do not appear in the Fourier transform of \cref{eqn:GAa}.

In a quantum gravity computation an exponentially growing unstable propagator would lead to huge uncontrollable effects. It seems therefore plausible that quantum fluctuations act in a way such that the quantum effective action does not lead to this type of unstable propagator. This may be achieved by strong renormalization effects for $V$.

\section{Mode functions in homogeneous and isotropic cosmology}
\label{sec:homogeneous and isotropic cosmology}

In this section we turn to the discussion of the metric correlation function in cosmology. We assume a homogeneous and isotropic background geometry with vanishing spacial curvature. For the propagating graviton the general solution of the propagator equation can be described in terms of mode functions. The normalization of the correlation function is only restricted by the inhomogeneous term in the propagator equation - no explicit quantum field operators and associated commutation relations are needed \cite{CWneu}, \cite{CW2}. The mode functions coincide with the solution of the linearized Einstein equations only in the case where the background geometry obeys the field equations. For the vector and scalar modes contained in the metric the propagator equation cannot always be solved by mode functions.

\subsection{Metric fluctuations for homogeneous and isotropic cosmology}

A homogeneous and isotropic background metric with zero spatial curvature can be written in the form
\begin{equation}\label{eqn:H1}
	\bar{g}_{\mu\nu}
	= a^2(\eta)\eta_{\mu\nu},
	\qquad
	\mathcal{H}(\eta) = \frac{\partial \ln a(\eta)}{\partial\eta},
\end{equation}
with $\eta$ conformal time and $a(\eta)$ the scale factor. Analytic continuation can be easily implemented by admitting a complex phase factor for $\eta_{00}$ such that $\eta_{00} = -1$ for Minkowski signature and $\eta_{00} = 1$ for euclidean signature.
We will assume that $\eta$ extends from $-\infty$ to $+\infty$, as typical for realistic cosmologies. We also assume that boundary terms can be neglected in the sense that partial integration for $\eta$-dependent functions can be performed.

For certain limiting cases as de Sitter space the scale factor may diverge at finite $\eta$, e.g.

\begin{equation}
	a
	= - \frac{1}{H \eta},
	\qquad
	\mathcal{H}
	= - \frac{1}{\eta}.
\end{equation}
In this case one may formally patch a de Sitter geometry with increasing $a^2$ for $\eta < 0 $ to one with decreasing $a^2$ for $\eta > 0$ by taking the limit $\epsilon \to 0$ for
\begin{equation}
	a^2
	= \frac{1}{H^2\left( \eta - i \epsilon \right)^2},
	\qquad
	\mathcal{H}
	= - \frac{1}{\eta - i \epsilon}.
\end{equation}
Since the propagator equation is local in $\eta$ and the correlation function is fixed by initial values for a differential equation, this formal continuation does not matter for the correlation function for arguments $\eta, \eta^\prime$ obeying $|\eta / \epsilon| \gg 1, |\eta^\prime/\epsilon|\gg 1$.

It is convenient to perform a Fourier transformation in the three space dimensions, e.g.
\begin{equation}\label{eqn:H2}
	f_{\mu\nu}(x) = f_{\mu\nu}(\eta,\vec{x}) = \int \frac{d^3 k}{(2\pi)^3}
	e^{i\vec{k}\vec{x}}f_{\mu\nu}(\eta,\vec{k}),
\end{equation}
and similar for other fields. We assume space-translation symmetry of the correlation functions. The propagator equation can then be decomposed into separate equations for each $k$-mode.

For a homogeneous and isotropic cosmology the scalar covariant Laplacian reads explicitly
\begin{equation}\label{eqn:P1}
	D^2\sigma
	= -a^{-2}(\partial^2_\eta + 2\mathcal{H}\partial_\eta + k^2)\sigma,
\end{equation}
with $k^2 = \delta^{mn}k_m k_n = a^2 k^m k_m$. For the components of $\sigma_{;\mu\nu}$ one finds
\begin{eqnarray}\label{eqn:64A}
	\sigma_{;00}
	&= (\partial^2_\eta-\mathcal{H}\partial_\eta)\sigma,
	\qquad
	\sigma_{;m0}
	= ik_m(\partial_\eta-\mathcal{H})\sigma,\\
	\sigma_{;mn}
	&= -(k_m k_n +\mathcal{H}\delta_{mn}\partial_\eta)\sigma.
\end{eqnarray}
For a symmetric traceless tensor $b_{\mu\nu}$ one has $D^2 b_{\mu\nu} = b_{\mu\nu;}{^\rho}{_\rho}$, with
\begin{eqnarray}\label{eqn:P2}
	D^2 b_{00}
	&= -\frac{1}{a^2}\Bigl\{ \bigl(\partial^2_\eta-2\mathcal{H}\partial_\eta-2\partial_\eta\mathcal{H}-8\mathcal{H}^2 + k^2\bigr)b_{00}\\
	&+4i\mathcal{H}\delta^{jl}k_j b_{l0}\Bigr\},\\
	D^2 b_{m0}
	&= -\frac{1}{a^2}\Bigl\{\bigl(\partial^2_\eta-2\mathcal{H}\partial_\eta-2\partial_\eta\mathcal{H}-6\mathcal{H}^2 + k^2)b_{m0}\\
	&+2i\mathcal{H}k_m b_{00}+2i\mathcal{H}\delta^{jl}k_j b_{ml}\Bigr\},\\
	D^2 b_{mn}
	&= -\frac{1}{a^2}\Bigl\{(\partial^2_\eta-2\mathcal{H}\partial_\eta-2\partial_\eta\mathcal{H}-2\mathcal{H}^2 + k^2)b_{mn}\\
	&-2\mathcal{H}^2\delta_{mn}b_{00}
	 + 2\mathcal{H}(ik_m b_{n0}+ik_n b_{m0})\Bigr\}.
\end{eqnarray}

The constraint $f_{\mu\nu;}{^\nu} = 0$ translates to the relations
\begin{equation}\label{eqn:64C}
	(\partial_\eta + 4\mathcal{H})b^0_\mu + ik_m b^m_\mu
	= -\frac{1}{4}\partial_\mu\sigma,
\end{equation}
or
\begin{equation}\label{eqn:64D}
	ik_j\delta^{jl}b_{\mu l}
	= -\frac{a^2}{4}\partial_\mu\sigma + (\partial_\eta + 2\mathcal{H})b_{\mu 0}.
\end{equation}
This can be inserted into the first two equations \labelcref{eqn:P2}. Relations for the action of $D^2$ on traceless divergence free tensors $t_{\mu\nu}$ can be computed from \cref{eqn:P2}, cf. \cref{sec:mode decomposition}.

The non-vanishing components of the curvature tensor, Ricci tensor and curvature scalar read
\begin{eqnarray}\label{eqn:64B}
	R_{0m0n}
	&= \frac{\partial_\eta\mathcal{H}}{a^2}g_{00}g_{mn},\\
	R_{mnkl}
	&= \frac{\mathcal{H}^2}{a^2}(g_{mk}g_{nl}-g_{ml}g_{nk}),\\
	R_{00}
	&= -3\partial_\eta\mathcal{H},
	\qquad
	R_{mn}
	= (2\mathcal{H}^2 + \partial_\eta\mathcal{H})\delta_{mn},\\
	R
	&= \frac{6}{a^2}(\mathcal{H}^2 + \partial_\eta\mathcal{H}).
\end{eqnarray}
Equipped with these relations one may try to solve the propagator equation for homogeneous isotropic geometries. The complications of this task arise from the mixing of different components, e.g. in \cref{eqn:P2}.

\subsection{Mode functions}

In cosmology correlation functions are often assumed to be a product of mode functions. These mode functions are solutions of the linearized Einstein equations or suitable generalizations. The normalization is then provided by free quantum fields in an appropriate vacuum. Given our basic formulation of the correlation function as a solution of the propagator equation we can investigate systematically the conditions for this ansatz to work or fail.

Space-translation invariant Green's functions take in Fourier space the form
\begin{equation}\label{eqn:M1}
	G_{\rho\tau\sigma\lambda}(\eta,\vec{k};\eta^\prime,\vec{k}^\prime) = G_{\rho\tau\sigma\lambda}
	(\vec{k},\eta,\eta^\prime)\delta(k-k^\prime),
\end{equation}
where
\begin{equation}\label{eqn:M1A}
	G_{\rho\tau\sigma\lambda}(\eta,\vec{k};\eta^\prime,\vec{k}^\prime) = \int_{x,y}e^{-i(\vec{k}\vec{x}-\vec{k}^\prime\vec{y})}
	G_{\rho\tau\sigma\lambda}(\eta,\vec{x};\eta^\prime,\vec{y})
\end{equation}
and $\delta(k-k^\prime) = (2\pi)^3\delta^3(\vec{k}-\vec{k}^\prime)$. Similarly, for a homogeneous background metric $\bar{g}_{\mu\nu}(\eta,\vec{x}) = \bar{g}_{\mu\nu}(\eta)$ the operator $\Gamma^{(2)}$ is diagonal in momentum space,
\begin{equation}\label{eqn:M2}
	\Gamma^{(2)\mu\nu\rho\tau}
	(\eta^\prime,\vec{k}^\prime;\eta,\vec{k}) = \Gamma^{(2)\mu\nu\rho\tau}(\vec{k},\eta^\prime,\eta)\delta(k-k^\prime).
\end{equation}
For our purposes it can be written as a differential operator $D_\eta$ acting on $\eta$ (cf. \cref{eqn:47A})
\begin{equation}\label{eqn:M3}
	\Gamma^{(2)\mu\nu\rho\sigma}(\vec{k}, \eta^\prime,\eta) = \delta(\eta-\eta^\prime)D_{(\eta)}^{\mu\nu\rho\tau}(\vec{k}).
\end{equation}
The propagator equation \labelcref{eqn:29} takes then the form
\begin{equation}\label{eqn:M4}
	D_{(\eta)}^{\mu\nu\rho\tau}(\vec{k})G_{\rho\tau\sigma\lambda}(\vec{k},\eta,\eta^\prime) = E^{\mu\nu}{_{\sigma\lambda}}
	(\vec{k},\eta,\eta^\prime),
\end{equation}
where the Fourier transform of $E^{\mu\nu}{_{\sigma\lambda}}(x,y)$ is given by $E^{\mu\nu}{_{\sigma\lambda}}(\vec{k}, \eta,\eta^\prime)\delta(k-k^\prime)$. This is a system of independent differential equations for each value of $\vec{k}$. Rotation symmetry imposes further constraints on $D_{(\eta)}$ and $G$.

If the inhomogeneous term in the propagator equation is proportional $\delta(\eta-\eta^\prime)$ we can solve this equation in terms of mode functions, as we will see below. For the physical metric fluctuations $E^{\mu\nu}{_{\sigma\lambda}}(\vec{k},\eta,\eta^\prime)$ is given by the projector $P^{(f)\mu\nu}{_{\sigma\lambda}}(\vec{k},\eta,\eta^\prime)$, which does not necessarily vanish for $\eta\neq\eta^\prime$, cf. \cref{eqn:155B}. (An exception is the graviton mode.) If the homogeneous term is not proportional to $\delta(\eta-\eta^\prime)$ an expression of the propagator as a sum of products of mode functions is, in general, not possible. Examples are the correlation functions \labelcref{eqn:86Q} in Minkowski space which cannot be represented as simple products of mode functions.

For solving the propagator equation in terms of mode functions we have two options. Either one works with unconstrained metric fluctuations and includes in $\bar{D}_\eta$ a gauge fixing term as described in \cref{sec:projectors and gauge fixing}. In this case the mode equation reads
\begin{equation}\label{eqn:174X}
	D^{\mu\nu\rho\tau}_{(\eta)}h_{\rho\tau}
	= 0,
\end{equation}
where the definition of $D_{(\eta)}$ in \cref{eqn:M3} includes the contribution to $\Gamma^{(2)}$ from the gauge fixing term. As an alternative, one may decompose the physical metric fluctuations into unconstrained representations of the rotation group. For some representations, the propagator equation may take the form
\begin{equation}\label{eqn:174A}
	D_{(\eta)}(\vec{k})G(\vec{k},\eta,\eta^\prime) = \delta(\eta-\eta^\prime).
\end{equation}
Here $D_{(\eta)}$ and $G$ are $N\times N$ matrices with $N$ the dimension of the representation, and we have not written explicitly the unit matrix on the r.h.s. For rotation symmetric correlation functions different irreducible $SO(3)$-representations do not mix and can be treated separately. (There may be, however, several representations of the same type.) The dependence $\sim\delta(\eta-\eta^\prime)$ of the inhomogeneous term in \cref{eqn:174A} does not follow from the decomposition itself but needs particular properties.

We next construct the solution of the propagator equation in terms of mode functions. For this purpose we assume a propagator equation of the type \labelcref{eqn:174A}. For $\eta\neq \eta^\prime$ the differential equation \labelcref{eqn:174A} is homogeneous. The (generalized) mode functions $w^{(\alpha)}(\vec{k},\eta)$ are a set of linearly independent solutions of the homogeneous equation
\begin{equation}\label{eqn:M5}
	D_{(\eta)} (\vec{k})w^{(\alpha)}(\vec{k},\eta) = 0,
\end{equation}
such that the most general solution to the mode equation
\begin{equation}\label{eqn:M6}
	D_{(\eta)}(\vec{k})\tilde{w}
	(\vec{k},\eta) = 0
\end{equation}
can be written in the form
\begin{equation}\label{eqn:M7}
	\tilde{w}
	(\vec{k},\eta) = \sum_\alpha c_\alpha(\vec{k})w^{(\alpha)}(\vec{k}, \eta).
\end{equation}
A given mode function $w^{(\alpha)}$ is a $N$-component vector and \crefrange{eqn:M5}{eqn:M7} are vector equations. For purely real or imaginary $D_{(\eta)}$ the functions $w^{(\alpha)\ast}$ also obey \cref{eqn:M5} and we can equally well expand the general solution of \cref{eqn:M6} in terms of $w^{(\alpha)\ast}$.

We will next show that the general solution of the propagator equation \labelcref{eqn:174A} can be expressed in terms of the mode functions as
\begin{eqnarray}\label{eqn:M8}
	G_\text{st}(\vec{k},\eta,\eta^\prime)
	&= \sum_{\alpha,\beta}\bigl\{ d_{\alpha\beta}(\vec{k})w^{(\alpha)}_s(\vec{k},\eta)w^{(\beta)\ast}_t(\vec{k},\eta^\prime)\Theta(\eta-\eta^\prime)\\
	&+d_{\beta\alpha}(\vec{k})w^{(\alpha)\ast}_s(\vec{k},\eta)w^{(\beta)}_t
	(\vec{k},\eta^\prime)\Theta(\eta^\prime-\eta)\bigr\}\\
	&-iF_\text{st}(\vec{k},\eta)\delta(\eta-\eta^\prime).
\end{eqnarray}
From \cref{eqn:M5} one infers $D_{(\eta)}G(\eta,\eta^\prime) = 0$ for $\eta\neq \eta^\prime$. (This shows that the mode function ansatz \labelcref{eqn:M8} cannot work if the inhomogeneous term in \cref{eqn:M4} differs from zero for $\eta\neq \eta^\prime$.) The coefficients $d_{\alpha\beta}(\vec{k})$ will be constrained by the inhomogeneous term in \cref{eqn:174A} and by symmetries. They can be viewed as a hermitean matrix
\begin{equation}\label{eqn:M8A}
	d^\ast_{\beta\alpha}(\vec{k}) = d_{\alpha\beta}(\vec{k}).
\end{equation}

The Green's function is symmetric,
\begin{equation}\label{eqn:M10}
	G(\vec{k},\eta,\eta^\prime) = G^{\transpose}(\vec{k},\eta^\prime,\eta).
\end{equation}
For $\eta\neq\eta^\prime$ it can be written as
\begin{eqnarray}\label{eqn:M11}
	&G(\vec{k},\eta,\eta^\prime) = G^{(s)}(\vec{k},\eta,\eta^\prime)\\
	&\qquad + G^{(a)}(\vec{k},\eta,\eta^\prime)\bigl(\theta(\eta-\eta^\prime)-\theta(\eta^\prime-\eta)\bigr),
\end{eqnarray}
with $G^{(s)}$ and $G^{(a)}$ symmetric and antisymmetric, respectively. Furthermore, the reality condition implies that for Minkowski signature both $G^{(s)}$ and $G^{(a)}$ are hermitean. Thus $G^{(s)}$ is real and $G^{(a)}$ purely imaginary. This implies \cref{eqn:M8A}. For $\eta\neq\eta^\prime$ both $G^{(s)}$ and $G^{(a)}$ obey separately the homogeneous equation
\begin{equation}\label{eqn:M13}
	D_{(\eta)}(\vec{k})G^{s,a}(\vec{k},\eta,\eta^\prime) = 0.
\end{equation}
Using \cref{eqn:M10} this implies a similar equation for the dependence on $\eta^\prime$
\begin{equation}\label{eqn:M14}
	D_{(\eta^\prime)}(\vec{k})G^{(s,a)}(\vec{k},\eta,\eta^\prime) = 0.
\end{equation}

It is now straightforward to show \cref{eqn:M8}. \Cref{eqn:M13,eqn:M6,eqn:M7} imply for $\eta>\eta^\prime$
\begin{equation}\label{eqn:M15}
	G_\text{st}(\vec{k},\eta,\eta^\prime) = \tilde{c}_{\alpha,t}(\vec{k},\eta^\prime)w^{(\alpha)}_s(\vec{k},\eta),
\end{equation}
where $\tilde{c}_{\alpha,t}$ are $N$-component vectors for each $\alpha$. Using \cref{eqn:M14} we infer from \cref{eqn:M6,eqn:M7}
\begin{equation}\label{eqn:M16}
	\tilde{c}_{\alpha,t}(\vec{k},\eta^\prime) = \sum_\beta
	d_{\alpha\beta}(\vec{k})w^{(\beta)\ast}_t(\vec{k},\eta^\prime).
\end{equation}
This establishes \cref{eqn:M8} for $\eta>\eta^\prime$. The behavior for $\eta^\prime<\eta$ follows by \cref{eqn:M10}. With \cref{eqn:M8A} we conclude for $F_\text{st} = 0$ that the real part of $G$ is continuous at $\eta = \eta^\prime$, while the imaginary parts jumps.

We next turn to the inhomogeneous term in the propagator equation \labelcref{eqn:M4}. We concentrate first on $F_\text{st} = 0$. The operator $D_{(\eta)}$ contains a factor $\sqrt{\bar{g}} = ia^4$ and can be written as $D_{(\eta)} = iD^{(R)}_{(\eta)}$, with $D^{(R)}_{(\eta)}$ a real differential operator. The inhomogeneous term on the r.h.s of \cref{eqn:M4} is real. It is therefore related to the behavior of the purely imaginary part of $G$, as given by $G^{(a)}$. Indeed, this imaginary part shows a discontinuity at $\eta = \eta^\prime$, cf. \cref{eqn:M11}, which can produce the inhomogeneous term. In contrast, the propagator equation for the real part $G^{(s)}$ is homogeneous for all $\eta$ and $\eta^\prime$. We conclude that $G^{(s)}$ and $G^{(a)}$ obey separate propagator equations
\begin{eqnarray}\label{eqn:187}
	\bar{D}_\eta G^{(s)}
	&= 0,\\
	\bar{D}_\eta\bigl\{G^{(a)} \sign(\eta-\eta^\prime)\bigr\}
	&= \delta(\eta-\eta^\prime).
\end{eqnarray}
The linear equation for $G^{(s)}$ does not fix its amplitude, allowing typically for a large variety of solutions of the propagator equation. The correlation function will therefore be uniquely determined only once boundary conditions are specified. The issue has been discussed extensively in ref. \cite{CW2}.

As an example we consider an operator of the form
\begin{eqnarray}\label{eqn:M18}
	D_{(\eta)}
	= i\mathcal{A}(\eta)\bigl[\partial^2_\eta + 2\mathcal{C}(\eta)\partial_\eta+\mathcal{B}(\eta)\bigr],
\end{eqnarray}
with real functions $\mathcal{A}(\eta),\mathcal{B}(\eta),\mathcal{C}(\eta)$. The equation
\begin{equation}\label{eqn:M19}
	D_{(\eta)}\Bigl\{G^{(a)}(\eta,\eta^\prime)\bigl[\theta(\eta-\eta^\prime)-\theta(\eta^\prime-\eta)\bigr]\Bigr\}
	= \delta(\eta-\eta^\prime)
\end{equation}
is obeyed for
\begin{equation}\label{eqn:M20}
	\partial_\eta G^{(a)}(\eta,\eta^\prime)_{|\eta = \eta^\prime}
	= -
	\frac{i}{2\mathcal{A}(\eta = \eta^\prime)}.
\end{equation}
Differential equations of this type apply for operators of the type \labelcref{eqn:47A}, whereby the functions $\mathcal{A},\mathcal{B},\mathcal{C}$ may differ for the different $SO(3)$-representations contained in $h_{\mu\nu}$ (see later). For the graviton the normalization of $d_{\alpha\beta}$ inferred from \cref{eqn:M20} corresponds to the normalization following from the commutator relations for free quantum fields. On the level of the quantum effective action it is a direct consequence of the basic identity \labelcref{eqn:29} and only involves properties of ``classical fields''. No operators and commutation relations are involved in our formalism.

For an operator $D_{(\eta)}$ containing two $\eta$-derivatives the general solution of the mode equation \labelcref{eqn:M6} involves two linearly independent mode functions for each component of the vector $w_s$. They can be related by complex conjugation. For irreducible representations of the rotation group the solutions are degenerate and related by symmetry. Our boundary conditions, that can be related to properties under analytic continuation \cite{CW2}, typically admit only one independent mode function for each irreducible representation. If we restrict the setting to the physical metric fluctuations the differential operator $D_{(\eta)}$ will typically admit more than one independent mode function for a given irreducible representation. The sum \labelcref{eqn:M8} can no longer be written as a product of a given mode function $w(\eta)w^\ast(\eta^\prime)$. An exception is the graviton for which $D_{(\eta)}$ remains second order.

The most general solution of the propagator equation may also contain a term $-iF_\text{st}\delta(\eta-\eta^\prime)$. This does not contribute for $\eta\neq\eta^\prime$, but it can contribute to the solution of the inhomogeneous equation. A simple example is
\begin{equation}\label{eqn:192AA}
	D_{(\eta)}
	= i\mathcal{A}(\eta)\delta_\text{st},
	\qquad
	F_\text{st}
	= \frac{1}{\mathcal{A}(\eta)}\delta_\text{st},
\end{equation}
where $\mathcal{A}(\eta)$ contains no $\eta$-derivatives, e.g. $\mathcal{A}(\eta) = k^2$. If $\mathcal{A}(\eta)$ has no zero, the mode equation $D{(\eta)}\tilde{w} = 0$ has only the trivial solution $w = 0$ such that for $\eta\neq\eta^\prime$ $G^{(s)} = G^{(a)} = 0$. The propagator \labelcref{eqn:M8} therefore only involves the term $-iF_\text{st}\delta(\eta-\eta^\prime)$. It cannot be written as a product of mode functions or a sum of such products. More generally, the contribution of $F_\text{st}$ to the inhomogeneous term is
\begin{equation}\label{eqn:192AB}
	-iD_{(\eta)}\bigl[F\delta(\eta-\eta^\prime)\bigr] = f(\eta,\eta^\prime)\delta(\eta-\eta^\prime).
\end{equation}

We conclude that the use of mode functions for the correlations of the physical metric fluctuations is more involved than for a single scalar field (or the graviton). Mode functions can only be employed if the inhomogeneous term is $\sim \delta (\eta - \eta^\prime)$, as realized for unconstrained fields. Expressing the propagator equation in terms of constrained physical fluctuations typically leads to higher order differential operators $D_{\left( \eta \right)}$. Due to the appearance of projectors the inhomogeneous term is often no longer proportional $\delta(\eta-\eta^\prime)$. As a consequence, it is not possible to use mode functions directly for constrained fluctuations.

An alternative possibility for the use of mode functions may be the explicit use of gauge fixing. Once a solution for the propagator equation is found in terms of mode functions for unconstrained fields, the projection onto constrained physical fluctuations is responsible both for the higher derivative terms and the deviation of the inhomogeneous term from the $\delta$-distribution due to the appearance of a projector. Still, the issue remains rather involved. The operator $\Gamma^{(2)}$ mixes the different components of $h_{\mu\nu}$ and one expects the presence of a large number of different mode functions in the sum \labelcref{eqn:M8}.

Finally, even for an inhomogeneous term $\sim\delta(\eta-\eta^\prime)$ sums of products of mode functions are not the only possible solution of the propagator equation. There can be additional terms in $G$ which are proportional to $\delta(\eta-\eta^\prime)$ themselves. These terms $(\sim F)$ cannot be expressed as products of mode functions. For non-derivative forms of $D_{(\eta)}$ without zero eigenvalues, the solution $G \sim \delta(\eta-\eta^\prime)$ is the only solution of the propagator equation.

\subsection{Mode functions with gauge fixing}

One possibility for constructing the correlation function from solutions of the mode equation employs an explicit gauge fixing term
\begin{equation}\label{eqn:GA}
	\Gamma_\text{gf}
	= \frac{1}{2\beta}\int_x\bar{g}^{1/2} h^\nu_{\mu;\nu}h^{\mu\rho}{_{;\rho}}.
\end{equation}
This adds to $\bar{\Gamma}^{(2)}$ as given by \cref{eqn:47A} a term
\begin{align}\label{eqn:204XA}
	&\begin{aligned}
	    \Gamma^{(2)\mu\nu\rho\tau}_\text{gf}
	    &= -\frac{1}{4\beta}\sqrt{\bar{g}}
    	\{\bar{g}^{\mu\rho}D^\nu D^\tau +\bar{g}^{\nu\rho}D^\mu D^\tau\\
    	&\hphantom{{}
	= {}}+ \bar{g}^{\mu\tau}D^\nu D^\rho + \bar{g}^{\nu\tau}D^\mu D^\rho\},
	\end{aligned}\\
	\label{eqn:289A}
	&\Gamma^{(2)}
	= \bar\Gamma^{(2)}+\Gamma^{(2)}_\text{gf}.
\end{align}
In the limit $\beta\to 0$ the second functional derivative $\Gamma^{(2)}$ is dominated by $\Gamma^{(2)}_\text{gf}$.

For and understanding of the structure of the mode equation $\Gamma^{(2)\mu\nu\rho\tau} h_{\rho\tau} = 0$, we split
\begin{eqnarray}\label{eqn:204XC}
	h_{\mu\nu}
	&= f_{\mu\nu}+a_{\mu\nu}\\
	f_{\mu\nu;}{^\nu}
	&= 0,
	\qquad
	a_{\mu\nu;}{^\nu}
	= A_\mu.
\end{eqnarray}
For our choice of a physical gauge fixing the leading order mode equation
\begin{equation}\label{eqn:204XD}
	\Gamma^{(2)\mu\nu\rho\tau}_\text{gf}f_{\rho\tau}
	= 0
\end{equation}
is obeyed for arbitrary physical metric fluctuations. For the ``gauge modes'' $a_{\rho\tau}$ the leading order mode equation
\begin{equation}\label{eqn:204XE}
	\Gamma^{(2)\mu\nu\rho\tau}_\text{gf}a_{\rho\tau}
	= 0
\end{equation}
implies
\begin{equation}\label{eqn:204XF}
	A_{\mu;\nu}+A_{\nu;\mu}
	= 0,
\end{equation}
or
\begin{equation}\label{eqn:204XG}
	a_{\mu\rho;}{^\rho}{_\nu}+a_{\nu\rho;}{^\rho}{_\mu}
	= 0.
\end{equation}

For a homogeneous and isotropic background one has in Fourier space
\begin{eqnarray}\label{eqn:204XH}
	A_0
	&= -\frac{1}{a^2}(\partial_\eta + \mathcal{H})a_{00}+\frac{\delta^{jk}}{a^2}(ik_j a_{k0}-\mathcal{H} a_{jk}),\\
	A_j
	&= -\frac{1}{a^2}(\partial_\eta + 2\mathcal{H})a_{j0}+i\frac{\delta^{lk}}{a^2}k_l a_{jk}.
\end{eqnarray}
The condition \labelcref{eqn:204XF} reads
\begin{eqnarray}\label{eqn:204XI}
	&(\partial_\eta-\mathcal{H})A_0 = 0,\\
	&(\partial_\eta-2\mathcal{H})A_j + ik_j A_0 = 0,\\
	&i(k_l A_j + k_j A_l)-2\mathcal{H}\delta_{jl}A_0 = 0.
\end{eqnarray}
The first \cref{eqn:204XI} is solved by
\begin{equation}\label{eqn:204XJ}
	A_0
	 = a(\eta)c_0(k),
\end{equation}
while for $A_j$ one has
\begin{eqnarray}\label{eqn:204XK}
    	&A_j = ik_j L + T_j,
	\qquad
	k^j T_j = 0,\\
    	&\left(\partial_\eta - 2 \mathcal{H} \right) L + A_0 = 0, \\
    	&\left( \partial_\eta - 2 \mathcal{H} \right) T_j = 0,\\
    	&i \left( k_l T_j + k_j T_l \right) - 2 \mathcal{H} \delta_{j l} A_0 - 2 k_l k_j L = 0.
\end{eqnarray}

Combining \cref{eqn:204XJ,eqn:204XK} one obtains for $\vec{k} \neq 0$
\begin{eqnarray}\label{eqn:235B}
	A_0 = 0, \quad L = 0, \quad T_j = 0.
\end{eqnarray}
Discarding the special case $\vec{k} = 0$ we conclude that the only solution is
\begin{eqnarray}\label{eqn:235c}
	A_{\mu} = 0.
\end{eqnarray}
The remaining leading order mode equation for the vector $a_{\mu}$,
\begin{eqnarray}\label{eqn:301}
	D^{2} a_{\mu} + D^{\nu} D_{\mu} a_{\nu} = 0,
\end{eqnarray}
obtains from \cref{eqn:204XH} by insertion of $a_{\mu\nu} = a_{\mu;\nu} + a_{\nu;\mu}$.

By virtue of \cref{eqn:204XD} the full mode equation becomes
\begin{equation}\label{eqn:full mode eqn}
	\bar\Gamma^{(2) \mu\nu\rho\tau} \left( f_{\rho \tau} + a_{\rho \tau} \right) + \Gamma^{(2) \mu\nu\rho\tau}_\text{gf} a_{\rho \tau}
	= 0,
\end{equation}
with $\bar{\Gamma}^{(2) \mu\nu\rho\tau}$ given by \cref{eqn:47A}. One set of solutions corresponds to the physical modes $f\neq 0,a = 0$,
\begin{equation}\label{eqn:303}
	\bar\Gamma^{(2)}f = 0,
\end{equation}
while the gauge modes, $f = 0,a\neq 0$, obey
\begin{equation}\label{eqn:303A}
	(\Gamma^{(2)}_\text{gf}+\bar\Gamma^{(2)})a = 0.
\end{equation}
The most general solutions are linear combinations of the solutions of \cref{eqn:303} with solutions of \cref{eqn:303A}. For small $\beta$ we can solve \cref{eqn:303A} iteratively
\begin{eqnarray}\label{eqn:303B}
	&a = a_0 + \beta a_1 +\dots,
	\qquad
	\Gamma^{(2)}_\text{gf} a_0
	= 0,\\
	&\bar{\Gamma}^{(2)} a_0 + \beta \Gamma^{(2)}_\text{gf}a_1 = 0.
\end{eqnarray}
With $\Gamma^{(2)}_\text{gf}\sim 1/\beta$ one infers that $a_1$ and $a_0$ have the same scaling with $\beta$. For $\beta\to 0$ one can neglect the term $\sim\beta a_1$ in the solutions of \cref{eqn:303A} and we recover the leading expression given by \cref{eqn:204XE,eqn:301}. For $\beta\to 0$ the general solution of the mode equation combines physical modes $w_f$ obeying \cref{eqn:303} with gauge modes $w_a$ obeying \cref{eqn:204XE}.

Assume now that the propagator can be represented as a product of mode functions,
\begin{equation}\label{eqn:DA}
	G(\eta,\eta^\prime) = w^-(\eta) w^+(\eta^\prime)\theta(\eta-\eta^\prime) + w^+(\eta) w^-(\eta^\prime)\theta(\eta^\prime-\eta),
\end{equation}
with
\begin{equation}\label{eqn:DB}
	w^\pm(\eta) = w^\pm_f(\eta) + w^\pm_a(\eta).
\end{equation}
Here $w^+$ and $w^-$ are vectors in the space of mode functions and $G$ is therefore a matrix in this space, cf. \cref{eqn:M8}. (An extension to a sum of such products will be straightforward.) The propagator equation
\begin{equation}\label{eqn:DC}
	\Gamma^{(2)}(\eta) G(\eta,\eta^\prime) = \delta(\eta-\eta^\prime)
\end{equation}
requires that $w^\pm$ has non-vanishing components $w^\pm_f$ and $w^\pm_a$. Multiplying from left and right with $P^{(f)}$ yields
\begin{equation}\label{eqn:DE}
	P^{(f)}(\bar{\Gamma}^{(2)}+\Gamma^{(2)}_\text{gf})
	(P^{(f)}+P^{(a)})
	G P^{(f)}
	= P^{(f)},
\end{equation}
or, using $P^{(f)}\Gamma^{(2)}_\text{gf} = 0$,
\begin{equation}\label{eqn:DF}
	P^{(f)}\bar{\Gamma}^{(2)}P^{(f)}G^{ff}+P^{(f)}\bar{\Gamma}^{(2)}P^{(a)}G^{af}
	= P^{(f)},
\end{equation}
with
\begin{equation}\label{eqn:DG}
	G^{ff}
	= P^{(f)}GP^{(f)},
	\qquad
	G^{af}
	= P^{(a)}GP^{(f)}.
\end{equation}
The projected pieces $G^{ff}$ and $G^{af}$involve appropriate factors of $w_f$ and $w_a$ in the products \labelcref{eqn:DA}, according to
\begin{eqnarray}\label{eqn:DH}
	&P^{(f)}w_f = w_f,
	\qquad
	&&P^{(f)}w_a = 0,\\
	&P^{(a)}w_f = 0,
	\qquad
	&&P^{(a)}w_a = w_a.
\end{eqnarray}

For $\beta\to 0$ the part $w_a$ in \cref{eqn:DB} has to be $\sim \beta^{1/2}$ or smaller, since otherwise the piece $\sim \Gamma^{(2)}_\text{gf}G$ would yield a divergent contribution multiplying $\delta(\eta-\eta^\prime)$. Therefore $G^{af}\lesssim \sqrt{\beta}$, and we end for $\beta\to 0$ with the propagator equation for constrained physical fluctuations
\begin{equation}\label{eqn:DI}
	\Gamma^{(2)}_f G^{ff}
	= P_f,
	\qquad
	\Gamma^{(2)}_f = P_f\Gamma^{(2)}P_f.
\end{equation}
(This corresponds in the more abstract discussion of \cref{sec:projectors and gauge fixing} to \cref{eqn:A.13}.) As a consequence, the propagator for the physical fluctuations is expressed in terms of the mode functions $w_f$ as
\begin{eqnarray}\label{eqn:DJ}
	G_{(ph)}
	&= G^{ff}(\eta,\eta^\prime)\\
	&= w^-_f(\eta) w^+_f(\eta^\prime)\theta(\eta-\eta^\prime) + (\eta\leftrightarrow\eta^\prime).
\end{eqnarray}
The mode functions $w^\pm_f$ are solutions of the homogeneous equation \labelcref{eqn:303}, with $\bar\Gamma^{(2)}$ involving up to two derivatives with respect to $\eta$. The higher derivatives in the inverse propagator for constrained fields $\Gamma^{(2)}_f$ act on the product of mode functions such that the inhomogeneous term of the propagator equation for constrained fields,
\begin{equation}\label{eqn:HK}
	\Gamma^{(2)}_f \, G_{ph}
	= P^{(f)},
\end{equation}
equals the projector $P^{(f)}$.

\subsection{Linearized Einstein equations}

The mode functions are usually associated to the solutions of the linearized Einstein equations. If the background metric $\bar{g}_{\mu\nu}$ is a solution of the ``background'' field equation this indeed coincides with the more general definition \labelcref{eqn:M5}. For background geometries not obeying the field equations this simple coincidence is no longer valid. We compute here the mode equations for the (constrained) physical fluctuations, assuming the effective action \labelcref{eqn:36}. We recall however, that the resulting mode functions in the scalar and vector channel cannot be used for the construction of the correlation function.

For the effective action \labelcref{eqn:36} the expansion of the Einstein equation,
\begin{equation}\label{eqn:F1}
	R_{\mu\nu}-\frac{1}{2} Rg_{\mu\nu}
	= \frac{1}{M^2}T_{\mu\nu},
\end{equation}
reads in linear order in $h_{\mu\nu}$,
\begin{eqnarray}\label{eqn:F2}
	G_{(1)\mu\nu}
	&= R_{\mu\nu(1)}-\frac{1}{2}(R g_{\mu\nu})_{(1)}
	= \frac{1}{M^2}T_{(1)\mu\nu},\\
	G_{(1)\mu\nu}
	&= \frac{1}{2}\Bigl\{h^\rho_{\mu;\nu\rho}+h^\rho_{\nu;\mu\rho}-h_{\mu\nu;}{^\rho}{_\rho}-h_{;\mu\nu}\\
	&-\bar{R} h_{\mu\nu}-(h^{\rho\tau}{_{;\rho\tau}}-h_;{^\rho}{_\rho}-\bar{R}^{\rho\tau}h_{\rho\tau})\bar{g}_{\mu\nu}\Bigr\}.
\end{eqnarray}
For the action \labelcref{eqn:36} and in the absence of any further contributions to $T_{\mu\nu}$ one has
\begin{equation}\label{eqn:63}
	T_{(1)\mu\nu}
	= -V h_{\mu\nu}.
\end{equation}
Solutions of the background field equation relate $\bar{R}_{\mu\nu}$ to $V$.

We may evaluate the mode equation \labelcref{eqn:174X} for physical metric fluctuations $f_{\mu\nu}$,
\begin{equation}\label{eqn:118A}
	\int_y\Gamma^{(2)\mu\nu\rho\tau}(x,y)f_{\rho\tau}(y) = 0,
\end{equation}
with $\Gamma^{(2)}$ given in \cref{eqn:47A}. This may be compared to the solution of the linearized Einstein equation \labelcref{eqn:F2,eqn:63}, also evaluated for $h_{\mu\nu} = f_{\mu\nu}$. We show in \cref{sec:mode equation} that the two equations for $f_{\mu\nu}$ only coincide if the background metric obeys the field equations. Otherwise, the solution of the mode equation \labelcref{eqn:118A} differs from the solution of the linearized Einstein equation. The origin of this difference arises from the relation of the first functional derivative $\delta\Gamma/\delta g_{\mu\nu}$ and the Einstein equation, which involves a factor $g^{1/2}g^{\mu\rho}g^{\nu\tau}$. The linearization of the first functional derivative has to take this factor into account.

For a homogeneous and isotropic background \labelcref{eqn:H1} we may further elaborate the linearized Einstein equation. For a vanishing Weyl tensor $\bar{C}_{\mu\nu\rho\tau} = 0$ one has for the physical metric fluctuations $f_{\mu\nu}$
\begin{eqnarray}\label{eqn:F4}
	G_{(1)\mu\nu}
	&= \frac{1}{2}\Bigl\{ 2\bar{R}^\rho_\mu f_{\nu\rho}+2\bar{R}^\rho_\nu f_{\mu\rho}-\bar{R}_{\mu\nu}f-\frac{4}{3} \bar{R}f_{\mu\nu}\\
	&+\frac{1}{3} \bar{R}f\bar{g}_{\mu\nu}-f_{\mu\nu;}{^\rho}{_\rho}-f_{;\mu\nu}+f_;{^\rho}{_\rho}\bar{g}_{\mu\nu}\Bigr\}.
\end{eqnarray}

We next split $f_{\mu\nu} = b_{\mu\nu}+\sigma\bar{g}_{\mu\nu}/4$, according to \cref{eqn:46C}.
\begin{eqnarray}\label{eqn:F5}
	G_{(1)\mu\nu}
	&= \frac{3}{8}\sigma_;{^\rho}{_\rho}\bar{g}_{\mu\nu}-\frac{1}{2}\sigma_{;\mu\nu}-\frac{1}{2} b_{\mu\nu;}{^\rho}{_\rho}\\
	&+\bar{R}^\rho_\mu b_{\nu\rho}+\bar{R}^\rho_{\nu} b_{\mu\rho}-\frac{2}{3}\bar{R} b_{\mu\nu},
\end{eqnarray}
with
\begin{eqnarray}\label{eqn:70A}
	G_{(1)00}
	&= \frac{1}{2a^2}(\partial^2_\eta + 2\mathcal{H}\partial_\eta + 2\partial_\eta\mathcal{H}-8\mathcal{H}^2 + k^2)b_{00}\\
	&+\frac{1}{8}(-\partial^2_\eta +6\mathcal{H}\partial_\eta + 3k^2)\sigma,\\
	G_{(1)m0}
	&= \frac{1}{2a^2}(\partial^2_\eta-2\partial_\eta\mathcal{H}-6\mathcal{H}^2 + k^2)b_{m0}\\
	&+ik_m\left(\mathcal{H}\frac{b_{00}}{a^2}-\frac{1}{2}\partial_\eta\sigma + \frac{1}{4}\mathcal{H}\sigma\right),\\
	G_{(1)mn}
	&= \frac{1}{2a^2}\bigl\{(\partial^2_\eta-2\mathcal{H}\partial_\eta-6\partial_\eta\mathcal{H}-2\mathcal{H}^2 + k^2)b_{mn}\\
	&-2\mathcal{H}^2\delta_{mn}b_{00}+2\mathcal{H}i(k_m b_{n0}+k_n b_{m0})\bigr\}\\
	&-\frac{1}{8}(3\partial^2_\eta + 2\mathcal{H}\partial_\eta + 3k^2)\delta_{mn}\sigma + \frac{1}{2} k_m k_n\sigma.
\end{eqnarray}
Here we recall that the covariant derivatives of $b_{\mu\nu}$ and $\sigma$ are related by \cref{eqn:46C}.

For the trace \cref{eqn:70A,eqn:64D} yield
\begin{eqnarray}\label{eqn:70B}
	a^2 G^{(1)}_{\mu\nu}\bar{g}^{\mu\nu}
	&= G_{(1)mn}\delta^{mn}-G_{(1)00}\\
	&= \frac{4}{a^2}(\mathcal{H}^2 -\partial_\eta\mathcal{H})b_{00}-(\partial^2_\eta+{2}\mathcal{H}\partial_\eta + k^2)\sigma.
\end{eqnarray}
In particular, for a de Sitter background one has $\partial_\eta\mathcal{H} = \mathcal{H}^2$ such that \cref{eqn:70B} only involves $\sigma$. Also the trace of $T_{(1)\mu\nu}$ depends only on $\sigma$
\begin{equation}\label{eqn:198A}
	a^2 T_{(1)\mu\nu} \, \bar{g}^{\mu\nu}
	= -a^2 V\sigma.
\end{equation}
For a solution of the background field equation,
\begin{equation}\label{eqn:190B}
	3\mathcal{H}^2 = \frac{a^2 V}{M^2},
\end{equation}
the mode equation for $\sigma $ therefore becomes
\begin{equation}\label{eqn:198C}
	(\partial_\eta^2 + 2\mathcal{H}\partial_\eta-3\mathcal{H}^2 + k^2)\sigma = 0.
\end{equation}

The mode equations for the other components of the metric are somewhat more involved. One first uses the relation $b_{\mu\nu} = t_{\mu\nu} + \tilde{s}_{\mu\nu}$, \cref{eqn:46G}, in order to combine \cref{eqn:F2,eqn:63,eqn:70A} into a coupled system of differential equations for $t_{\mu\nu}$ and $\sigma$. Finding solutions of this linearized Einstein equation will be facilitated if we decompose the metric fluctuations into irreducible representations of the rotation group.

\subsection{Projectors}

Projectors are needed for the definition of the constrained fluctuations or for the projection of unconstrained fluctuations onto the physical fluctuations. The projectors $P^{(f)}$ and $P^{(a)}$ on physical or gauge fluctuations are in Fourier space functions of $k_m$, and involve a unit matrix $\delta (k-k^\prime)$. They depend on two time arguments $\eta$ and $\eta^\prime$, e.g.
\begin{equation}\label{eqn:PR1}
	\int_{\eta^\prime}P^{(a)}_{\mu\nu}{^{\rho\tau}}(\eta,\eta^\prime)h_{\rho\tau}(\eta^\prime) = a_{\mu\nu}(\eta).
\end{equation}
For $P^{(a)}$ we write, similar to \cref{eqn:54A},
\begin{eqnarray}\label{eqn:PR1a}
	P^{(a)}_{\mu\nu}{^{\rho\tau}}(\eta,\eta^\prime)
	&= \frac{1}{2} D_{(\eta)\mu}N_\nu{^\rho}(\eta,\eta^\prime)D^\tau_{(\eta^\prime)} \\
    &\hphantom{={}}+(\mu\leftrightarrow \nu) + (\rho \leftrightarrow \tau),
\end{eqnarray}
where $D_{(\eta)}$ acts on $\eta$ and $D_{(\eta^\prime)}$ on $\eta^\prime$. The projector property
\begin{equation}\label{eqn:PR2}
	\int_{\eta^{\prime\prime}}P^{(a)}_{\mu\nu}{^{\rho\tau}}(\eta,\eta^{\prime\prime})P^{(a)\lambda\sigma}_{\rho\tau}(\eta^{\prime\prime},\eta^\prime) = P^{(a)\lambda\sigma}_{\mu\nu}(\eta,\eta^\prime)
\end{equation}
is realized if $N_\nu{^\rho}$ obeys
\begin{equation}\label{eqn:PR3}
	(D^2\delta^\nu_\mu + D^\nu D_\mu)N_\nu{^\rho}
	= \delta^\rho_\mu\delta(\eta-\eta^\prime).
\end{equation}
Here the covariant derivatives act on $\eta$ and are taken as acting only on the index $\nu$, since $\rho$ and $\tau$ are contracted with $h_{\rho\tau}$ in \cref{eqn:PR1}.

We can view $N$ as the inverse of the derivative operator $D^2\delta^\nu_\mu + D^\nu D_\mu$.
In flat space one has
\begin{equation}\label{eqn:PR5}
	N_\nu{^\rho}
	= \frac{1}{D^2}\delta^\rho_\nu-\frac{1}{2D^4}D^\rho D_\nu,
\end{equation}
but this form gets modified once covariant derivatives no longer commute.

The projector onto the physical metric fluctuations $P^{(f)}$ is determined by
\begin{equation}
	P^{(f) \rho \tau}_{\mu\nu} = \frac{1}{2} \left( \delta^{\rho}_{\mu} \delta^{\tau}_{\nu} + \delta^{\tau}_{\mu} \delta^{\rho}_{\nu} \right) - P^{(a) \rho \tau}_{\mu\nu}.
\end{equation}
One verifies
\begin{eqnarray}\label{eqn:PR4}
	D^\mu f_{\mu\nu}
	&= D^\mu P^{(f)}_{\mu\nu}{^{\rho\tau}}h_{\rho\tau}\\
	&= D^\mu h_{\mu\nu}-D^\mu P^{(a)}_{\mu\nu}{^{\rho\tau}}h_{\rho\tau}\\
	&= h_{\nu\mu;}{^\mu}-(D^2\delta^\mu_\nu + D^\mu D_\nu)A_\mu{^\rho} h_{\rho\tau;}{^\tau}
	= 0.
\end{eqnarray}
Finding the explicit form of the projectors is not expected to be an easy task.

\section{Mode decomposition}
\label{sec:mode decomposition}

The solution of mode equations or the expression of the effective action in terms of unconstrained fields is facilitated if we decompose the metric fluctuations into representations of the rotation group. We proceed here to a separate decomposition of the physical metric fluctuations and the gauge fluctuations. The connection to other, perhaps more familiar decompositions is established in \cref{sec:unconstrained metric}. In this appendix we also display the relation between the vector and scalar parts of the physical metric fluctuations to the Bardeen potentials.

\subsection{Decomposition of physical metric fluctuations into \texorpdfstring{$SO(3)$}{SO(3)}-representations}

Similar to flat space, we decompose the physical metric fluctuations into irreducible representations of the rotation group $SO(3)$,
\begin{equation}\label{eqn:192A}
	f_{\mu\nu}
	= t_{\mu\nu}+\hat{S}_{\mu\nu}\sigma = t_{\mu\nu}+s_{\mu\nu},
\end{equation}
with
\begin{eqnarray}\label{eqn:Z1a}
	t_{00}
	&= {a}^2\kappa,\\
	t_{m0}
	&= {a}^2\left[W_m -\frac{ik_m}{{k}^2}(\partial_{\eta}+4\mathcal{H})\kappa\right],\\
	t_{mn}
	&= {a}^2\left[\gamma_{mn}-\frac{i}{{k}^2}(\partial_{\eta}+4\mathcal{H})(k_m W_n +k_n W_m)\right.\\
		&+\frac{1}{2k^2}(\partial_{\eta}+4\mathcal{H})^2\left(\delta_{mn}-\frac{3k_m k_n}{k^2}\right)\kappa\\
	&+\left.\frac{1}{2}\left(\delta_{mn}-\frac{k_m k_n}{k^2}\right)\kappa\right],
\end{eqnarray}
and
\begin{equation}\label{eqn:Z2a}
	\gamma_{mn}\delta^{mn}
	= 0,
	\qquad
	k^{m}\gamma_{mn}
	= 0,
	\qquad
	k^{m}W_m
	= 0.
\end{equation}

This decomposition is consistent with $t_{\mu;\nu}^{\nu} = 0$, $t_{\mu}^{\mu} = 0$, $f_{\mu;\nu}^{\nu} = 0$, if $s_{\mu\nu} = \hat{S}_{\mu\nu}\sigma$ is symmetric and obeys, cf. \cref{eqn:46F},
\begin{eqnarray}\label{eqn:Z3a}
	s^m_m + s^0_0
	&= \sigma,\\
	(\partial_{\eta}+4\mathcal{H})s^0_0 + ik_m s^m_0
	&= \mathcal{H}\sigma,\\
	(\partial_{\eta}+4\mathcal{H})s^0_m + ik_n s^n_m
	&= 0.
\end{eqnarray}
The physical metric fluctuations contain a traceless divergence free tensor $\gamma_{mn}$, a divergence free vector $W_m$ and two scalars $\sigma$ and $\kappa$. The decomposition in the scalar sector is not unique, since \cref{eqn:Z3a} has no unique solution. Different solutions correspond to different definitions of $\sigma$.

Two possible simple choices for $s_{\mu\nu}$ are
\begin{eqnarray}\label{eqn:Z4a}
	s^{(1)}_{00}
	&= 0,
	\qquad
	s^{(1)}_{m0}
	= -\frac{ik_m}{k^2}a^2\mathcal{H}\sigma,\\
	s^{(1)}_{mn}
	&= \frac{a^2}{2k^2}\Big[ 2k_m k_n\\
	&+\left(\delta_{mn}-\frac{3k_m k_n}{k^2}\right)(k^2 + \mathcal{H} \partial_\eta + \partial_\eta\mathcal{H} +4\mathcal{H}^2)\Big]\sigma,
\end{eqnarray}
and
\begin{eqnarray}\label{eqn:Z5a}
	s^{(2)}_{00}
	&= -a^2\sigma,
	\qquad
	s^{(2)}_{m0}
	= \frac{ik_m}{k^2}a^2(\partial_\eta + 3\mathcal{H})\sigma,\\
	s^{(2)}_{mn}
	&= -\frac{a^2}{2k^2}\left(\delta_{mn}-\frac{3k_m k_n}{k^2}\right)
	(\partial_\eta + 4\mathcal{H})(\partial_\eta + 3\mathcal{H})\sigma.
\end{eqnarray}
They differ by a traceless divergence free tensor
\begin{equation}\label{eqn:Z6a}
	\Delta s_{\mu\nu}
	= s^{(1)}_{\mu\nu}-s^{(2)}_{\mu\nu},
\end{equation}
with
\begin{eqnarray}\label{eqn:Z7a}
	\Delta s_{00}
	&= a^2\sigma,
	\qquad
	\Delta s_{m0}
	= -\frac{ik_m}{k^2}a^2(\partial_\eta + 4\mathcal{H})\sigma,\\
	\Delta s_{mn}
	&= \frac{a^2}{2}\left\{\delta_{mn}-\frac{k_m k_n}{k^2}\right.\\
	&+\left.\left(\delta_{mn}-\frac{3k_m k_m}{k^2}\right)\frac{(\partial_\eta + 4\mathcal{H})}{k^2}\sigma\right\},
\end{eqnarray}
which has the same properties as $t_{\mu\nu}$, e.g.
\begin{equation}\label{eqn:Z8a}
	\Delta s^\mu_\mu = 0,
	\qquad
	\Delta s^\nu_{\mu;\nu}
	= 0.
\end{equation}

The general solution of \cref{eqn:Z3a} involves an arbitrary scalar field $\epsilon$ with
\begin{eqnarray}\label{eqn:Z9a}
	s_{\mu\nu}
	&= s^{(1)}_{\mu\nu}+s^{(\epsilon)}_{\mu\nu},\\
	s^{(\epsilon)}_{00}
	&= a^2\epsilon,
	\qquad
	s^{(\epsilon)}_{m0}
	= -\frac{ik_m}{k^2}a^2(\partial_\eta + 4\mathcal{H})\epsilon,\\
	s^{(\epsilon)}_{mn}
	&= \frac{a^2}{2}
	\left\{\delta_{mn}-\frac{k_m k_n}{k^2}\right.\\
	&\left.+
	\left(\delta_{\mu\nu}-\frac{3k_m k_n}{k^2}\right)\frac{(\partial_\eta + 4\mathcal{H})^2}{k^2}\right\}\epsilon.
\end{eqnarray}
The freedom in the choice of the decomposition associated to $\epsilon$ can be used in order to simplify the effective action. We have already discussed in \cref{sec:correlation function} a choice for maximally symmetric spaces that makes $\Gamma^{(2)}$ diagonal. This choice amounts to
\begin{equation}\label{eqn:Z9A}
	\epsilon = \frac{1}{a^2}(k^2 + 3\mathcal{H}\partial_\eta-3\mathcal{H}^2)
	(3D^2 + \bar{R})^{-1}\sigma,
\end{equation}
which entails for de Sitter space the relation
\begin{equation}\label{eqn:Z9B}
	(k^2 + \partial^2_\eta + 6\mathcal{H}\partial_\eta + 12\mathcal{H}^2)\epsilon
	= -\frac{1}{3}(k^2 + 3\mathcal{H}\partial_\eta + 9\mathcal{H}^2)\sigma.
\end{equation}

\subsection{Effective action for graviton, vector and scalars in de Sitter space}

Let us concentrate on the background geometry of de Sitter space,
\begin{eqnarray}\label{eqn:S1}
	\bar{R}_{\mu\nu}
	&= \frac{\bar{R}}{4}\bar{g}_{\mu\nu}
	= \frac{V}{M^2}\bar{g}_{\mu\nu}
	= \frac{3\mathcal{H}^2}{a^2}\bar{g}_{\mu\nu},\\
	\mathcal{H}
	&= -\frac{1}{\eta}
	= Ha,
	\qquad
	\partial_\eta\mathcal{H}
	= \mathcal{H}^2.
\end{eqnarray}
Here we have assumed that the background metric obeys the field equations. For de Sitter space we will use the definition \labelcref{eqn:46H}, \labelcref{eqn:46R} for $s_{\mu\nu}$. For a background geometry solving the field equation the quadratic effective action $\Gamma_1 = \Gamma^{(t)}_2 + \Gamma^{(\sigma)}_2$ is then given in momentum space by \cref{eqn:46W},
\begin{equation}\label{eqn:SN1}
	\Gamma^{(t)}_2
	= -\frac{iM^2}{8}\int_{\eta,k}a^4 t^{\mu\nu\ast}
	\left(D^2 -\frac{\bar{R}}{6}\right)t_{\mu\nu},
\end{equation}
and \cref{eqn:46V}
\begin{equation}\label{eqn:SN2}
	\Gamma_{2}^{(\sigma)}
	= \frac{iM^2}{12}\int_{\eta,k}a^4\sigma^\ast
	\frac{(D^2 + \frac{1}{4}\bar{R})^2}{D^2 + \frac{1}{3}\bar{R}}\sigma.
\end{equation}
For the scalar $\sigma$ the covariant Laplacian is given by \cref{eqn:P1} and $\bar{R}$ is constant.

From \cref{eqn:P2} we infer the relation
\begin{eqnarray}\label{eqn:TD1}
	\left(D^2 -\frac{\bar{R}}{6}\right)t_{00}
	&= -\frac{1}{a^2}(\partial_\eta^2 + 2\mathcal{H}\partial_\eta + k^2)t_{00}\\
	&= -(\partial_\eta^2 + 6\mathcal{H}\partial_\eta + 10\mathcal{H}^2 + k^2)\kappa,
\end{eqnarray}
and similarly
\begin{eqnarray}\label{eqn:TD2}
	\bigl(D^2 &-\tfrac{\bar{R}}{6}\bigr)t_{m0}\\
	&= -\frac{1}{a} (\partial_\eta^2 + 2\mathcal{H}\partial_\eta + k^2)(a W_m)\\
	&\hphantom{={}}+\frac{ik_m}{a^2 k^2}\partial_\eta(\partial_\eta^2 + 2\mathcal{H}\partial_\eta + k^2)t_{00}\\
	&= -(\partial_\eta^2 + 4\mathcal{H}\partial_\eta + 4\mathcal{H}^2 + k^2)W_m\\
	&\hphantom{={}}+\frac{ik_m}{k^2}(\partial_\eta + 2\mathcal{H})(\partial_\eta^2 + 6\mathcal{H}\partial_\eta + 10\mathcal{H}^2 + k^2)\kappa,
\end{eqnarray}
while
\begin{eqnarray}\label{eqn:TD3}
	&\bigl(D^2 -\tfrac{\bar{R}}{6}\bigl) t_{mn}\\
	&= -(\partial_\eta^2 + 2\mathcal{H}\partial_\eta + k^2)\gamma_{mn} - \frac{\delta_{mn}}{3a^2}(\partial_\eta^2 + 2\mathcal{H}\partial_\eta + k^2)t_{00}\\
	&\hphantom{={}}+ \frac{1}{2a^2 k^2}
	\left(\frac{k_m k_n}{k^2}-\frac{\delta_{mn}}{3}\right) (3\partial_\eta^2 + k^2) (\partial_\eta^2 + 2\mathcal{H}\partial_\eta + k^2)t_{00}\\
	&\hphantom{={}}+\frac{i}{ak^2}(\partial_\eta + \mathcal{H}) (\partial_\eta^2 + 2\mathcal{H}\partial_\eta + k^2)
	\bigl[a(k_m W_n + k_n W_m)\bigr].
\end{eqnarray}
Solutions of the equation
\begin{equation}
	\bigl(D^2 - \tfrac{\bar{R}}{6}\bigr) t_{\mu\nu}
	= 0
\end{equation}
imply that the three mode functions $t_{00}$, $\tilde{W}_m = aW_m$ and $\gamma_{mn}$ obey all the same mode equation
\begin{equation}\label{eqn:TD4}
	(\partial_\eta^2 + 2\mathcal{H}\partial_\eta + k^2)w = 0.
\end{equation}
In the presence of a gauge fixing the vector and scalar mode equations will have an additional source term according to \cref{eqn:full mode eqn}. This will permit additional solutions with \mbox{$a_{\mu\nu}\neq 0$.}

We next compute for a de Sitter geometry the effective action for the physical modes. Inserting the expressions \labelcref{eqn:TD1}-\labelcref{eqn:TD3}, \cref{eqn:SN1} decomposes as
\begin{equation}\label{eqn:SN3}
	\Gamma^{(t)}_2 = \Gamma^{(\gamma)}_2 + \Gamma^{(W)}_2 + \Gamma^{(\kappa)}_2.
\end{equation}
The graviton part reads
\begin{eqnarray}\label{eqn:SN4}
	\Gamma^{(\gamma)}_2 &=\frac{iM^2}{8}\int_{\eta,k}a^2 \left\{k^2\gamma^\ast_{mn}\gamma_{pq}-\partial_\eta \gamma^\ast_{mn}\partial_\eta \gamma_{pq}\right\} P^{(\gamma)mnpq} \\
    &= \frac{i M^2}{8} \int a^2 \gamma^\ast_{m n} \hat{D} \gamma_{p q} P^{\left( \gamma \right) m n p q}.
\end{eqnarray}
Here we use the shorthand
\begin{equation}\label{eqn:A16-b}
	\hat{D} = \partial_\eta^2 + 2\mathcal{H}\partial_\eta + k^2.
\end{equation}
The projector $P^{\left( \gamma \right)}$ is given by \cref{eqn:86B9,eqn:670}.

For the vector part one finds
\begin{equation}\label{eqn:SN5}
	\Gamma^{(W)}_2
	= -\frac{iM^2}{4}\int_{\eta,k}a^2 k^2\Omega^\ast_m\Omega_n Q^{mn},
\end{equation}
with
\begin{equation}\label{eqn:SN6}
	\Omega_m = \frac{1}{ak^2}\hat{D}(aW_m)
\end{equation}
the gauge invariant vector fluctuation.
Finally, the scalar part obtains as
\begin{equation}\label{eqn:SN7}
	\Gamma^{(\kappa)}_2 = \frac{3iM^2}{16}
	\int_{\eta,k}\frac{1}{a^2}
	(k^2\rho^\ast\rho-\partial_\eta\rho^\ast\partial_\eta\rho),
\end{equation}
where
\begin{equation}\label{eqn:SN8}
	\rho = \frac{\hat{D}}{k^2}
	(a^2\kappa).
\end{equation}
Comparing these results with flat space we find correspondence with \cref{eqn:86B8} if we set $a = 1$ and replace $q^2\to \partial_\eta^2 + 2\mathcal{H}\partial_\eta + k^2 = \hat{D}$.

\subsection{Decomposition for gauge fluctuations}

The gauge fluctuations $a_{\mu\nu} = a_{\mu;\nu} + a_{\nu;\mu}$ can be obtained from a vector $a_{\mu}$. We decompose $a_{\mu}$ into two scalars $a_0$ and $r$ and a divergence free vector $U_m$,
\begin{equation}
	a_m = i k_m r + U_m, \quad k^m U_m = 0.
\end{equation}
This yields for $a_{\mu\nu}$
\begin{eqnarray}
	  & a_{00} = 2 D_0 a_0 = 2 \left( \partial_{\eta} - \mathcal{H} \right) a_0,\\
	  & a_{m 0} = i k_m \left[ a_0 + \left( \partial_{\eta} - 2 \mathcal{H} \right) r \right] + \left( \partial_{\eta} - 2 \mathcal{H} \right) U_m,  \\
	  & a_{m n} = i \left( k_m U_n + k_n U_m \right) - 2 k_m k_n r - 2 \mathcal{H} \delta_{m n} a_0,\\
	  & \delta^{m n} a_{m n} = - 2 k^2 r -6 \mathcal{H} a_0.
\end{eqnarray}

The leading order mode equations for $a_0$, $r$ and $U_m$ follow from \cref{eqn:204XH} with $A_{\mu} = 0$,
\begin{eqnarray}
	\left( 2 \partial_{\eta}^2 - 8 \mathcal{H}^2 - 2 \partial_{\eta} \mathcal{H} + k^2 \right) a_0 + \left( \partial_{\eta} - 4 \mathcal{H} \right) k^2 r = 0,    \\
	\left( \partial_{\eta} + 4 \mathcal{H} \right) a_0 + \left( \partial_{\eta}^2 - 4 \mathcal{H}^2 - 2 \partial_{\eta} \mathcal{H} + 2 k^2 \right) r = 0,  \\
	\left( \partial_{\eta}^2 - 4 \mathcal{H}^2 - 2 \partial_{\eta} \mathcal{H} + k^2 \right) U_m = 0.
\end{eqnarray}

\section{Graviton correlation}
\label{sec:graviton correlation}

In this section we discuss the on-shell graviton propagator in a de Sitter geometry. The graviton corresponds to the traceless and divergence free metric fluctuations $\gamma_{mn}$. If the background obeys the field equations we recover the standard results of perturbation theory for linear cosmic fluctuations. This section therefore links directly the formal concepts developed in the present paper to cosmological observation and earlier theoretical work. The graviton correlation can be constructed from mode functions.

The metric component corresponding to the graviton obtains from a general metric fluctuation by a particularly simple projection
\begin{equation}
	h^{\left( \gamma \right)}_{\mu\nu}
	= P^{\left( \gamma \right) \rho \tau}_{\mu\nu} h_{\rho \tau}.
\end{equation}
The projector $P^{\left( \gamma \right)}$ is given by \cref{eqn:86B9} if all indices are spacelike and vanishes for all other index combinations. Its time dependence is a simple unit matrix $\delta (\eta -\eta^\prime)$. Indeed, one has for arbitrary metrics of the form \labelcref{eqn:H1} the relations
\begin{eqnarray}
	D^{\mu} h^{\left( \gamma \right)}_{\mu n}
	&= i k^{m} h^{\left( \gamma \right)}_{m n},\\
	D^{\mu} h^{\left( \gamma \right)}_{\mu 0}
	&= - \mathcal{H} h^{\left( \gamma \right) m}_m.
\end{eqnarray}
By virtue of the relations
\begin{equation}
	k^m P^{\left( \gamma \right) p q}_{m n} = 0,
	\qquad
	\delta^{m n} P^{\left( \gamma \right) p q}_{m n}
	= 0,
\end{equation}
or
\begin{equation}
	k^m h^{\left( \gamma \right)}_{m n}
	= 0,
	\qquad
	\delta^{m n} h^{\left( \gamma \right)}_{m n}
	= 0,
\end{equation}
one establishes that $h^{\left( \gamma \right)}_{m n}$ belongs to the physical metric fluctuations, $D^{\mu} h^{\left( \gamma \right)}_{\mu\nu} = 0$. Furthermore, $h^{\left( \gamma \right)}_{m n}$ is divergence free and traceless. We can therefore identify $h^{\left( \gamma \right)}_{m n} = a^2 \gamma_{mn}$. The simple time dependence of the graviton projector $P^{\left( \gamma \right)}$ is the reason why the graviton contribution to the metric correlation is much simpler than those from vector and scalar modes.

\subsection{Evolution equation for graviton propagator}

We first derive the general propagator equation for the graviton correlation. The most general graviton correlation is specified by initial values for the solution of this differential equation. The effective action \labelcref{eqn:SN4} for $\gamma_{m n}$ involves only two time derivatives and one finds directly the propagator equation for the graviton fluctuations
\begin{equation}\label{eqn:304}
	\frac{i M^2 a^2}{4} \left( \partial_{\eta}^2 + 2 \mathcal{H} \partial_{\eta} + k^2 \right) G_{m n p q}^{\gamma \gamma}
	= P_{m n p q}^{(\gamma)} \delta ( \eta - \eta^\prime).
\end{equation}
Rotation symmetry implies for a traceless and divergence free symmetric tensor $(k = |\vec{k}|)$
\begin{eqnarray}\label{eqn:S8}
	G^{\gamma\gamma}_{mnpq}(\vec{k},\eta,\eta^\prime) = P_{mnpq}^{(\gamma)} G_\text{grav}(k,\eta,\eta^\prime).
\end{eqnarray}
The function $G_{grav}$ obeys the evolution equation
\begin{equation}\label{eqn:S9}
	(\partial^2_\eta + 2\mathcal{H}\partial_\eta + k^2)G_{grav}(k,\eta,\eta^\prime)
	= -
	\frac{4i}{M^2 a^2}\delta(\eta-\eta^\prime).
\end{equation}
It is the same as for a massless scalar. (This holds up to an overall normalization factor $4/M^2$ on the r.h.s. of \cref{eqn:S9} which could be absorbed by a rescaling of $G_{grav}$.)

The propagator equation \labelcref{eqn:304} can also be found by projecting the inverse propagator \labelcref{eqn:47A} on the tensor structure of the graviton. In \cref{sec:metric fluctuations decomposition} we decompose $\Gamma^{(2)}$ into a traceless and trace parts. The graviton $\gamma_{mn}$ does not contribute to the trace of the metric, $h = \bar{g}^{\mu\nu}h_{\mu\nu} = 0$, nor does it contribute to the divergence, $h^\nu_{\mu;\nu} = 0$. The relevant part of $\Gamma^{(2)}$ is given by the last equation \labelcref{eqn:S2}, e.g.
\begin{equation}\label{eqn:224A}
	\Gamma_{bb}^{(2)\mu\nu\rho\tau}
	= -\frac{i M^2 a^4}{4}
	\left(D^2 -\frac{\bar{R}}{6}\right)
	P^{(b)\mu\nu\rho\tau},
\end{equation}
with projector onto the traceless part
\begin{equation}\label{eqn:225A}
	P^{(b)}_{\mu\nu}{^{\rho\tau}}
	= \frac{1}{2}
	\left(\delta^\rho_\mu\delta^\tau_\nu + \delta^\tau_\mu\delta^\rho_\nu\right)
	-\frac{1}{4}\bar{g}_{\mu\nu}\bar{g}^{\rho\tau}.
\end{equation}

For the graviton only the space components contribute, such that the operator $D^2$ from \cref{eqn:P2} reads
\begin{equation}\label{eqn:S5}
	D^2
	= -\frac{1}{a^2(\eta)}
	(\partial^2_\eta-2\mathcal{H}\partial_\eta -2\partial_\eta\mathcal{H}-2\mathcal{H}^2 + k^2).
\end{equation}
For the graviton propagator we employ $h_{mn} = a^2\gamma_{mn}$, such that
\begin{equation}\label{eqn:S5a}
	\langle h_{mn}(\eta,\vec{k})h^\ast_{pq}(\eta^\prime,\overrightarrow{k}^\prime\rangle_c =a^2(\eta)a^2(\eta^\prime)\langle \gamma_{mn}(\eta,\vec{k})\gamma^\ast_{pq}(\eta^\prime,\vec{k}^\prime)\rangle_c.
\end{equation}
The differential operator acting on $\gamma_{\mu\nu}$ is given for a de Sitter geometry by
\begin{equation}
	-\bigl(D^2 - \tfrac{\bar{R}}{6}\bigr) a^2 = \hat{D}
	= \partial^2_{\eta} + 2 \mathcal{H} \partial_{\eta} + k^2.
\end{equation}

Projecting on the traceless part of $b_{m n}$ replaces $a^4 P^{(b) \mu\nu\rho\tau}$ in \cref{eqn:224A} by $P^{\left( \beta \right) m n p q}$, with projector
\begin{equation}\label{eqn:158A}
	P^{(\beta)}_{mnpq}
	= \frac{1}{2}(\delta_{mp}\delta_{nq}+\delta_{mq}\delta_{np})-\frac{1}{3}\delta_{mn}\delta_{pq},
\end{equation}
and indices of $P^{\left( \beta \right)}$ raised with $\delta^{m n}$.

We finally have to project onto the transversal part of $b_{mn}$ by imposing $k^m b_{mn} = 0$. This replaces the projector $P^{(\beta)}_{mnpq}$ by $P^{(\gamma)}_{mnpq}$, as given by \cref{eqn:86B9}, with $Q_{m n}$ given by \cref{eqn:670}. The projector $P^{\left( \gamma \right)}$ obeys
\begin{eqnarray}\label{eqn:159Aa}
	&P^{(\gamma)}_{mnpq}Q^q_s = P^{(\gamma)}_{mnps},
	\qquad
	P^{(\gamma)}_{mnpq}Q^{pq}
	= 0,
\end{eqnarray}
and
\begin{equation}\label{eqn:160A}
	P^{(\gamma)}_{mn}{^{rs}}P^{(\beta)}_{rsuv}
	P^{(\gamma)uv}{_{pq}}
	= P^{(\gamma)}_{mnpq}.
\end{equation}

The various projections of $\Gamma^{(2)}$ result in the differential operator
\begin{equation}
	\Gamma^{(2) m n p q}_{\gamma \gamma} a^2
	= \frac{i M^2}{4} P^{\left( \gamma \right) m n p q} \hat{D}.
\end{equation}
(Here a factor $a^4$ is absorbed by our index convention for $P^{\left( \gamma \right)}$.) The propagator equation for $G^{\gamma \gamma}$ becomes
\begin{eqnarray}\label{eqn:307B}
	\Gamma^{(2) m n p q}_{\gamma \gamma} a^2 \left( \eta \right) a^2 \left( \eta^\prime \right) G^{\gamma \gamma}_{p q r s} \left( \eta, \eta^\prime \right)
	= P^{(\gamma) m n}_{\qquad r s}.
\end{eqnarray}
With $P^{\left( \gamma \right) p q}_{m n} \, G^{\gamma \gamma}_{p q r s} = G^{\gamma \gamma}_{m n r s}$ \cref{eqn:307B} coincides with \cref{eqn:304}.

\subsection{General solution for graviton correlation in de Sitter space}

The general solution of \cref{eqn:S9} has been discussed extensively in ref. \cite{CWneu}, \cite{CW2}. For $\eta>\eta^\prime$ it reads
\begin{eqnarray}\label{eqn:S10}
	&G_\text{grav}(k,\eta,\eta^\prime) = \frac{2\bigl(\alpha(k) + 1\bigr)}{M^2}
	w^-_k(\eta)w^+_k(\eta^\prime)\\
	&\quad +\frac{2\bigl(\alpha(k)-1\bigr)}{M^2}w^+_k(\eta)w^-_k(\eta^\prime)\\
	&+\quad \frac{4\zeta(k)}{M^2}w^+_k(\eta)w^+_k(\eta^\prime)+
	\frac{4\zeta^\ast(k)}{M^2}w^-_k(\eta)w^-_k(\eta^\prime),
\end{eqnarray}
with mode functions given by the solution of the mode equation
\begin{equation}
    (\partial_\eta^2 + 2 \mathcal{H} \partial_\eta + k^2) \, w_k^\pm(\eta)
    = 0.
\end{equation}
For de Sitter space, $\mathcal{H} = -1/\eta$, one has
\begin{equation}\label{eqn:S12}
	w^-_k(\eta) = \bigl(w^+_k(\eta)\bigr)^\ast = \frac{1}{a(\eta)\sqrt{2k}}\left(1-\frac{i}{u}\right)e^{-iu},
\end{equation}
and
\begin{equation}\label{eqn:S12a}
	u = k\eta =-\frac{k}{\mathcal{H}(\eta)}
	= -\frac{k}{a(\eta)H}.
\end{equation}

For Bunch-Davies initial conditions \cite{BD}, which correspond to the scaling correlation of ref. \cite{CW2}, one has $\alpha(k) = 1,\zeta(k) = 0$, such that
\begin{equation}\label{eqn:134}
	G_\text{grav} (k,\eta,\eta^\prime) = \frac{4}{M^2}w^-_k(\eta)w^+_k(\eta^\prime).
\end{equation}
In the limit $u, u^\prime \to -\infty$ $(\eta, \eta^\prime \to -\infty)$ the graviton correlation becomes
\begin{equation}\label{eqn:165A}
	G_\text{grav}(k,\eta,\eta^\prime) = \frac{2}{M^2 ka(\eta)a(\eta^\prime)}
	e^{-ik(\eta-\eta^\prime)}.
\end{equation}
For $a(\eta) = a(\eta^\prime) = 1$ this coincides with the flat space correlation \labelcref{eqn:86K}. In the opposite limit $u, u^\prime\to 0$ the graviton propagator reaches a constant amplitude
\begin{equation}\label{eqn:244A}
	G_\text{grav}(k,\eta,\eta^\prime) = \frac{2H^2}{M^2 k^3}e^{-ik|\eta-\eta^\prime|}.
\end{equation}

The equal time correlation $(\eta^\prime = \eta)$ reads
\begin{eqnarray}\label{eqn:135}
	G_\text{grav}(k,\eta)
	&= G_\text{grav}(k,\eta,\eta)
	= \frac{4}{M^2}|w^-_k(\eta)|^2\\
	&= \frac{2}{M^2 a^2 k}\left(1 + \frac{1}{k^2\eta^2}\right).
\end{eqnarray}
For a de Sitter geometry this becomes
\begin{equation}
	G_\text{grav}
	= \frac{2H^2}{M^2 k^3}(1 + k^2\eta^2).
\end{equation}
This yields the tensor power spectrum which is defined by
\begin{equation}\label{eqn:136}
	\Delta^2_T(k,\eta) = \frac{k^3}{\pi^2}G_\text{grav}(k,\eta).
\end{equation}
Correspondingly, the tensor spectral index obeys
\begin{equation}\label{eqn:137}
	n_T
	= \frac{\partial \ln \Delta^2_T}{\partial\ln k}.
\end{equation}
For modes far outside the horizon, $k^2\eta^2\ll 1$, the spectral index vanishes and the tensor spectrum is proportional to $H^2_0$,
\begin{equation}\label{eqn:138}
	\Delta^2_T = \frac{2H^2}{\pi M^2},
	\qquad
	n_T = 0.
\end{equation}
(For geometries neighbouring de Sitter space the mode functions and therefore the power spectrum and $n_T$ are modified.) As long as $k/a$ remains much smaller than $H$ the time independent power spectrum

\begin{equation}
	G_\text{grav}
	= \frac{2 H^2}{M^2 k^3}
\end{equation}
remains unmodified. Once a given $k$-mode ``enters the horizon'', $k/a \gg H$, it starts again the damped oscillation \labelcref{eqn:165A}. The resulting tensor power spectrum is accessible to observation if the amplitude is large enough.

The formulae after \cref{eqn:134} are the standard ones used in cosmology. We have displayed them here in order to demonstrate that for appropriate initial conditions the graviton correlation, as obtained by a solution of the exact propagator equation \labelcref{eqn:I3}, coincides with the one obtained from the normalization of a free quantum field in a Bunch Davies vacuum. The vector and scalar part of the metric correlation has no such simple interpretation in terms of free quantum fields.

The explicit form of the vector and scalar propagator for the physical metric fluctuations in a de Sitter geometry still need to be worked out. They may be inferred from the general results for the metric correlator in de Sitter space in ref. \cite{MTW2}. Alternatively, explicit knowledge of the projectors would be useful for an extraction from \cref{eqn:46X2} by use of \cref{eqn:PPA}.

\section{Conclusions}
\label{sec:conclusions}

This paper addresses mainly the conceptual issues of the metric correlation function in quantum gravity and cosmology. The recipe mainly employed in cosmology, namely the construction of the correlation function as a product of mode functions or a sum of such products, cannot be applied in general. There are simple cases where the mode functions vanish in the vector and scalar channel, while the correlation function differs from zero. We therefore have to build our discussion from a more basic level, using the defining differential equation for the Green's function. The differential operator in this equation is given by the second functional derivative of the quantum effective action $\Gamma^{(2)}$. The relation between $\Gamma^{(2)}$ and the correlation function $G$ is exact. Approximations only concern the precise form of the effective action.

The first question to address concerns the physical meaning of the metric correlation function. In a gauge fixed version of quantum gravity this correlation function depends manifestly on the chosen gauge fixing. One may therefore question to which extent the metric correlation is a meaningful physical object. We propose here to distinguish between physical metric fluctuations that couple to a conserved energy momentum tensor, and gauge fluctuations that are generated by gauge transformations of a given cosmological solution. The physical metric fluctuations are conceptually similar to the Bardeen potentials, in the sense that they are invariant with respect to infinitesimal diffeomorphism transformations of the ``background metric''. The physical metric fluctuations are directly formulated on the level of the metric in a covariant way. This differs from the Bardeen potentials. We establish the formal relations between the physical metric fluctuations and the Bardeen potentials.

The object of our interest is the correlation function for the physical metric fluctuations. It can be obtained by restricting in the functional integral the sources to ``physical sources'' that correspond to a conserved energy momentum tensor. The argument of the effective action involves then only the physical metric fluctuations, not the gauge fluctuations. As a result, the relation between physical sources and physical metric fluctuations is invertible and the effective action can be constructed in a standard way. No gauge fixing is needed for the inversion of the second functional derivative $\Gamma^{(2)}$. The correlation function for the physical metric fluctuations can also be found using a standard procedure with gauge fixing. The gauge fixing is not arbitrary, however. It has to enforce the condition for physical metric fluctuations, $h^{\nu}_{\mu;\nu} = 0$. We show explicitly the equivalence between the restriction to physical sources and fields on one side, and the appropriate gauge fixed formalism on the other side.

With all quantities well defined the metric correlation function $G$ obtains as a solution of the propagator equation \labelcref{eqn:I3}. Conceptually, its computation amounts to the inversion of the differential operator $\Gamma^{(2)}$. We are interested to solve this inversion problem for geometries corresponding to realistic cosmologies. Even for a rather simple form of the effective action, as given by the Einstein-Hilbert action with a cosmological constant, the inversion of $\Gamma^{(2)}$ is a complex task. The operator $\Gamma^{(2)}$ contains only up to two time derivatives $\partial_{\eta}$. It is, however, a matrix in the space of physical metric fluctuations. Making it block diagonal involves projections.

In quantum gravity computations the inversion of $\Gamma^{(2)}$ is a standard task. It is usually done for some particular gauge fixing (unfortunately often not compatible with the projection on physical metric fluctuations) and for simple Euclidean geometries as the sphere or flat space, or geometries close to those \cite{DSZ}. We are interested to obtain the metric correlations for geometries close to de Sitter space, as relevant for cosmology. This needs a computation with Minkowski signature, for which the solution of the propagator equation becomes an initial value problem. One may think of obtaining the metric propagator in de Sitter space by analytic continuation from a corresponding Euclidean geometry. This corresponding geometry is the maximally symmetric space with negative curvature \cite{CW2}, and one needs the metric correlation in such a hyperbolic space. (The analytic continuation of the sphere is anti de Sitter space, both having a discrete spectrum differing qualitatively from the continuous spectrum in de Sitter space.) \\
\- The present paper provides a formalism for the computation of the metric correlation in homogeneous and isotropic cosmologies. The extensive discussion of the metric correlation in flat space establishes several important features in an explicit way. The scalar and vector part of the metric correlation function does not vanish despite the fact that the solution of the linearized Einstein equation leads to vanishing scalar and vector Bardeen potentials. The vector and scalar part of the metric correlation cannot be found from solutions of the linearized Einstein equations. They rather involve the inversion of operators with up to four (vector) or up to six (scalar) derivatives. This results in a secular behavior \labelcref{eqn:86Q}. For realistic cosmologies we will be interested in initial values of the metric correlations for which the high momentum tail is given by the time-translation invariant correlations in flat space. Our discussion of the flat space correlation functions provides those initial conditions.

For realistic homogeneous and isotropic cosmologies the graviton part of the on-shell metric correlation is rather well understood. The remaining task concerns the scalar and vector parts of the metric correlation. For this purpose several different, but equivalent, strategies may be followed. One may derive the propagator equation for $W_m$ and $\kappa$ from \cref{eqn:SN5,eqn:SN7}. This is straightforward, and the unit operator in the corresponding function space is $\sim \delta (\eta -\eta^\prime)$. (In the case of $W_m$ it involves a $k$-dependent projector.) The complexity in this approach arises from the fact that the differential operators to be inverted involve up to four ($W_m$) or six ($\kappa$) derivatives. As an alternative, one may compute the explicit form of projectors and solve \cref{eqn:46X2}. The inhomogeneous term on the right hand side involves now projectors that depend on $\eta$ and $\eta^\prime$ in the case of vector and scalar modes. Finally, one may employ a gauge fixed version and solve \cref{eqn:full mode eqn}. The complexity arises here from the high number of coupled modes - two vectors and four scalars.

The merit of such a calculation will be to shed light on the infrared structure of the physical metric propagator in realistic cosmologies. This should help to understand better several important issues in quantum gravity, as related to locality, anomalies or the possible existence \cite{CWIRFP} of an infrared fixed point. Quantum gravity computations of the quantum effective action, from which the field equations and correlation functions can be derived, involve the off-shell propagator for the metric fluctuations. It will be interesting to learn the impact of the particular properties of on-shell propagators for physical fluctuations as one approaches solutions of the field equations in the space of configurations.

\paragraph{Acknowledgment} This work is supported by ERC-AdG 290623.

\appendix
\renewcommand{\theequation}{\Alph{section}.\arabic{equation}}
\numberwithin{equation}{section}

\section{Projectors and gauge fixing}
\label{sec:projectors and gauge fixing}

In this appendix we recall a few general features of projectors and gauge fixing that are useful for our discussion.
Let us consider two matrices $D$ and $G$ obeying
\begin{equation}\label{eqn:A.1}
	DG = 1.
\end{equation}
Here $D$ corresponds to $\Gamma^{(2)}$ and $G$ to the correlation function. Assume further the existence of a projector $P$,
\begin{equation}\label{eqn:A.2}
	P^2 = P,
	\qquad
	(1-P)^2 = (1-P),
	\qquad
	P(1-P) = 0.
\end{equation}
We may then decompose
\begin{equation}\label{eqn:A.2a}
	D = D_{++}+D_{+-}+D_{-+}+D_{--},
\end{equation}
with
\begin{eqnarray}\label{eqn:A.3}
	D_{++}
	&= P^{\dagger}DP,
	\qquad
	D_{+-}
	= P^{\dagger}D(1-P),\\
	D_{-+}
	&= (1-P^{\dagger})DP,
	\qquad
	D_{--}
	= (1-P^{\dagger})D(1-P),
\end{eqnarray}
and similarly for $G$
\begin{eqnarray}\label{eqn:A.4A}
	&G_{++}
	= P G P^{\dagger},\\
	&G_{+ -}
	= P G \left(1 - P^{\dagger}\right),\\
	&G_{- +}
	= \left( 1 - P \right) G P^{\dagger},\\
	&G_{- -}
	= \left(1 - P\right) G \left(1 - P^{\dagger}\right).
\end{eqnarray}
Insertion into \cref{eqn:A.1} and multiplying \cref{eqn:A.1} with suitable factors $P$ and $(1-P)$ from left and right yields the relations
\begin{eqnarray}\label{eqn:A.4}
	&D_{++}G_{++}+D_{+-}G_{-+}
	= P^{\dagger},\\
	&D_{--}G_{--}+D_{-+}G_{+-}
	= 1-P^{\dagger},\\
	&D_{++}G_{+-}+D_{+-}G_{--}
	= 0,\\
	&D_{--}G_{-+}+D_{-+}G_{++}
	= 0.
\end{eqnarray}

For our discussion two simple cases are of importance. For the first $D$ is block diagonal, $D_{+-} = D_{-+} = 0$,
\begin{equation}\label{eqn:A.5}
	D = D_{++}+D_{--}.
\end{equation}
Then $G_{++}$ and $G_{--}$ obey
\begin{eqnarray}\label{eqn:A.6}
	&D_{++}G_{++}
	= P^{\dagger},\\
	&D_{--}G_{--}
	= 1-P^{\dagger}.
\end{eqnarray}
If $D_{++}$ is invertible once projected on the appropriate subspace, the projected propagator $G_{++}$ is its inverse. The remaining equations
\begin{equation}\label{eqn:A.7}
	D_{++}G_{+-}
	= 0,
	\qquad
	D_{--}G_{-+}
	= 0
\end{equation}
have the solution
\begin{equation}\label{eqn:A.8}
	G_{+-}
	= 0,
	\qquad
	G_{-+}
	= 0,
\end{equation}
for which $G$ is block diagonal. If $D$ is regular and therefore $G$ is unique, the solution \labelcref{eqn:A.7} is the only solution. In contrast, if $D$ is a differential operator for which a unique solution of \cref{eqn:A.1} requires the specification of initial values or boundary conditions, \cref{eqn:A.8} does not necessarily hold.

For a second important case we consider a family of matrices $D_\beta$ for which $D_{--}$ is multiplied by a factor $\frac{1}{\beta}$. We keep the notation $G_{+-},G_{--}$ etc. for the solutions of \cref{eqn:A.1} for $\beta = 1$, and denote the solutions for arbitrary $\beta$ with $G^{(\beta)}_{+-}$, $G^{(\beta)}_{-+}$ etc. Consider now small $\beta$. The components $G^{(\beta)}_{--}$ and $G^{(\beta)}_{-+}$ have to scale $\sim\beta$ (or they vanish). In the limit $\beta\to0$ we can neglect them, resulting in
\begin{equation}\label{eqn:A.9}
	D_{++}G^{(\beta)}_{++}
	= P^{\dagger},
\end{equation}
and
\begin{equation}\label{eqn:A.10}
	D_{++}G^{(\beta)}_{+-}
	= 0.
\end{equation}
If $G^{(\beta)}$ is unique the second equation implies that $G^{(\beta)}_{+-}$ vanishes. One ends with
\begin{equation}\label{eqn:A.11}
	G^{(\beta)}
	= G^{(\beta)}_{++}.
\end{equation}
The correlation function differs from zero only in the subspace of eigenvalues of $P$ with unit eigenvalue.

In a gauge fixed version of the effective action one adds to $\Gamma^{(2)}$ a gauge fixing term $(1/\beta)\bar{\Gamma}^{(2)}_\text{gf}$. Let us assume for simplicity that $\Gamma^{(2)} = \bar{\Gamma}^{(2)}+(1/\beta)\bar{\Gamma}^{(2)}_\text{gf}$ is regular. We further assume that $\bar{\Gamma}^{(2)}_\text{gf}$ can be written in terms of a projector
\begin{equation}\label{eqn:A.12}
	\bar{\Gamma}^{(2)}_\text{gf}
	= (1-P^{\dagger})\tilde{D}_{--}(1-P).
\end{equation}
In the limit $\beta\to0$ one therefore arrives at
\begin{equation}\label{eqn:A.13}
	G = G_{++},
	\qquad
	D_{++}G_{++}
	= P^{\dagger}.
\end{equation}
In other words, the non-vanishing part of the correlation function involves only the modes with eigenvalues one of $P$. They can be viewed as fluctuations obeying a constraint. The operator $D_{++}$ corresponds to $\Gamma^{(2)}$ subject to this constraint, and $G_{++}$ is the inverse of $\Gamma^{(2)}$ on the subspace of the constrained fluctuations. The formulation in terms of the physical metric fluctuations $f_{\mu\nu}$ obeying the constraint $f^\nu_{\mu;\nu} = 0$, that we employ in this paper, is equivalent to a gauge fixed version in the limit $\beta\to0$. In case where the projector $P^{(f)}$ on $f_{\mu\nu}$ is not known explicitly (or in case of ambiguities) we will define the correlation function $G_{++}$ by the limit $\beta\to0$ of a gauge fixed version.

In more detail we consider
\begin{equation}\label{eqn:AB1}
	\Gamma_\text{gf}
	= \frac{1}{2\beta} \int_x\bar{g}^{1/2} h^\nu_{\mu;\nu}h^{\mu\rho}_{;\rho}.
\end{equation}
It adds to $\Gamma^{(2)}$ a term of the type discussed before, e.g.
\begin{equation}\label{eqn:AB2}
	\Delta\Gamma^{(2)}
	= \frac{1}{\beta}\bigl(1-P^{(f)\dagger}\bigr) \tilde{D}_{--}\bigl(1-P^{(f)}\bigr),
\end{equation}
with $P^{(f)}$ the projector onto covariantly conserved metric fluctuations $f_{\mu\nu}$,
\begin{equation}\label{eqn:AB3}
	f^\nu_{\mu;\nu}
	= 0,
	\qquad
	P^{(f)}_{\mu\nu}{^{\rho\tau}}h_{\rho\tau}
	= f_{\mu\nu}.
\end{equation}
In the presence of the gauge fixing the second functional derivative $\Gamma^{(2)}$ is invertible if suitable boundary conditions are imposed. (We discard here the zero-momentum modes $k = 0$ which would need a separate discussion.) Also $\tilde{D}_{--}$ is invertible on the subspace of ``longitudinal fluctuations'', defined by the modes with zero eigenvalues of $P^{(f)}$. We can then take the limit $\beta \to 0$ and obtain the correlation function $G_{++}$ for the ``physical fluctuations'' $f_{\mu\nu}$. The ``gauge parts'' of the correlation function vanish in this limit, justifying the restriction to the physical fluctuations.

These general considerations can easily be followed explicitly in flat space. In momentum space $P = P^{(f)}$ is real, with
\begin{equation}\label{eqn:A18}
	(P^{\transpose})_{\mu\nu}^{\quad \rho \tau} = P_{\rho \tau}^{\quad \mu\nu} = P^{\mu\nu}_{\quad \rho \tau}.
\end{equation}

\section{Local gauge symmetries}
\label{sec:local gauge symmetries}

In this appendix we discuss the precise implementation of the gauge symmetry of general coordinate or diffeomorphism transformations. This will justify the use of a gauge invariant effective action in the main text.

The source term
\begin{equation}
	\int_x \hat{g}_{\mu\nu} K^{\mu\nu} = \int_x \bar{g}_{\mu\nu} K^{\mu\nu} + \int_x f_{\mu\nu} K^{\mu\nu}
\end{equation}
is invariant under a simultaneous diffeomorphism transformation of $\bar{g}_{\mu\nu}$, the source transformation \labelcref{eqn:3}, and a transformation of $f_{\mu\nu}$ as a tensor,
\begin{equation}
	\delta_{\xi} f_{\mu\nu} = - \partial_{\mu} \xi^{\rho} f_{\rho \nu} - \partial_{\nu} \xi^{\rho} f_{\mu \rho} - \xi^{\rho} \partial_{\rho} f_{\mu\nu}.
\end{equation}
Since the relation \labelcref{eqn:22} between sources and fields is covariant ($\delta \Gamma / \delta \hat{g}_{\mu\nu} = \delta \Gamma / \delta f_{\mu\nu}$) we conclude that the effective action \labelcref{eqn:21} is invariant under simultaneous transformations of $\bar{g}_{\mu\nu}$ and $f_{\mu\nu}$.

Furthermore, we may multiply the source constraint \labelcref{eqn:9} with $\xi_{\mu} = \bar{g}_{\mu\nu} \xi^{\nu}$ and integrate over $x$,
\begin{equation}
	\int_x \xi_{\mu} \left( \partial_{\nu} K^{\mu\nu} + \bar{\Gamma}_{\nu \rho}^{\quad \mu} K^{\nu \rho} \right) = 0.
\end{equation}
By partial integration this yields
\begin{equation}
	\int_x K^{\mu\nu} \xi_{\mu;\nu} = 0.
\end{equation}
Insertion of \cref{eqn:22} results in
\begin{equation}\label{eqn:39e}
	\int_x  \frac{\partial \Gamma}{\partial f_{\mu\nu}} \tilde{\delta}_{\mu\nu} = 0, \quad \tilde{\delta}_{\mu\nu} = - \left( \xi_{\mu;\nu} + \xi_{\nu;\mu} \right).
\end{equation}
Formally, this can be interpreted as invariance under a local gauge transformation of $f_{\mu\nu}$, with infinitesimal transformation $\tilde{\delta} f_{\mu\nu} = \tilde{\delta}_{\mu\nu}$.

The transformation $\tilde{\delta} f_{\mu\nu} = \tilde{\delta}_{\mu\nu}$ is, however, not compatible with the constraint \labelcref{eqn:constraint32}, since $D^{\nu} \tilde{\delta}_{\mu\nu} \neq 0$. We can extend the effective action to be a functional of unconstrained metric fluctuations $h_{\mu\nu}$ by replacing $f_{\mu\nu}$ by $h_{\mu\nu}$, $\Gamma \left[ f_{\mu\nu}, \bar{g}_{\mu\nu} \right] \rightarrow \Gamma \left[ h_{\mu\nu}, \bar{g}_{\mu\nu} \right]$. The extended effective action depends now on two unconstrained metrics $\bar{g}_{\mu\nu}$ and
\begin{equation}
	g_{\mu\nu} = \bar{g}_{\mu\nu} + h_{\mu\nu},
\end{equation}
i.e.
\begin{equation}\label{eqn:39g}
	\Gamma \left[ g_{\mu\nu}, \bar{g}_{\mu\nu} \right] = \Gamma \left[ f_{\mu\nu} \rightarrow h_{\mu\nu}, \bar{g}_{\mu\nu} \right].
\end{equation}
By virtue of \cref{eqn:39e} it is invariant under the infinitesimal gauge transformation
\begin{equation}\label{eqn:39h}
	\tilde{\delta} h_{\mu\nu} = - \left( \xi_{\mu;\nu} + \xi_{\nu;\mu} \right) = \delta_{\xi} \bar{g}_{\mu\nu}.
\end{equation}
The transformation \labelcref{eqn:39h} is taken at fixed $\bar{g}_{\mu\nu}$. It expresses the fact that $\Gamma$ depends actually only on $f_{\mu\nu}$ and not on the gauge fluctuations $a_{\mu\nu}$.

The two local transformations,
\begin{equation}\label{eqn:39i}
	\int_x \left( \frac{\delta \Gamma}{\delta g_{\mu\nu}}\biggr|_{\bar{g}} \delta_{\xi} h_{\mu\nu} + \frac{\delta \Gamma}{\delta \bar{g}_{\mu\nu}}\biggr|_h \delta_{\xi} \bar{g}_{\mu\nu} \right) = 0.
\end{equation}
and
\begin{equation}\label{eqn:39j}
	\int_x \frac{\delta \Gamma}{\delta g_{\mu\nu}}_{|\bar{g}} \delta_{\xi} \bar{g}_{\mu\nu} = 0,
\end{equation}
imply the invariance of $\Gamma$ under simultaneous diffeomorphism transformations of $g_{\mu\nu}$ and $\bar{g}_{\mu\nu}$,
\begin{equation}
	\int_x \left( \frac{\delta \Gamma}{\delta g_{\mu\nu}}\biggr|_{\bar{g}} \delta_{\xi} g_{\mu\nu} + \frac{\delta \Gamma}{\delta \bar{g}_{\mu\nu}}\biggr|_h \delta_{\xi} \bar{g}_{\mu\nu} \right) = 0.
\end{equation}
(Recall $\delta \Gamma / \delta g_{\mu\nu}|_{\bar{g}} = \delta \Gamma / \delta h_{\mu\nu}|_{\bar{g}}$.)

Instead of the variables $h_{\mu\nu}$ and $\bar{g}_{\mu\nu}$ it is convenient to use $g_{\mu\nu}$ and $h_{\mu\nu}$,
\begin{eqnarray}\label{eqn:39l}
	\Gamma^\prime \left[ g_{\mu\nu}, h_{\mu\nu} \right] &= \Gamma \left[h_{\mu\nu},\bar{g}_{\mu\nu} = g_{\mu\nu} - h_{\mu\nu} \right]\\
	&= \Gamma \left[g_{\mu\nu},\bar{g}_{\mu\nu} = g_{\mu\nu} - h_{\mu\nu} \right].
\end{eqnarray}
With
\begin{eqnarray}
	\frac{\partial \Gamma^\prime}{\partial h_{\mu\nu}}\biggr|_g &= \frac{\partial \Gamma}{\partial h_{\mu\nu}}\biggr|_{\bar{g}} - \frac{\partial \Gamma}{\partial \bar{g}_{\mu\nu}}\biggr|_h, \\
	\frac{\partial \Gamma^\prime}{\partial g_{\mu\nu}}\biggr|_h &= \frac{\partial \Gamma}{\partial \bar{g}_{\mu\nu}}\biggr|_h,
\end{eqnarray}
the symmetry relations \labelcref{eqn:39i,eqn:39j} read
\begin{equation}\label{eqn:39n}
	\int_x \left( \frac{\partial \Gamma^\prime}{\partial g_{\mu\nu}}\biggr|_h \delta_{\xi} g_{\mu\nu} + \frac{\partial \Gamma^\prime}{\partial h_{\mu\nu}}\biggr|_g \delta_{\xi} h_{\mu\nu} \right) = 0,
\end{equation}
and
\begin{equation}\label{eqn:39o}
	\int_x \frac{\partial \Gamma^\prime}{\partial h_{\mu\nu}}\biggr|_g \left( \delta_{\xi} g_{\mu\nu} - 2 \delta_{\xi} h_{\mu\nu} \right) = \int_x \frac{\partial \Gamma^\prime}{\partial g_{\mu\nu}}\biggr|_h \delta_{\xi} h_{\mu\nu}.
\end{equation}

We may expand $\Gamma^\prime$ in powers of $h$,
\begin{eqnarray}\label{eqn:39p}
	\Gamma^\prime\left[ g, h \right] &= \bar{\Gamma} \left[ g \right] + \int_x M_1^{\rho \tau} \left[ g \right] h_{\rho \tau}\\
	&+ \frac{1}{2} \int_x h_{\lambda \sigma} M_{2}^{\lambda \sigma \rho \tau} \left[ g \right] h_{\rho \tau} + \cdots,
\end{eqnarray}
with $M_2$ typically involving derivative operators. Nonvanishing $M_1, M_2$ reflect the residual dependence of $\Gamma$ on the background metric $\bar{g}_{\mu\nu}$ for fixed $g_{\mu\nu}$. The symmetry \labelcref{eqn:39n} of simultaneous diffeomorphism transformations of $g_{\mu\nu}$ and $h_{\mu\nu}$ is obeyed if $\bar{\Gamma} \left[ g \right]$ is a gauge invariant functional of $g_{\mu\nu}$ and $M_1, M_2$ transform as appropriate tensor densities. The local gauge symmetry \labelcref{eqn:39o} constrains the possible form of $M_1$ and $M_2$, but is not sufficient to enforce that these quantities vanish.

In the presence of gauge fixing the term $\sim M_2$ can be made to diverge in an appropriate limit of zero gauge fixing parameter $\alpha\to 0$. (This corresponds to Landau gauge in quantum electrodynamics, see refs. \cite{CWGFE,CWNN}.) For an appropriate choice of the gauge this divergent part will only involve $a_\mu\nu$, and not $f_\mu\nu$. Typically, this is the only divergent part for $\alpha\to 0$, with $M_1$ remaining finite. The insertion into $\Gamma^\prime$ eliminates all terms involving $a_\mu\nu$, such that $h_\mu\nu\to f_\mu\nu$ in \cref{eqn:39p}. The residual terms $\sim M_1,M_2$ reflect the explicit background field dependence through the projectors. In our approximation they are neglected. By a modified choice of the covariant derivatives in the projectors one can achieve that the projection of $M_2$ on the physical metric fluctuations vanishes \cite{CWGFE,CWNN}. In this case one has $\Gamma^{\prime(2)} = \bar\Gamma^{(2)}$, as used for our practical computations.

In the formulation with constrained fields and sources the effective action $\Gamma[g_{\mu\nu};\bar{g}_{\mu\nu}]$ depends on $g_{\mu\nu}$ directly, and further on $\bar{g}_{\mu\nu}$ which enters the constraints for the physical sources $K^{\mu\nu}$ and the physical metric $g_{\mu\nu}$. Due to the source constraint $\Gamma$ actually only depends on physical metric fluctuations and $\bar{g}_{\mu\nu}$. Replacing $g_{\mu\nu}$ by $\, \hat{g}_{\mu\nu} = \bar{g}_{\mu\nu} + f_{\mu\nu}$ we can write $\Gamma[b_{\mu\nu},\sigma, \bar{g}_{\mu\nu}]$, where the decomposition \labelcref{eqn:18} is performed for fixed $\bar{g}_{\mu\nu}$. We recall that no ``gauge part'' of $g_{\mu\nu}$ appears due to the restriction to physical sources, i.e. $v_{\mu} = 0, \tau = 0$. This will allow for an invertibility of $\Gamma^{(2)}$ on an appropriate space of functions and for appropriate boundary conditions.

Consider next the transversal split transformation $\bar{g}_{\mu\nu}\to \bar{g}_{\mu\nu}+u_{\mu\nu}$, $f^\prime_{\mu\nu}\to f^\prime_{\mu\nu}-u_{\mu\nu}$, with $u^\nu_{\mu;\nu} = 0$. (This is complementary to the longitudinal split transformation $s_{\mu\nu} = \xi_{\mu;\nu}+\xi_{\nu;\mu}$ discussed in \cref{sec:qea}.) The transversal split symmetry is violated only by the constraints on $K^{\mu\nu}$ and $\hat{g}^\prime_{\mu\nu}$. If we neglect effects of this explicit breaking the effective action becomes invariant under the split transformation $\bar{g}_{\mu\nu}\to \bar{g}_{\mu\nu}+u_{\mu\nu}$, $f_{\mu\nu}\to f_{\mu\nu}-u_{\mu\nu}$. It is then a gauge invariant functional of the unique metric $\hat{g}_{\mu\nu} = \bar{g}_{\mu\nu}+f_{\mu\nu}$. The transversal split symmetry implies $M_{1,2} = 0$ in the expansion \labelcref{eqn:39p}. Extending again the argument of $\Gamma$ to arbitrary metric fluctuations $h_{\mu\nu}$, $g_{\mu\nu} = \bar{g}_{\mu\nu} + h_{\mu\nu}$, the effective action becomes a diffeomorphism invariant functional of $g_{\mu\nu}$, corresponding to $\bar{\Gamma}\left[ g \right]$ in \cref{eqn:39p}. We will adopt this approximation, neglecting corrections due to the explicit $\bar{g}_{\mu\nu}$-dependence of the constraints.

\section{Decomposition of metric fluctuations into trace and traceless parts}
\label{sec:metric fluctuations decomposition}

In this appendix we decompose the metric fluctuations into a trace and traceless part, and correspondingly the inverse propagator $\Gamma^{(2)}$ and the correlation function $G$. This is done both for unconstrained metric fluctuations $h_{\mu\nu}$ and for the physical metric fluctuations $f_{\mu\nu}$. In the second case one has to keep track that the constraint $f_{\mu\nu;}^{\quad\nu} = 0$ mixes trace and traceless parts.

\subsection{Decomposition of unconstrained metric fluctuations}

The unconstrained metric fluctuations $h_{\mu\nu}$ can be decomposed into the trace $h$ and a traceless part $\tilde{b}_{\mu\nu}$,
\begin{equation}\label{eqn:T1}
	h_{\mu\nu}
	= \tilde{b}_{\mu\nu}+\frac{1}{4} h\bar{g}_{\mu\nu},
	\qquad
	\bar{g}^{\mu\nu}\tilde{b}_{\mu\nu}
	= 0.
\end{equation}
For the physical metric fluctuations, with $v_\mu = 0$, $\tau = 0$ in \cref{eqn:18}, one has $\tilde{b}_{\mu\nu} = b_{\mu\nu}$, $h = \sigma$, and we will turn to this case later. According to the decomposition \labelcref{eqn:T1} we write
\begin{eqnarray}\label{eqn:31A}
	\Gamma^{(2)\mu\nu\rho\tau}
	&= \Gamma^{(2)\mu\nu\rho\tau}_{bb}+\Gamma^{(2)\mu\nu}_{bh}\bar{g}^{\rho\tau}\\
	&\bar{g}^{\mu\nu}\Gamma^{(2)\rho\tau}_{h b}+\bar{g}^{\mu\nu}\Gamma^{(2)}_{hh}\bar{g}^{\rho\tau},
\end{eqnarray}
where
\begin{eqnarray}\label{eqn:38A}
	&\bar{g}_{\mu\nu}\Gamma^{(2)\mu\nu\rho\tau}_{bb}
	= 0,
	\qquad
	&&\Gamma^{(2)\mu\nu\rho\tau}_{bb}\bar{g}_{\rho\tau}
	= 0\\
	&\bar{g}_{\mu\nu}\Gamma^{(2)\mu\nu}_{bh}
	= 0,
	\qquad
	&&\Gamma^{(2)\rho\tau}_{h b}\bar{g}_{\rho\tau}
	= 0.
\end{eqnarray}
Similarly, we define the correlation functions
\begin{eqnarray}\label{eqn:32}
	&G^{bb}_{\mu\nu\rho\tau}
	= \langle \tilde{b}_{\mu\nu}(x) \tilde{b}_{\rho\tau}(y)\rangle_c,\\
	&G^{b h}_{\mu\nu}(x,y)
	= \langle \tilde{b}_{\mu\nu}(x)h (y)\rangle_c,\\
	&G^{h b}_{\mu\nu}(x,y)
	= \langle h(x)\tilde{b}_{\mu\nu}(y)\rangle_c = G^{b h}_{\mu\nu}(y,x),\\
	&G^{hh}(x,y)
	= \langle h(x) h (y)\rangle_c,
\end{eqnarray}
such that the propagator decomposes as
\begin{eqnarray}\label{eqn:33}
	G^{(2)}_{\mu\nu\rho\tau}(x,y) = G^{bb}_{\mu\nu\rho\tau}(x,y)&+\frac{1}{4}G^{bh}_{\mu\nu}(x,y)\bar{g}_{\rho\tau}(y)\\
	 + \frac{1}{4}\bar{g}_{\mu\nu}(x)G^{hb}_{\rho\tau}(x,y)&+\frac{1}{16}G^{hh}(x,y)\bar{g}_{\mu\nu}(x)\bar{g}_{\rho\tau}(y).
\end{eqnarray}
The propagator equation reads
\begin{eqnarray}\label{eqn:34}
	&\Gamma^{(2)\mu\nu\rho\sigma}_{bb}G^{bb}_{\rho\sigma\tau\lambda}+\Gamma^{(2)\mu\nu}_{b h }G^{hb}_{\tau\lambda}\\
	&\qquad = \frac{1}{2}(\delta^{\mu}_{\tau}\delta^{\nu}_{\lambda}+\delta^{\nu}_{\tau}\delta^{\mu}_{\lambda})-\frac{1}{4}\bar{g}^{\mu\nu}\bar{g}_{\tau\lambda},\\
	&\Gamma^{(2)\mu\nu\rho\sigma}_{bb} G^{b h}_{\rho\sigma}+\Gamma^{(2)\mu\nu}_{b h}G^{hh}
	= 0,\\
	&\Gamma^{(2)\mu\nu}_{hb} G^{bb}_{\mu\nu\tau\lambda} + \Gamma^{(2)}_{hh}G^{hb}_{\tau\lambda}
	= 0,\\
	&\Gamma^{(2)\mu\nu}_{hb} G^{bh}_{\mu\nu}+\Gamma^{(2)}_{hh}G^{hh}
	= 1,
\end{eqnarray}
where we have omitted the coordinates and associated $\delta(x-z)$ factors.

The projection on the traceless part can be performed by using the projection operator
\begin{equation}\label{eqn:A6A}
	P^{(b)}_{\mu\nu}{^{\rho\tau}}
	= \frac{1}{2}(\delta^\rho_\mu\delta^\tau_\nu + \delta^\tau_\mu\delta^\rho_\nu)-\frac{1}{4}\bar{g}_{\mu\nu}\bar{g}^{\rho\tau}.
\end{equation}
It obeys
\begin{eqnarray}\label{eqn:A.6B}
	P^{(b)}_{\mu\nu}{^{\rho\tau}}\bar{g}_{\rho\tau}
	&= 0,
	\qquad
	\bar{g}^{\mu\nu}P^{(b)}_{\mu\nu}{^{\rho\tau}}
	= 0,\\
	P^{(b)}_{\mu\nu}{^{\rho\tau}}P^{(b)}_{\rho\tau}{^{\sigma\lambda}}
	&= P^{(b)}_{\mu\nu}{^{\sigma\lambda}},
\end{eqnarray}
such that
\begin{equation}\label{eqn:A.6C}
	P^{(b)}_{\mu\nu}{^{\rho\tau}}h_{\rho\tau}
	= \tilde{b}_{\mu\nu}.
\end{equation}
The corresponding projection on the trace reads
\begin{eqnarray}\label{eqn:A.6D}
	P^{(h)}_{\mu\nu}{^{\rho\tau}}
	&= \frac{1}{2}(\delta^\rho_\mu\delta^\tau_\nu + \delta^\tau_\mu\delta^\rho_\nu)-P^{(b)}_{\mu\nu}{^{\rho\tau}}\\
	&= \frac{1}{4}\bar{g}_{\mu\nu}\bar{g}^{\rho\tau}.
\end{eqnarray}

The different pieces of the inverse propagator are computed as projections from \cref{eqn:47A}, supplemented by contributions from the gauge fixing term. We display here only the physical part corresponding to \cref{eqn:47A}. One finds
\begin{eqnarray}\label{eqn:T2}
	\Gamma^{(2)}_{hh}
	&= \frac{1}{16}\bar{g}_{\mu\nu}\bar{g}_{\rho\tau}\Gamma^{(2)\mu\nu\rho\tau}\\
	&= \frac{3M^2}{32}\sqrt{\bar{g}}D^2 + \frac{V}{8}\sqrt{\bar{g}}
\end{eqnarray}
and
\begin{eqnarray}\label{eqn:T3}
	\Gamma^{(2)\mu\nu}_{bh}
	&= \frac{1}{4}\Gamma^{(2)\mu\nu\rho\tau}\bar{g}_{\rho\tau}-\bar{g}^{\mu\nu}\Gamma^{(2)}_{hh}\\
	&= \frac{M^2}{32}\sqrt{\bar{g}}\bigl[D^2\bar{g}^{\mu\nu}-2(D^\mu D^\nu + D^\nu D^\mu)\bigr],\\
	\Gamma^{(2)\rho\tau}_{h b}
	&= \frac{M^2}{32}\sqrt{\bar{g}}\bigl[D^2\bar{g}^{\rho\tau}-2(D^\rho D^\tau + D^\tau D^\rho)\bigr].
\end{eqnarray}
The pure traceless part obtains by subtracting these pieces from $\Gamma^{(2)}$,
\begin{eqnarray}\label{eqn:T4}
	&\Gamma^{(2)\mu\nu\rho\tau}_{bb}
	= -\frac{M^2}{32}\sqrt{\bar{g}}
	\Bigl\{\bigl[4(\bar{g}^{\mu\rho}\bar{g}^{\nu\tau}+\bar{g}^{\mu\tau}\bar{g}^{\nu\rho})-3\bar{g}^{\mu\nu}\bar{g}^{\rho\tau}\bigr]D^2\\
	&\qquad + 2\bar{g}^{\mu\nu}(D^\tau D^\rho +D^\rho D^\tau) + 2\bar{g}^{\rho\tau}
	(D^\mu D^\nu + D^\nu D^\mu)\\
	&\qquad-4(\bar{g}^{\mu\rho}D^\tau D^\nu + \bar{g}^{\nu\rho}D^\tau D^\mu + \bar{g}^{\mu\tau} D^\rho D^\nu + \bar{g}^{\nu\tau} D^\rho D^\mu)\\
	&\qquad + 4\bar{R}(\bar{g}^{\mu\nu}\bar{g}^{\rho\tau}-\bar{g}^{\mu\rho}\bar{g}^{\nu\tau}-\bar{g}^{\mu\tau}\bar{g}^{\nu\rho})\\
	&\qquad + 8(\bar{R}^{\mu\rho}\bar{g}^{\nu\tau}+\bar{R}^{\nu\rho}\bar{g}^{\mu\tau}+\bar{R}^{\mu\tau}\bar{g}^{\nu\rho}+\bar{R}^{\nu\tau}\bar{g}^{\mu\rho})\\
	&\qquad-8(\bar{R}^{\mu\nu}\bar{g}^{\rho\tau}+\bar{R}^{\rho\tau}\bar{g}^{\mu\nu})\Bigr\}\\
	&\qquad + \frac{V}{8}\sqrt{\bar{g}}\Bigl\{\bar{g}^{\mu\nu}\bar{g}^{\rho\tau}-2(\bar{g}^{\mu\rho}\bar{g}^{\nu\tau}+\bar{g}^{\mu\tau}\bar{g}^{\nu\rho})\Bigr\}.
\end{eqnarray}

Eq \labelcref{eqn:T4} simplifies for a vanishing Weyl tensor
\begin{eqnarray}\label{eqn:A.9A}
	&\bar{C}_{\mu\nu\rho\tau}
	= \bar{R}_{\mu\nu\rho\tau}+\frac{1}{6}\bar{R}(\bar{g}_{\mu\rho}\bar{g}_{\nu\tau}-\bar{g}_{\mu\tau}\bar{g}_{\nu\rho})\\
	&-\frac{1}{2}(\bar{g}_{\mu\rho}\bar{R}_{\nu\tau}+\bar{g}_{\nu\tau}\bar{R}_{\mu\rho}-\bar{g}_{\mu\tau}\bar{R}_{\nu\rho}-\bar{g}_{\nu\rho}\bar{R}_{\mu\tau}) = 0.
\end{eqnarray}
Using appropriate commutators for covariant derivatives yields
\begin{eqnarray}\label{eqn:T5}
	&\Gamma^{(2)\mu\nu\rho\tau}_{bb}
	= -\frac{M^2}{32}\sqrt{\bar{g}}
	\Bigl\{ \bigl[4(\bar{g}^{\mu\rho}\bar{g}^{\nu\tau}+\bar{g}^{\mu\tau}\bar{g}^{\nu\rho})-3\bar{g}^{\mu\nu}\bar{g}^{\rho\tau}\bigr]D^2\\
	&\qquad + 2\bar{g}^{\mu\nu}(D^\tau D^\rho +D^\rho D^\tau) + 2\bar{g}^{\rho\tau}(D^\mu D^\nu + D^\nu D^\mu)\\
	&\qquad-4(\bar{g}^{\mu\rho}D^\nu D^\tau + \bar{g}^{\nu\rho}D^\mu D^\tau + \bar{g}^{\mu\tau}D^\nu D^\rho + \bar{g}^{\nu\tau} D^\mu D^\rho)\\
	&\qquad + \frac{4}{3} \bar{R}\bigl[\bar{g}^{\mu\nu}\bar{g}^{\rho\tau}-2(\bar{g}^{\mu\rho}\bar{g}^{\nu\tau}+\bar{g}^{\mu\tau}\bar{g}^{\nu\rho})\bigr]\Bigr\}\\
	&\qquad + \frac{V}{8}\sqrt{\bar{g}}\Bigl\{\bar{g}^{\mu\nu}\bar{g}^{\rho\tau}-2(\bar{g}^{\mu\rho}\bar{g}^{\nu\tau}+\bar{g}^{\mu\tau}\bar{g}^{\nu\rho})\Bigr\}.
\end{eqnarray}
When applied on the traceless field $\tilde{b}_{\rho\tau}$ the terms $\sim \bar{g}^{\rho\tau}$ in \cref{eqn:T5} do not contribute.

When acting on physical metric fluctuations the pieces $\sim D^\rho$ or $\sim D^\tau$ do not contribute, such that the different pieces \labelcref{eqn:T2,eqn:T3,eqn:T5} read
\begin{eqnarray}\label{eqn:S2}
	\Gamma^{(2)}_{hh}
	&= \frac{3M^2}{32}\sqrt{\bar{g}}
	\left(D^2 + \frac{4\mathcal{H}^2}{a^2}\right),\\
	\Gamma^{(2)\mu\nu}_{bh}
	&= \frac{M^2}{32}\sqrt{\bar{g}}
	(\bar{g}^{\mu\nu}D^2 -2D^\mu D^\nu-2D^\nu D^\mu),\\
	\Gamma^{(2)\rho\tau}_{hb}
	&= 0,\\
	\Gamma^{(2)\mu\nu\rho\tau}_{bb}
	&= -\frac{M^2}{4}\sqrt{\bar{g}}P^{(b)\mu\nu\rho\tau}
	\left(D^2 -\frac{\bar{R}}{6}\right).
\end{eqnarray}

\subsection{Decomposition of physical metric fluctuations}

The decomposition into trace and traceless parts remains valid if we impose the constraint $h^\nu_{\mu;\nu} = 0$ for the metric physical fluctuations. This replaces in \cref{eqn:T1} $h_{\mu\nu}\to f_{\mu\nu}$, $\tilde{b}_{\mu\nu}\to b_{\mu\nu}$, $h\to \sigma$, with $b_{\mu\nu} = P^{(b)}_{\mu\nu}{^{\rho\tau}}f_{\rho\tau}$. The fields $b_{\mu\nu}$ and $\sigma$ are no longer independent, however, due to the relation $b^\nu_{\mu;\nu} = -\partial_\mu\sigma/4$.

We first insert the decomposition $h_{\mu\nu} = b_{\mu\nu}+\sigma\bar{g}_{\mu\nu}/4$, $\bar{g}^{\mu\nu}b_{\mu\nu} = 0$, directly inside the effective action. Taking account of the constant $h^\nu_{\mu;\nu} = 0$ one finds
\begin{eqnarray}\label{eqn:N1}
	&-\int_x\frac{M^2}{2}(\sqrt{g}R)_{(2)}
	= \int_x\frac{M^2}{2}\sqrt{\bar{g}}
	\left\{\frac{1}{4} b^{\mu\nu;\rho}b_{\mu\nu;\rho}\right.\\
	&\qquad-\frac{3}{16}\sigma_;{^\mu}\sigma_{;\mu}
	 + \frac{1}{4}\bar{R} b_{\mu\nu} b^{\mu\nu}\\
	&\qquad\left.-\frac{1}{2}\bar{R}^\rho_\mu b^{\mu\nu}b_{\nu\rho}-\frac{1}{2}\bar{R}_{\mu\rho\nu\sigma} b^{\mu\nu} b^{\rho\sigma}\right\}
\end{eqnarray}
and
\begin{equation}\label{eqn:N2}
	\int_x(\sqrt{g} V)_{(2)}
	= V\int_x\sqrt{\bar{g}}
	\left(\frac{1}{16}\sigma^2 -\frac{1}{4} b_{\mu\nu}b^{\mu\nu}\right).
\end{equation}
With
\begin{equation}\label{eqn:N3}
	\bar{R}_{\mu\rho\nu\sigma}b^{\mu\nu}b^{\rho\sigma}
	= \bar{C}_{\mu\rho\nu\sigma} b^{\mu\nu}b^{\rho\sigma}-\bar{R}^\rho_\mu b^{\mu\nu}b_{\nu\rho}+\frac{1}{6}\bar{R}b^{\mu\nu} b_{\mu\nu}
\end{equation}
one obtains for a vanishing Weyl tensor $\bar{C}_{\mu\rho\nu\sigma} = 0$ the simple expression
\begin{eqnarray}\label{eqn:N4}
	\Gamma_2
	&= \int_x\sqrt{\bar{g}}
	\left\{\frac{M^2}{8}b^{\mu\nu}\left(-D^2 + \frac{2}{3}\bar{R}-\frac{2V}{M^2}\right)b_{\mu\nu}\right.\\
	&\left. + \frac{3M^2}{32}\sigma\left(D^2 + \frac{2V}{3M^2}\right)\sigma\right\}.
\end{eqnarray}

We next write
\begin{equation}\label{eqn:A.15A}
	b_{\mu\nu}
	= t_{\mu\nu}+\tilde{s}_{\mu\nu},
	\qquad
	t^\nu_{\mu;\nu}
	= 0,
	\qquad
	t^\mu_\mu = 0,
\end{equation}
where $\tilde{s}_{\mu\nu}$ is a function of $\sigma$ as given by \cref{eqn:46H,eqn:46EA}, and $t_{\mu\nu}$ is the independent traceless and divergence free tensor field. The part $\tilde{s}_{\mu\nu}$ obeys
\begin{equation}\label{eqn:A.15B}
	\tilde{s}^\nu_{\mu;\nu}
	= -\frac{1}{4}\partial_\mu\sigma.
\end{equation}
We decompose
\begin{equation}\label{eqn:A.15E}
	\Gamma_2 = \Gamma^{(t)}_2 + \Gamma^{(\sigma)}_2 + \Gamma^{(\sigma t)}_2,
\end{equation}
with transversal traceless part
\begin{equation}\label{eqn:A.15F}
	\Gamma^{(t)}_2 = \frac{M^2}{8}\int_x\sqrt{\bar{g}}t^{\mu\nu}
	\left(-D^2 + \frac{2}{3}\bar{R}-\frac{2V}{M^2}\right)t_{\mu\nu},
\end{equation}
trace part
\begin{eqnarray}\label{eqn:A.15G}
	\Gamma^{(\sigma)}_2
	&= \frac{M^2}{32}\int_x\sqrt{\bar{g}}
	\left\{\sigma\left(3D^2 + \frac{2V}{M^2}\right)\sigma\right.\\
	&\left. + 4\tilde{s}^{\mu\nu}
	\left(-D^2 + \frac{2}{3}\bar{R}-\frac{2V}{M^2}\right)\tilde{s}_{\mu\nu}\right\},
\end{eqnarray}
and mixed term
\begin{equation}\label{eqn:A.15H}
	\Gamma^{(t\sigma)}_2 = \frac{M^2}{4}\int_x\sqrt{\bar{g}}t^{\mu\nu}
	\left(-D^2 + \frac{2}{3}\bar{R}-\frac{2V}{M^2}\right)\tilde{s}_{\mu\nu}.
\end{equation}

In comparison, we can employ $\Gamma^{(2)}$, as given by \cref{eqn:T2,eqn:T3,eqn:T4} or \labelcref{eqn:T5} and apply it to the physical metric fluctuations $f_{\mu\nu}$,
\begin{eqnarray}\label{eqn:31}
	\Gamma_2
	&= \frac{1}{2}\int_{x,y} f_{\mu\nu}(x)\Gamma^{(2)\mu\nu\rho\tau}(x,y)f_{\rho\tau}(y)\\
	&= \frac{1}{2}\int_{x,y}\Bigl\{b_{\mu\nu}(x)\Gamma^{(2)\mu\nu\rho\tau}_{bb}(x,y)b_{\rho\tau}(y)\\
	&\hphantom{= \frac{1}{2}\int_{x,y} \Bigl\{} +b_{\mu\nu}(x)\Gamma^{(2)\mu\nu}_{bh}(x,y)\sigma(y)\\
	&\hphantom{= \frac{1}{2}\int_{x,y} \Bigl\{}+\sigma(x)\Gamma^{(2)\rho\tau}_{h b}(x,y)b_{\rho\tau}(y)\\
	&\hphantom{= \frac{1}{2}\int_{x,y} \Bigl\{} +\sigma(x)\Gamma^{(2)}_{hh}(x,y)\sigma(y)\Bigr\}.
\end{eqnarray}
Employing again the decomposition \labelcref{eqn:A.15A} we observe that the mixed terms $\sim \Gamma^{(2)}_{bh},\Gamma^{(2)}_{hb}$ only contribute to parts involving $\tilde{s}_{\mu\nu}$, and not $t_{\mu\nu}$. The part $\Gamma^{(t)}_2$ for the traceless divergence free tensor can be extracted from \cref{eqn:A.9A} by omitting all terms where $D^\rho$ or $D^\tau$ act on the right. The resulting expression reads
\begin{equation}\label{eqn:T6}
	\Gamma^{(2)\mu\nu\rho\tau}_{bb}
	= -\frac{M^2}{4}\sqrt{\bar{g}}
	P^{(b)\mu\nu\rho\tau}
	\left(D^2 -\frac{2}{3}\bar{R} + \frac{2V}{M^2}\right),
\end{equation}
such that $\Gamma^{(t)}$ coincides with \cref{eqn:A.15F}.

The trace part $\Gamma^{(\sigma)}_2$ obtains contributions from $\Gamma^{(2)}_{hh}$, as well as from $\Gamma^{(2)}_{bb},\Gamma^{(2)}_{bh}$ and $\Gamma^{(2)}_{hb}$, with $b_{\mu\nu}$ replaced by $\tilde{s}_{\mu\nu}$. The sum of all contributions equals indeed \cref{eqn:A.15G}, and we see that the off-diagonal terms are necessary for this result. While the inverse $t-t$-propagator \labelcref{eqn:T6} can be directly extracted from \cref{eqn:A.15F}, the inverse $\sigma-\sigma$ propagator needs the term $\sim \tilde{s}^{\mu\nu}F\tilde{s}_{\mu\nu}$. The inverse propagator for $\sigma$ does not coincide with the inverse propagator $\Gamma^{(2)}_{hh}$ for the unconstrained field $h$ in \cref{eqn:T2}.

\subsection{Scalar fluctuations}

The scalar part of the physical metric fluctuations involves the trace $\sigma$ and a second scalar contained in $t_{\mu\nu}$. Their precise definition involves the contribution of $\sigma$ to $b_{\mu\nu}$, i.e. the form of $\tilde{s}_{\mu\nu}$. In \cref{sec:correlation function} we have discussed the form of $\tilde{s}_{\mu\nu}$ for background geometries with constant curvature scalar. Alternatively, we may try the ansatz
\begin{equation}\label{eqn:A.15HA}
	\tilde{s}_{\mu\nu}
	= D_\mu s_\nu + D_\nu s_\mu-\frac{1}{2} D^\rho s_\rho\bar{g}_{\mu\nu}.
\end{equation}
The vector $s_\mu$ has to be chosen such that \cref{eqn:A.15B} is obeyed. Combining \cref{eqn:A.15HA,eqn:A.15B} one has
\begin{eqnarray}\label{eqn:A.15Ha}
	F_\mu{^\nu}s_\nu
	&= -\frac{1}{4}\partial_\mu\sigma,\\
	F_\mu{^\nu}
	&= D^2\delta^\nu_\mu + D^\nu D_\mu-\frac{1}{2} D_\mu D^\nu\\
	&= D^2\delta^\nu_\mu + \bar{R}^\nu_\mu + \frac{1}{2} D_\mu D^\nu.
\end{eqnarray}
Here we have used the commutator relation
\begin{equation}\label{eqn:A.15I}
	[D^\nu,D_\mu]s_\nu = \bar{R}^\nu_\mu s_\nu.
\end{equation}
We need the inverse of the operator $F_\mu{^\nu}$
\begin{eqnarray}\label{eqn:A.15J}
	C_\rho{^\mu}F_\mu{^\nu}
	&= \delta^\nu_\rho,\\
	s_\rho
	&= -\frac{1}{4} C_\rho{^\mu}\partial_\mu\sigma.
\end{eqnarray}
For a general background geometry the explicit computation of $C_\rho{^\mu}$ is not easy due to the non-commuting properties of the covariant derivatives.

Let us consider first the ansatz $C_\rho{^\mu} = \bar{C}_\rho{^\mu}$,
\begin{equation}\label{eqn:A.15K}
	\bar{C}_\rho{^\mu}
	= D^{-2}\delta^\mu_\rho-\frac{1}{3} D^{-2}D_\rho D^{-2}D^\mu,
\end{equation}
which implies
\begin{equation}\label{eqn:A.15L}
	\bar{s}_\mu
	= -\frac{1}{6} D^{-2}D_\mu\sigma
\end{equation}
and
\begin{equation}\label{eqn:A.15M}
	\bar{s}_{\mu\nu}
	= -\frac{1}{6}
	(D_\mu D^{-2}D_\nu + D_\nu D^{-2}D_\mu-\frac{1}{2} \bar{g}_{\mu\nu}
	D^\rho D^{-2}D_\rho)\sigma.
\end{equation}

For flat space this solves \cref{eqn:A.15J}, and the result \labelcref{eqn:A.15M} is in accordance with \cref{eqn:67B}. For more general geometries we observe
\begin{eqnarray}\label{eqn:A.15N}
	\bar{C}_\rho{^\mu}F^\nu_\mu
	&= \delta^\nu_\rho + D^{-2}\bar{R}^\nu_\rho\\
	&-\frac{1}{3} D^{-2}D_\rho D^{-2}
	\bigl([D^\nu,D^2] + \bar{R}^\nu_\mu D^\mu + \bar{R}^\nu_{\mu;}{^\mu}\bigr)\\
	&= \delta^\nu_\rho + D^{-2}
	\left(\delta^\mu_\rho-\frac{2}{3} D_\rho D^{-2}D^\mu\right)
	\bar{R}^\nu_\mu = D^{-2}A_\rho{^\nu},
\end{eqnarray}
where we employ
\begin{equation}\label{eqn:A.15O}
	[D^\nu, D^2]s_\nu = \bar{R}^{\mu\nu}D_\mu s_\nu + \bar{R}^{\mu\nu}{_{;\mu}}s_\nu.
\end{equation}
The solution for $C$ is therefore
\begin{eqnarray}\label{eqn:A15P}
	C_\rho{^\mu}
	&= (A^{-1})_{\rho}{^\nu}D^2\bar{C}_\nu{^\mu}\\
	&= (A^{-1})_\rho^\nu\left(\delta^\mu_\nu-\frac{1}{3} P^{(l)\mu}_\nu\right),
\end{eqnarray}
where
\begin{eqnarray}\label{eqn:A.15Q}
	A_{\rho}{^\nu}
	&= D^2\delta_\rho^{\nu}+B_\rho{^\mu}\bar{R}_\mu^{\nu},\\
	B_\rho{^\mu}
	&= \delta^\mu_\rho-\frac{2}{3} P^{(l)\mu}_\rho.
\end{eqnarray}
The longitudinal propagator $P^{(l)}$,
\begin{equation}\label{eqn:A.15R}
	P^{(l)}_\nu{^\mu}
	= D_\nu D^{-2}D^\mu,
\end{equation}
obeys
\begin{equation}\label{eqn:A.15S}
	P^{(l)\rho}_\nu P^{(l)\mu}_\rho = P^{(l)\mu}_\nu,
	\qquad
	P^{(l)\mu}_\nu D_\mu F = D_\nu F.
\end{equation}
We infer
\begin{equation}\label{eqn:A.15T}
	s_\mu
	= -\frac{1}{6}(A^{-1})_\mu{^\nu}\partial_\nu\sigma,
\end{equation}
such that the operator $A^{-1}$ replaces $D^{-2}$ in \cref{eqn:A.15L}. The task is now the inversion of $A$.

For making contact with \cref{sec:correlation function} we can specialize to
\begin{equation}\label{eqn:A.15Ta}
	\bar{R}^\nu_\mu = \frac{1}{4}\bar{R}\delta^\nu_\mu,
\end{equation}
with constant $\bar{R}$. We employ
\begin{equation}\label{eqn:A.15U}
	A_\rho{^\nu}
	= \left(D^2 + \frac{1}{4}\bar{R}\right)\delta^{\nu}_\rho-\frac{1}{6}\bar{R} P^{(l)\nu}_\rho.
\end{equation}
The inverse is found easily
\begin{eqnarray}\label{eqn:A.15V}
	&(A^{-1})_\mu{^\nu}
	= \left(D^2 + \frac{\bar{R}}{4}\right)^{-1}\delta^\nu_\mu\\
	&\qquad +\frac{\bar{R}}{6}\left(D^2 + \frac{\bar{R}}{4}\right)^{-1}\left(D^2 + \frac{\bar{R}}{12}\right)^{-1}
	P^{(l)\nu}_\mu
\end{eqnarray}
such that
\begin{equation}\label{eqn:A.15W}
	s_\mu
	= -\frac{1}{6}
	\left(D^2 + \frac{R}{12}\right)^{-1}\partial_\mu\sigma.
\end{equation}
Using the commutator relation
\begin{equation}\label{eqn:A.15X}
	\bigl[D_\nu,(3D^2 + \bar{R})^{-1}\bigr] = \frac{\bar{R}}{4}
	\left(D^2 + \frac{\bar{R}}{12}\right)^{-1}
	(3D^2 + \bar{R})^{-1}D_\nu
\end{equation}
one has
\begin{equation}\label{eqn:A.15Y}
	D_\nu(3D^2 + \bar{R})^{-1}\sigma = \frac{1}{3}
	\left(D^2 + \frac{\bar{R}}{12}\right)^{-1}\partial_\nu\sigma
	= -2s_\nu.
\end{equation}
This establishes that $\tilde{s}_{\mu\nu}$, as computed from \cref{eqn:A.15HA}, indeed coincides with \cref{eqn:46R}. \Cref{eqn:A.15HA,eqn:A.15T} can be used for an expansion in the vicinity of maximally symmetric geometries.

\section{Mode equation and linearized Einstein equation}
\label{sec:mode equation}

We show in this appendix that the mode functions obey Einstein's field equations for small fluctuations around a background field, provided that the background field is itself a solution of the field equations. If not, the mode functions do not obey the linearized Einstein equations. We restrict here the metric fluctuations to the physical fluctuations $f_{\mu\nu}$ and we do not include contributions to the field equations from possible gauge fixing terms.

We start with the defining equation \labelcref{eqn:M5} for the mode functions, with $D_{(\eta)}$ related to the second functional derivative $\Gamma^{(2)}$ by \cref{eqn:M3}. The equivalence of the field equation \labelcref{eqn:26} for $\Delta K^{\mu\nu} = 0$, i.e. the mode equation,
\begin{equation}\label{eqn:BA}
	\Gamma^{(2)\mu\nu\rho\tau}f_{\rho\tau}
	= 0,
\end{equation}
with the linearized field equation around a background that solves the field equation is not restricted to a homogeneous and isotropic situation. We therefore keep general $f_{\rho\tau}$ and the general form \labelcref{eqn:47A} for $\Gamma^{(2)}$. Since in \cref{eqn:BA} $\Gamma^{(2)}$ acts on $f_{\rho\tau}$ the terms involving $D^\rho$ or $D^\tau$ positioned at right do not contribute. Furthermore, one can use the general commutator relation
\begin{equation}\label{eqn:47B}
	[D^\rho,D^\mu]A_{\rho\tau}
	= R^{\mu\rho}A_{\rho\tau}-R^{\mu\rho}{_\tau}{^\lambda}A_{\rho\lambda}.
\end{equation}
For the mode equation \labelcref{eqn:BA} one therefore has
\begin{eqnarray}\label{eqn:47C}
	&\Gamma^{(2)\mu\nu\rho\tau}f_{\rho\tau}
	= -\frac{M^2}{4}\sqrt{\bar{g}}\Bigl\{ f^{\mu\nu}{_;}{^\rho}{_\rho}-f_;{^\rho}{_\rho}\bar{g}^{\mu\nu}+f_;{^{\mu\nu}}\\
	& +\frac{1}{2}\bar{R} f\bar{g}^{\mu\nu}-\bar{R} f^{\mu\nu}+\bar{R}^{\mu\rho}f^\nu_\rho + \bar{R}^{\nu\rho}f^\mu_\rho-\bar{R}^{\mu\nu}f\\
	& -\bar{R}^{\rho\tau}f_{\rho\tau}\bar{g}^{\mu\nu}+2\bar{R}^{\mu\rho\nu\tau}f_{\rho\tau}\Bigr\}\\
	& + \frac{V}{4}\sqrt{\bar{g}}\{ f\bar{g}^{\mu\nu}-2 f^{\mu\nu}\}
	= 0.
\end{eqnarray}

In comparison, we next evaluate the linearized Einstein equation \labelcref{eqn:F2,eqn:63} for the physical metric $f_{\mu\nu}$. With $h_{\mu\nu;}{^\nu} = 0$ \cref{eqn:F2} simplifies to
\begin{eqnarray}\label{eqn:F3}
	G_{(1)\mu\nu}
	&= \frac{1}{2}\Bigl\{\bar{R}^\rho_\mu f_{\nu\rho}+\bar{R}^\rho_\nu f_{\mu\rho}-2\bar{R}_{\mu\rho\nu\tau}f^{\rho\tau}-\bar{R} f_{\mu\nu}\\
	&\hphantom{= \frac{1}{2}\Bigl\{}+\bar{R}^{\rho\tau}f_{\rho\tau}\bar{g}_{\mu\nu}
	-f_{\mu\nu;}{^\rho}{_\rho}-f_{;\mu\nu}+f_;{^\rho}{_\rho}\bar{g}_{\mu\nu}\Bigr\}.
\end{eqnarray}
Comparing with \cref{eqn:47C} one finds for the difference between the linearized Einstein equation and the mode equation \labelcref{eqn:BA}
\begin{eqnarray}\label{eqn:68A}
	&G_{(1)\mu\nu}+\frac{V}{M^2}f_{\mu\nu}-\frac{2}{M^2\sqrt{\bar{g}}}\Gamma^{(2)\rho\tau}_{\mu\nu}f_{\rho\tau}
	= \\
	&\left(\bar{R}^\rho_\mu-\frac{1}{2}\bar{R}\delta^\rho_\mu + \frac{V}{\bar{M}^2}\delta^\rho_\mu\right)f_{\rho\nu}+
	\left(\bar{R}^\rho_\nu-\frac{1}{2}\bar{R}\delta^\rho_\nu + \frac{V}{M^2}\right)f_{\rho\mu}\\
	&\qquad -\frac{1}{2} \left(\bar{R}_{\mu\nu}-\frac{1}{2}\bar{R}\bar{g}_{\mu\nu}+\frac{V}{M^2}\right) f.
\end{eqnarray}
If the background metric obeys the field equation,
\begin{equation}\label{eqn:68B}
	\bar{R}_{\mu\nu}-\frac{1}{2}\bar{R}\bar{g}_{\mu\nu}+\frac{V}{M^2}\bar{g}_{\mu\nu}
	= 0,
\end{equation}
the r.h.s. of \cref{eqn:68A} vanishes, such that \cref{eqn:47C} indeed yields the linearized Einstein equation for small deviations from the background solution.

We emphasize, however, that for background metrics not obeying the field equation \labelcref{eqn:68B} the linearized Einstein equations \labelcref{eqn:F2} should not be used for the definition of the mode functions. The correct equation, which also carries the information on the partial normalization of $G$, is the propagator equation \labelcref{eqn:M4} which entails \cref{eqn:M5} or \cref{eqn:BA}. The difference results from the fact that the first functional derivative of the effective action is given by
\begin{equation}\label{eqn:B.9}
	\frac{\delta\Gamma}{\delta g_{\mu\nu}}
	= -\frac{M^2}{2}g^{1/2}g^{\mu\rho}g^{\nu\tau}
	\left(R_{\rho\tau}-\frac{1}{2} R g_{\rho\tau}+\frac{V}{M^2}g_{\rho\tau}\right).
\end{equation}
The linearization of this expression yields the mode equation \labelcref{eqn:BA} or \labelcref{eqn:47C}. The linearized Einstein equation only involves the linearization of the last factor. Away from background geometries that solve the field equations the linearization of
\begin{equation}\label{eqn:B.10}
	(g^{1/2}g^{\mu\rho}g^{\nu\tau})_{(1)}
	= \frac{f}{2}
	\bar{g}^{\mu\rho}\bar{g}^{\nu\tau}
	-f^{\mu\rho}\bar{g}^{1/2}\bar{g}^{\nu\tau}-f^{\nu\tau}\bar{g}^{1/2}\bar{g}^{\mu\rho}
\end{equation}
contributes additional terms that account for the r.h.s. of \cref{eqn:68A}.

A simplification of the mode equation \labelcref{eqn:47C} occurs for background geometries with a vanishing Weyl tensor. For this purpose we express $\bar{R}_{\mu\rho\nu\sigma}$ in terms of the Weyl tensor $\bar{C}_{\mu\rho\nu\sigma}$,
\begin{eqnarray}\label{eqn:47D}
	\bar{R}_{\mu\rho\nu\tau}
	&= \bar{C}_{\mu\rho\nu\tau}+\frac{1}{2}(\bar{g}_{\mu\nu}\bar{R}_{\rho\tau}+\bar{g}_{\rho\tau}\bar{R}_{\mu\nu}\\
	&-\bar{g}_{\mu\tau}\bar{R}_{\nu\rho}-\bar{g}_{\nu\rho}\bar{R}_{\mu\tau})-\frac{1}{6}\bar{R}(\bar{g}_{\mu\nu}\bar{g}_{\rho\tau}-\bar{g}_{\mu\tau}\bar{g}_{\nu\rho}).
\end{eqnarray}
For a vanishing Weyl tensor, $\bar{C}_{\mu\rho\nu\sigma} = 0$, we can then replace in \cref{eqn:47A}
\begin{eqnarray}\label{eqn:50A}
	&\bar{g}^{\nu\tau} D^\rho D^\mu \to \bar{g}^{\nu\tau}\bar{R}^{\mu\rho}-\bar{R}^{\mu\rho\nu\tau}\\
	&\to \bar{g}^{\nu\tau}\bar{R}^{\mu\rho}-\frac{1}{2} (\bar{g}^{\mu\nu}\bar{R}^{\rho\tau}+\bar{g}^{\rho\tau}\bar{R}^{\mu\nu}-\bar{g}^{\mu\tau}\bar{R}^{\nu\rho}-\bar{g}^{\nu\rho}\bar{R}^{\mu\tau})\\
	&\qquad + \frac{1}{6}\bar{R}(\bar{g}^{\mu\nu}\bar{g}^{\rho\tau}-\bar{g}^{\mu\tau}\bar{g}^{\nu\rho}),
\end{eqnarray}
where the contribution $\sim D^\mu D^\rho$ is omitted. With this simplification the action of $\Gamma^{(2)}$ on $f_{\rho\tau}$ becomes
\begin{eqnarray}\label{eqn:50B}
	\Gamma^{(2)\mu\nu\rho\tau}
	&= -\frac{M^2}{8}\sqrt{\bar{g}}\bigl\{(\bar{g}^{\mu\rho}\bar{g}^{\nu\tau}+\bar{g}^{\nu\rho}\bar{g}^{\mu\tau}-2\bar{g}^{\mu\nu}\bar{g}^{\rho\tau})D^2\\
	&+\bar{g}^{\rho\tau}(D^\mu D^\nu +D^\nu D^\mu)\\
	&+\frac{1}{3}\bar{R}(\bar{g}^{\mu\nu}\bar{g}^{\rho\tau}-2\bar{g}^{\mu\rho}\bar{g}^{\nu\tau}-2\bar{g}^{\mu\tau}\bar{g}^{\nu\rho})\\
	&-\frac{2V}{M^2}(\bar{g}^{\mu\nu}\bar{g}^{\rho\tau}-\bar{g}^{\mu\rho}\bar{g}^{\nu\tau}-\bar{g}^{\mu\tau}\bar{g}^{\nu\rho})\bigr\}.
\end{eqnarray}

\section{Decomposition of unconstrained metric fluctuations into \texorpdfstring{$SO(3)$}{SO(3)} representations}
\label{sec:unconstrained metric}

For a homogeneous and isotropic background geometry \labelcref{eqn:H1} the unconstrained metric fluctuations $h_{\mu\nu}$ decompose with respect to the $SO(3)$-rotation group as four scalars, two divergence free vectors and the graviton. We have discussed in \cref{sec:mode decomposition} the decomposition of the physical metric fluctuations (two scalars, one vector and the graviton) and the gauge fluctuations (two scalars and one vector) separately. In this appendix we display more familiar decompositions of $h_{\mu\nu}$ and establish the connection to the decomposition employed in the present paper. We also describe the Bardeen potentials within the familiar decomposition and establish their connection to the physical metric fluctuations in the scalar and vector sector.

\subsection{Decomposition}

We start from the familiar decomposition of general metric fluctuations $h^\nu_\mu$ with respect to the rotation group. In Fourier space it is given by
\begin{eqnarray}\label{eqn:H4}
	h^0_0
	&= 2A,\\
	h^j_i
	&= a^2(\gamma^j_i + ik^j V_i + ik_i V^j -2k_i k^j B) + 2C\delta^j_i\\
	h^j_0
	&= a^2(W^j + ik^j D),
	\qquad
	h^0_j
	= -(W_j + ik_j D),
\end{eqnarray}
with
\begin{equation}\label{eqn:H5}
	\gamma^j_j = 0,
	\qquad
	k_j\gamma^j_i = 0,
	\qquad
	k_j V^j = 0,
	\qquad
	k_j W^j = 0.
\end{equation}
A restriction to fluctuations obeying $h^\nu_{\mu;\nu} = 0$ will be done later.

For an explicit relation between the four scalars $A,B,C,D$ and the metric components one may use the relations
\begin{eqnarray}\label{eqn:A7E}
	h^0_0
	&= 2A,
	\qquad
	ik_m h^m_0
	= -k^2 D,\\
	h^m_m
	&= 6C-2k^2 B,
	\qquad
	a^2 k^m k_j h^j_m = 2k^2(C-k^2 B),
\end{eqnarray}
such that
\begin{eqnarray}\label{eqn:AFF}
	C
	&= \frac{1}{4}\left(h^m_m -a^2\frac{k^m k_j}{k^2}h^j_m\right)\\
	B
	&= \frac{1}{4k^2}\left(h^m_m -3a^2\frac{k^m k_j}{k^2}h^j_m\right).
\end{eqnarray}
In particular, one has
\begin{equation}\label{eqn:A7G}
	h = 2A + 6C-2k^2 B,
	\qquad
	\tilde{b}^0_0 = \frac{3}{2}(A-C) + \frac{k^2}{2}B.
\end{equation}

In order to identify the ``gauge invariant part'' of the decomposition \labelcref{eqn:H4} we consider the inhomogeneous part of the gauge transformation
\begin{equation}\label{eqn:H6}
	\delta_\text{inh}h^\nu_\mu
	= -(\xi_{\mu;}{^\nu}+\xi^\nu{_{;\mu}}),
\end{equation}
which amounts to
\begin{eqnarray}\label{eqn:H7}
	\delta_\text{inh} h^0_0
	&= -2\partial_\eta\xi^0 -2\mathcal{H}\xi^0,\\
	\delta_\text{inh} h^j_0
	&= -\partial_\eta\xi^j -ik^j\xi_0,\\
	\delta_\text{inh} h^j_i
	&= -i(k_i\xi^j + k^j\xi_i)-2\mathcal{H}\delta^j_i\xi^0.
\end{eqnarray}
Comparison with \cref{eqn:H4} yields
\begin{eqnarray}\label{eqn:H8}
	&\delta A
	= -(\partial_\eta+\mathcal{H})\xi^0,
	\qquad
	&&\delta C
	= -\mathcal{H}\xi^0,\\
	&\delta D
	= \xi^0 -\partial_\eta\xi_L,
	\qquad
	&&\delta B
	= -\xi_L,\\
	&\delta V^i
	= -\xi^i_T,
	\quad
	\delta \gamma^j_i = 0,
	\qquad
	&&\delta W^i
	= -(\partial_\eta +2\mathcal{H})\xi^i_T,
\end{eqnarray}
with
\begin{equation}\label{eqn:H9}
	\xi_L
	= -i\frac{k_j\xi^j}{k^2},
	\qquad
	\xi^i = a^2(\xi^i_T + ik^i\xi_L),
	\qquad
	k_j\xi^j_T = 0.
\end{equation}
Writing
\begin{eqnarray}\label{eqn:H14}
	A
	&= \Psi +(\partial_\eta +\mathcal{H})(\partial_\eta B-D),\\
	C
	&= \Phi +\mathcal{H}(\partial_\eta B-D),\\
	W^j
	&= \Omega^j + (\partial_\eta + 2\mathcal{H})V^j
\end{eqnarray}
one observes that the Bardeen potentials \cite{Bar} $\Psi$ and $\Phi$, as well as $\Omega_j = W_j -\partial_\eta V_j$ and $\gamma^j_i$, are invariant under the inhomogeneous gauge transformations.

\subsection{Einstein equation}

In terms of these fields the linearized Einstein equations \labelcref{eqn:F2} involve the first variation of the Einstein tensor
\begin{eqnarray}\label{eqn:XA}
	&G_{(1)00}
	= 2k^2\phi + 6\mathcal{H}\partial_\eta C,\\
	&G_{(1)m0}
	= \frac{1}{2} k^2\Omega_m -2i k_m(\partial_\eta\phi-\mathcal{H}\Psi)\\
	&-(2\partial_\eta\mathcal{H}+\mathcal{H}^2)W_m -ik_m
	\bigl[2(\partial_\eta\mathcal{H}-\mathcal{H}^2)\partial_\eta B + 3\mathcal{H}^2 D\bigr],\\
	&G_{(1)mn}
	= \frac{1}{2}(\partial^2_\eta + 2\mathcal{H}\partial_\eta-4\partial_\eta\mathcal{H}-2\mathcal{H}^2 + k^2)\gamma_{mn}\\
	&\quad-\frac{i}{2}(\partial_\eta + 2\mathcal{H})(k_m\Omega_n + k_n\Omega_m)\\
	&\quad-i(2\partial_\eta\mathcal{H}+\mathcal{H}^2)(k_m V_n +k_n V_m) + g_1\delta_{mn}+g_2 k_m k_n,
\end{eqnarray}
with
\begin{eqnarray}\label{eqn:XB}
	g_1
	&= (2\mathcal{H}\partial_\eta + 4\partial_\eta\mathcal{H}+2\mathcal{H}^2 -k^2)\Psi\\
	&-(2\partial^2_\eta + 4\mathcal{H}\partial_\eta + 4\partial_\eta\mathcal{H}+2\mathcal{H}^2 + k^2)\phi\\
	&+2(\partial^2_\eta\mathcal{H}+\mathcal{H}\partial_\eta\mathcal{H})(D-\partial_\eta B),\\
	g_2
	&= \phi + \Psi + 2(2\partial_\eta\mathcal{H}+\mathcal{H}^2)B.
\end{eqnarray}
For a background geometry obeying the field equations (cf. \cref{eqn:68B,eqn:64B})
\begin{equation}\label{eqn:XC}
	\frac{4V}{M^2}
	= \bar{R} = \frac{6}{a^2}(\mathcal{H}^2 + \partial_\eta\mathcal{H})
\end{equation}
the linearized Einstein equation becomes (cf. \cref{eqn:63})
\begin{equation}\label{eqn:XD}
	G_{(1)\mu\nu}
	= -\frac{V}{M^2} h_{\mu\nu}
	= -\frac{3}{2a^2}
	(\mathcal{H}^2 + \partial_\eta\mathcal{H})\bar{g}_{\nu\rho} h^\rho_\mu.
\end{equation}

The different components read
\begin{align}\label{eqn:XE}
	&G_{(1)00}-3(\mathcal{H}^2 + \partial_\eta\mathcal{H})A = 0,\\
	&G_{(1)m0}+\frac{3}{2}(\mathcal{H}^2 + \partial_\eta\mathcal{H})(W_m + ik_m D) = 0,\label{eqn:XF}\\
	&G_{(1)mn}+\frac{3}{2}(\mathcal{H}^2 + \partial_\eta\mathcal{H})(\gamma_{mn}+ik_m V_n + ik_n V_m\\
	&\qquad-2k_m k_n B + 2C\delta_{mn}) = 0.\label{eqn:XG}
\end{align}
The first equation \labelcref{eqn:XE} yields
\begin{eqnarray}\label{eqn:XH}
	&2k^2\phi + 6\mathcal{H}\partial_\eta\phi-3(\mathcal{H}^2 + \partial_\eta\mathcal{H})\Psi\\
	&\quad +3(\mathcal{H}^2 -\partial_\eta\mathcal{H})(\partial_\eta-\mathcal{H})(\partial_\eta B-D) = 0.
\end{eqnarray}
For $V>0$ the solution of the field equation \labelcref{eqn:68B} is de Sitter space for which $\partial_\eta\mathcal{H} = \mathcal{H}^2$. For a background metric obeying the field equation the linearized Einstein equation \labelcref{eqn:XE} involves only the invariant Bardeen potentials,
\begin{equation}\label{eqn:XI}
	2k^2\phi + 6\mathcal{H}(\partial_\eta\phi-\mathcal{H}\Psi) = 0.
\end{equation}
Also \cref{eqn:XF} involves only ``gauge invariant fluctuations'',
\begin{equation}\label{eqn:XJ}
	\frac{1}{2} k^2\Omega_m -2ik_m(\partial_\eta\phi-\mathcal{H}\Psi) = 0,
\end{equation}
and similarly for \cref{eqn:XG},
\begin{eqnarray}\label{eqn:XK}
	&\frac{1}{2}(\partial^2_\eta + 2\mathcal{H}\partial_\eta + k^2)\gamma_{mn}-\frac{i}{2}
	(\partial_\eta + 2\mathcal{H})(k_m\Omega_n + k_n\Omega_m)\\
	&+\bigl[(2\mathcal{H}\partial_\eta + 6\mathcal{H}^2 -k^2)\Psi-(2\partial^2_\eta + 4\mathcal{H}\partial_\eta + k^2)\phi\bigr]\delta_{mn}\\
	&+(\phi + \Psi)k_m k_n = 0.
\end{eqnarray}

The solution of this system of equations is rather simple. (We only consider $k_m\neq 0$ here.) Multiplying \cref{eqn:XJ} with $k^m$ yields
\begin{equation}\label{eqn:XL}
	\partial_\eta\phi = \mathcal{H}\Psi,
\end{equation}
such that \cref{eqn:XI,eqn:XJ} require
\begin{equation}\label{eqn:XM}
	\phi = 0,
	\qquad
	\Psi = 0,
	\qquad
	\Omega_m = 0.
\end{equation}
The only non-vanishing mode is the graviton $\gamma_{mn}$ which obeys the standard field equation for massless excitations
\begin{equation}\label{eqn:XN}
	(\partial^2_\eta + 2\mathcal{H}\partial_\eta + k^2)\gamma_{mn}
	= 0.
\end{equation}
This mode equation is the same as for a massless scalar field and has been discussed extensively in the literature \cite{STA,GP,BST,AW,STAT,RSV}.

\subsection{Physical metric fluctuations}

For the ``physical degrees of freedom'' we impose $h^\nu_{\mu;\nu} = 0$. The components of the constraints for $f_{\mu\nu}$ are
\begin{eqnarray}\label{eqn:A7A}
	f^\nu_{0;\nu}
	&= \partial_\eta f^0_0 + ik_m f^m_0 + 3\mathcal{H}f^0_0 -\mathcal{H}f^m_m\\
	&= 2\partial_\eta A + 6\mathcal{H}(A-C) + 2\mathcal{H}k^2 B-k^2 D,
\end{eqnarray}
and
\begin{eqnarray}\label{eqn:A7B}
	f^\nu_{j;\nu}
	&= (\partial_\eta + 3\mathcal{H})f^0_j + ik_m f^m_j -\mathcal{H}\delta_{jm}f^m_0\\
	&= 2ik_j(C-k^2 B)-k^2 V_j -(\partial_\eta + 4\mathcal{H})(W_j + ik_j D).
\end{eqnarray}
This yields two constraints for the scalar fields
\begin{eqnarray}\label{eqn:A7C}
	(\partial_\eta + 4\mathcal{H})D
	&= 2(C-k^2 B)\\
	(\partial_\eta + 3\mathcal{H})A
	&= 3\mathcal{H}C-\frac{k^2}{2}(2\mathcal{H}B-D).
\end{eqnarray}
Together with the two defining equations \labelcref{eqn:H14} for $\phi$ and $\Psi$ they allow us to express $A,B,C$ and $D$ in terms of $\phi$ and $\Psi$.

The vector constraint from \cref{eqn:A7A} reads
\begin{equation}\label{eqn:A7D}
	k^2 V_j
	= -(\partial_\eta + 4\mathcal{H})W_j,
	\qquad
	k^2 V^j
	= -(\partial_\eta + 6\mathcal{H})W^j.
\end{equation}
This expresses the gauge invariant vector fluctuation $\Omega_j$ in terms of $W_j$,
\begin{equation}\label{eqn:F27A}
	\Omega_j = \frac{1}{k^2}[k^2 + \partial_\eta(\partial_\eta + 4\mathcal{H})]W_j.
\end{equation}

The relation between the Bardeen potentials and the scalar metric fluctuations $A,B,C,D$ is rather complex for the constraint $h^\nu_{\mu;\nu} = 0$. One has to eliminate two of the fields by using the constraint, and subsequently establish the relation between the two remaining scalar fluctuations and the gauge invariant potentials $\Phi$ and $\Psi$.

In the presence of the constraints \labelcref{eqn:A7C} we can relate $D$ to $A$ and $C$ as
\begin{equation}\label{eqn:M1a}
	D = 2\bigl[k^2 + 2\mathcal{H}(\partial_\eta + 4\mathcal{H})\bigr]^{-1}
	\bigl[(\partial_\eta + 3\mathcal{H})A-2\mathcal{H} C\bigr],
\end{equation}
and similar for $B$,
\begin{eqnarray}\label{eqn:M2a}
	B
	&= \frac{1}{k^2}\bigl[k^2 + 2\mathcal{H}\partial_\eta + 8\mathcal{H}^2 + 2\partial_\eta\mathcal{H}\bigr]^{-1}\\
	&\Bigl\{\bigl[k^2 + 6(\partial_\eta + 4\mathcal{H})\mathcal{H}\bigr]C\\
	&-2(\partial_\eta + 4\mathcal{H})(\partial_\eta + 3\mathcal{H})A\Bigr\}.
\end{eqnarray}
Inversely, $C$ and $A$ can be expressed in terms of $B$ and $D$,
\begin{eqnarray}\label{eqn:M3a}
	C
	&= k^2 B + \frac{1}{2}(\partial_\eta + 4\mathcal{H})D\\
	A
	&= (\partial_\eta + 3\mathcal{H})^{-1}\Bigl\{2\mathcal{H} k^2 B\\
	&\frac{1}{2}\bigl[k^2 + 3\mathcal{H}(\partial_\eta + 4\mathcal{H})\bigr]D\Bigr\}.
\end{eqnarray}
Using these expressions we can write the potentials $\Phi$ and $\Psi$ as functions of $B$ and $D$, e.g.
\begin{equation}\label{eqn:M4a}
	\Phi = (k^2 -\mathcal{H}\partial_\eta)B + \frac{1}{2}(\partial_\eta + 6\mathcal{H})D,
\end{equation}
with $\Psi$ a more lengthly expression.

Perhaps the most convenient setting keeps $A$ and $B$ as independent variables, with
\begin{equation}\label{eqn:M5a}
	D = 2\bigl[k^2 + 3\mathcal{H}(\partial_\eta + 4\mathcal{H})\bigr]^{-1}
	\bigl\{(\partial_\eta + 3\mathcal{H})A-2\mathcal{H} k^2 B\bigr\}
\end{equation}
and
\begin{eqnarray}\label{eqn:M6a}
	&C = \bigl[k^2 + 3\mathcal{H}(\partial_\eta + 4\mathcal{H}) + 3\partial_\eta\mathcal{H}\bigr]^{-1}\\
	&\times\Bigl\{(\partial_\eta +4\mathcal{H})(\partial_\eta +3\mathcal{H})A + \bigl[k^2 + (\partial_\eta +4\mathcal{H})\mathcal{H}\bigr]k^2 B\Bigr\}.
\end{eqnarray}
We can then express the Bardeen potentials $\Phi$ and $\Psi$ in terms of the metric components $A$ and $B$,
\begin{eqnarray}\label{eqn:M9a}
	\Phi
	&= \bigl[k^2 + 3\mathcal{H}(\partial_\eta + 4\mathcal{H}) + 3(\partial_\eta \mathcal{H})\bigr]^{-1}\\
	&\times\Bigl\{(\partial_\eta + 6\mathcal{H})(\partial_\eta +3\mathcal{H})A\\
	&+\bigl[k^4 + (\partial_\eta\mathcal{H})k^2 -3\mathcal{H}^2\partial_\eta^2 -6\mathcal{H}(\partial_\eta\mathcal{H}+2\mathcal{H}^2)\partial_\eta\bigr]B\Bigr\}
\end{eqnarray}
and
\begin{eqnarray}\label{eqn:M10a}
	\Psi
	&= \bigl[k^2 + 3 \mathcal{H}(\partial_\eta + 4 \mathcal{H}) + 3(\partial_\eta\mathcal{H})\bigr]^{-1}\\
	&\hphantom{={}}\times \Bigl\{\bigl[k^2 + 2\partial_\eta^2 + 11 \mathcal{H}\partial_\eta + 18 \mathcal{H}^2 + 9 (\partial_\eta\mathcal{H})\bigr] A\\
	&\hphantom{=\times\Bigl\{}- \bigl[k^2(\partial_\eta + \mathcal{H})(\partial_\eta + 4 \mathcal{H}) + 3\bigl\{\mathcal{H}\partial_\eta^2\\
	&\hphantom{=\times\Bigl\{}+ (5 \mathcal{H}^2 + \partial_\eta\mathcal{H})\partial_\eta + 4\mathcal{H}^3 + 2\mathcal{H}\partial_\eta\mathcal{H}\bigr\}\partial_\eta\bigr]B\Bigr\}.
\end{eqnarray}
These equations can be inverted in order to obtain $A$ and $B$, and consecutively als $C$ and $D$ as functions of $\Phi$ and $\Psi$. In contrast to longitudinal or Newtonian gauge, where $B = D = 0$, $\Phi = C$, $\Psi = A$, the relation between the metric components and the gauge invariant Bardeen potentials is rather complex since inversions of differential operators are needed. This makes the reconstruction of the metric correlation from the correlations of $\Phi$ and $\Psi$ rather cumbersome for the covariant gauge $h^{\nu}_{\mu;\nu} = 0$.

\subsection{Relation between decomposition of physical metric fluctuations and unconstrained metric fluctuations}

The relation with the decomposition of the physical metric in sect.VII can be made by the identifications
\begin{eqnarray}\label{eqn:R1}
	f_{00}
	&= -2a^2 A = a^2 [{\kappa} +\epsilon],\\
	f_{0m}
	&= a^2(W_m + ik_m D)\\
	&= a^2\left[W_m -\frac{ik_m}{k^2}(\partial_\eta + 4\mathcal{H})({\kappa}+\epsilon)-\frac{ik_m}{k^2}\mathcal{H}\sigma\right],\\
	f_{mn}
	&= a^2(\gamma_{mn}+ik_m V_n + ik_n V_m -2k_m k_n B + 2\delta_{mn}C)\\
	&= a^2 \biggl[\gamma_{mn}-\frac{i}{k^2}(\partial_\eta + 4\mathcal{H})(k_m W_n + k_n W_m)\\
	&\hphantom{= a_2 \biggl[}+ \frac{1}{2k^2}(\partial_\eta + 4\mathcal{H})^2\left(\delta_{mn}-\frac{3k_m k_n}{k^2}\right)({\kappa}+\epsilon)\\
	&\hphantom{= a_2 \biggl[}+ \frac{1}{2}(\delta_{mn}-\frac{k_m k_n}{k^2})({\kappa}+\epsilon)\\
	&\hphantom{= a_2 \biggl[}+ \frac{1}{2k^2}\Bigl(2 k_m k_n + \bigl(\delta_{mn}-\tfrac{3k_m k_n}{k^2}\bigr)\\
	&\hphantom{= a_2 \biggl[\frac{1}{2k^2}\Bigl(}\times (k^2 + \mathcal{H}\partial_\eta + \partial_\eta\mathcal{H}+4\mathcal{H}^2)\Bigr)\sigma\biggr],
\end{eqnarray}
We infer
\begin{equation}\label{eqn:R3}
	V_m
	= -\frac{1}{k^2}(\partial_\eta + 4\mathcal{H})W_m,
\end{equation}
while in the scalar sector one has
\begin{eqnarray}\label{eqn:R4}
	A
	&= -\frac{1}{2}(\kappa + \epsilon),\\
	B
	&= \frac{1}{4k^4}\bigl\{[k^2 + 3(\partial_\eta + 4\mathcal{H})^2](\kappa + \epsilon)\\
	&+[k^2 + 3(\partial_\eta + 4\mathcal{H})\mathcal{H}]\sigma\bigr\},\\
	C
	&= \frac{1}{4k^2}\{[k^2 + (\partial_\eta + 4\mathcal{H})^2](\kappa +\epsilon)\\
	&+[k^2 + (\partial_\eta + 4\mathcal{H})\mathcal{H}]\sigma\},\\
	D
	&= -\frac{1}{k^2}\{(\partial_\eta + 4\mathcal{H})(\kappa + \epsilon) + \mathcal{H}\sigma\}.
\end{eqnarray}
One can verify that \cref{eqn:R4} obeys the constraints \labelcref{eqn:A7C}.

For the computation of the Bardeen potentials $\Phi$ and $\Psi$ and one needs the combination
\begin{eqnarray}\label{eqn:R5}
	\partial_\eta B-D
	= \frac{1}{4k^4}\bigl\{&[3\partial_\eta(\partial_\eta + 4\mathcal{H})^2 + 4k^2 (\partial_\eta + 4\mathcal{H})\\
	&+ k^2\partial_\eta](\kappa + \epsilon) + [3\partial_\eta(\partial_\eta + 4\mathcal{H})\mathcal{H}\\
	&+k^2(\partial_\eta + 4\mathcal{H})]\sigma\bigr\}.
\end{eqnarray}
The relation between $\Phi$ and $\Psi$ and the fields $\kappa$ and $\sigma$ depends on the choice of $\epsilon$,
\begin{eqnarray}\label{eqn:F42A}
	\Phi
	= \frac{1}{4k^4}\bigl\{&[k^4 + k^2(\partial_\eta\mathcal{H}) + k^2\partial_\eta(\partial_\eta + 3\mathcal{H})\\
	&- 3\mathcal{H}\partial_\eta(\partial_\eta + 4\mathcal{H})^2](\kappa +\epsilon)\\
	&+[k^4 + k^2(\partial_\eta\mathcal{H})-3\mathcal{H}\partial_\eta(\partial_\eta + 4\mathcal{H})\mathcal{H}]\sigma\bigr\},
\end{eqnarray}
and
\begin{eqnarray}\label{eqn:R6}
	\Psi
	= -\frac{1}{4k^4}\bigl\{&[2k^4 + k^2(\partial_\eta + \mathcal{H})(5\partial_\eta + 16\mathcal{H})\\
	&+ 3(\partial_\eta + \mathcal{H})\partial_\eta(\partial_\eta + 4\mathcal{H})^2](\kappa + \epsilon)\\
	&+ (\partial_\eta + \mathcal{H})[3\partial_\eta(\partial_\eta + 4\mathcal{H})\mathcal{H}+k^2(\partial_\eta + 4\mathcal{H})]\sigma\bigr\}.
\end{eqnarray}
For maximally symmetric spaces we may use the choice \labelcref{eqn:Z9A} for $\epsilon$ and employ the relation \labelcref{eqn:Z9B}. In any case, the relation between the Bardeen potentials and the metric components $\kappa$ and $\sigma$ remains rather involved.

\section{Second functional derivative and propagator equation}
\renewcommand{\theequation}{F.\arabic{equation}}

In this appendix we recall a few properties of second functional derivatives that are useful for the derivation of the propagator equation. In particular, we address the effects of a change in the field basis for the propagator equation.

For a given complex field $\varphi_a(\eta,\vec{k})$ in Fourier space the quadratic effective action takes the general form
\begin{equation}\label{eqn:M1-a}
	\Gamma_2 = \int_{\eta,k}\varphi^\ast_a(\eta,\vec{k})\mathcal{A}_{ab}
	(\vec{k};\eta,\eta^\prime)\varphi_b(\eta^\prime,\vec{k}).
\end{equation}
The $k$-integral comprises for every $\vec{k}$ both contributions from $\varphi(\vec{k})$ and $\varphi(-\vec{k})$, and we have to remember that these fields are not independent, i.e. $\varphi_a(\eta,-\vec{k}) = \varphi_a^\ast(\eta,\vec{k})$. As a consequence, the second functional derivative reads
\begin{eqnarray}\label{eqn:M2-a}
	\Gamma^{(2)}_{ab}(\eta,\vec{k};\eta^\prime,\vec{k}^\prime)
	&= \frac{\partial^2\Gamma_2}{\partial\varphi^\ast_a(\eta,\vec{k})\partial\varphi_b(\eta^\prime,\vec{k}^\prime)}\\
	&= \delta(k-k^\prime)
	\bigl[\mathcal{A}_{ab}(\vec{k};\eta,\eta^\prime)+\mathcal{A}_{ba}(-\vec{k};\eta^\prime,\eta)\bigr].
\end{eqnarray}
For $\mathcal{A}_{ab}(-\vec{k},\eta,\eta^\prime) = \mathcal{A}_{ab}(\vec{k},\eta,\eta^\prime)$ only the symmetric part of $\mathcal{A}$ contributes to $\Gamma^{(2)}$. This is realized in our case where $\mathcal{A}$ only involves even powers of $k_m$ through projectors onto the modes $\gamma_{mn},W_m,\kappa$ and $\sigma$, multiplied with operators that depend on $k^2$.

Furthermore, $\mathcal{A}_{ab}$ turns out to be a purely imaginary differential operator,
\begin{equation}\label{eqn:M3-a}
	\mathcal{A}(\vec{k};\eta^\prime,\eta) = i\delta(\eta-\eta^\prime)D_{ab}(\vec{k}, \eta,\partial_\eta).
\end{equation}
Writing
\begin{equation}\label{eqn:251Aa}
	D_{ab}(\vec{k},\eta,\partial_\eta) = D^{(0)}_{ab}(\vec{k},\eta) + D^{(1)}_{ab}
	(\vec{k},\eta)\partial_\eta + D^{(2)}_{ab}(\vec{k},\eta)\partial^2_\eta + \dots
\end{equation}
one infers
\begin{equation}\label{eqn:247A}
	\Gamma^{(2)}_{ab}(\eta,\vec{k};\eta^\prime,\vec{k}^\prime) = i\delta(\eta-\eta^\prime)\delta(k-k^\prime)\tilde{D}_{ab}(\vec{k},\eta,\eta^\prime),
\end{equation}
with
\begin{eqnarray}\label{eqn:251B}
	&\tilde{D}_{ab}(\vec{k},\eta,\eta^\prime) = \bigl[D^{(0)}_{ab}(\vec{k},\eta) + D^{(0)}_{ba}(\vec{k}, \eta)\bigr]\\
	&\qquad +\bigl[D^{(1)}_{ab}(\vec{k},\eta)-D^{(1)}_{ba}(\vec{k},\eta)\bigr]\partial_\eta-\partial_\eta D^{(1)}_{ba}(\vec{k},\eta)\\
	&\qquad +\bigl[D^{(2)}_{ab}(\vec{k},\eta) + D^{(2)}_{ba}(\vec{k},\eta)\bigr]\partial^2_\eta + 2\partial_\eta D^{(2)}_{ba}
	(\vec{k},\eta)\partial_\eta\\
	&\qquad + \partial^2_\eta D^{(2)}_{ba}(\vec{k},\eta) + \dots
\end{eqnarray}
This structure can be most easily visualized as a result of partial integration of the terms involving $\varphi^\ast_b(-\vec{k})\varphi_a(-\vec{k})$. For example, one has the following associations
\begin{eqnarray}\label{eqn:251C}
	&D
	= a^2\partial^2_\eta
	&&\to &&\tilde{D}
	= a^2\bigl(\partial^2_\eta + (\partial_\eta + 2\mathcal{H})^2\bigr),\\
	&D
	= 2\mathcal{H} a^2\partial_\eta
	&&\to &&\tilde{D}
	= -a^2 (4\mathcal{H}^2 + 2\partial_\eta\mathcal{H}),\\
	&D
	= a^2 (\partial^2_\eta + 2\mathcal{H}\partial_\eta)
	&&\to &&\tilde{D}
	= 2a^2(\partial^2_\eta + 2\mathcal{H}\partial_\eta).
\end{eqnarray}

The propagator equation $\Gamma^{(2)} \, G = E$ can be solved in an arbitrary field basis. Consider unconstrained fields $\varphi_a$ where $E = \delta^b_a$,
\begin{equation}\label{eqn:A18-1}
	\frac{\partial^2\Gamma}{\partial\varphi_a\partial\varphi_b}\langle \varphi_b\varphi_c\rangle_c = \delta^a_c.
\end{equation}
For a linear regular transformation
\begin{equation}\label{eqn:A18-2}
	\varphi_a
	= \mathcal{A}_{ab}\psi_b
\end{equation}
this translates to
\begin{equation}\label{eqn:A18-3}
	\frac{\partial^2\Gamma}{\partial\psi_a\partial\psi_b}
	\langle \psi_b\psi_c\rangle_c = \delta^a_c.
\end{equation}

We will encounter field transformations
\begin{equation}\label{eqn:251D}
	\varphi_a(\vec{k},\eta)
	= \mathcal{B}_{ac}(\vec{k},\eta,\partial_\eta)\psi_c(\vec{k},\eta)
\end{equation}
that are not necessarily regular. We still can first evaluate the correlation function in the $\psi$-basis and subsequently use
\begin{eqnarray}\label{eqn:251E}
	&G^{(\varphi)}_{ab}(\eta,\vec{k};\eta^\prime,\vec{k}^\prime) = \langle \varphi_a(\eta,\vec{k})\varphi^\ast_b(\eta^\prime,\vec{k})\rangle_c\\
	&= \mathcal{B}_{ac}(\vec{k},\eta,\partial_\eta)\mathcal{B}^\ast_{bd}(\vec{k},\eta^\prime,\partial_{\eta^\prime})
	\langle\psi_c(\eta,\vec{k})\psi^\ast_\alpha(\eta^\prime,\vec{k}^\prime)\rangle_c\\
	&= \mathcal{B}_{ac}(\vec{k},\eta,\partial_\eta)\mathcal{B}^\ast_{bd}
	(\vec{k},\eta^\prime,\partial_{\eta^\prime})G^{(\psi)}_{cd}(\eta,\vec{k};\eta^\prime,\vec{k}^\prime).
\end{eqnarray}
This situation is realized if we want to compute the metric correlation from the propagators of the physical fluctuations $\gamma_{m n}, W, \kappa$ and $\sigma$. The latter are represented here by $\psi$, while the relation between the metric components $\phi$ and $\psi$ is given by $h_{\mu\nu} = t_{\mu\nu} + s_{\mu\nu}$ and the expansion \labelcref{eqn:Z1a}.

\printbibliography

\end{multicols}

\end{document}